\documentclass[pdflatex,sn-basic,sort,compress]{sn-jnl}

\usepackage{graphicx}%
\usepackage{multirow}%
\usepackage{amsmath,amssymb,amsfonts}%
\usepackage{amsthm}%
\usepackage{mathrsfs}%
\usepackage[title]{appendix}%
\usepackage{xcolor}%
\usepackage{textcomp}%
\usepackage{manyfoot}%
\usepackage{booktabs}%
\usepackage{algorithm}%
\usepackage{algorithmicx}%
\usepackage{algpseudocode}%
\usepackage{listings}%
\usepackage{bm}%

\theoremstyle{thmstyleone}%
\theoremstyle{thmstyletwo}%

\theoremstyle{thmstylethree}%

\begin{document}

\title[Spin-orbit coupling]{Spin-orbit coupling of the primary body in a binary asteroid system}


\author[1,2]{\fnm{Hanlun} \sur{Lei}}\email{leihl@nju.edu.cn}

\affil[1]{\orgdiv{School of astronomy and space science}, \orgname{Nanjing University}, \orgaddress{\city{Najing}, \postcode{210023}, \country{China}}}

\affil[2]{\orgdiv{Key Laboratory of Modern Astronomy and Astrophysics in Ministry of Education}, \orgname{Nanjing University}, \orgaddress{\city{Nanjing}, \postcode{210023}, \country{China}}}


\abstract{Spin-orbit coupling is widespread in binary asteroid systems and it has been widely studied for the case of ellipsoidal secondary. Due to angular momentum exchange, dynamical coupling is stronger when the orbital and rotational angular momenta are closer in magnitudes. Thus, the spin-orbit coupling effects are significantly different for ellipsoidal secondaries and primaries. In the present work, a high-order Hamiltonian model in terms of eccentricity is formulated to study the effects of spin-orbit coupling for the case of ellipsoidal primary body in a binary asteroid system. Our results show that the spin-orbit coupling problem for the ellipsoidal primary holds two kinds of spin equilibrium, while there is only one for the ellipsoidal secondary. In particular, 1:1 and 2:3 spin-orbit resonances are further studied by taking both the classical pendulum approximation as well as adiabatic approximation (Wisdom's perturbative treatment). It shows that there is a critical value of total angular momentum, around which the pendulum approximation fails to work. Dynamical structures are totally different when the total angular momentum is on two sides of the critical value.}





\maketitle

\section{Introduction}
\label{Sect1}

Spin-orbit resonance happens if the rotational period of a considered object is commensurable with its orbital period \citep{Goldreich1966Spin,peale1977rotation}. In our Solar system, spin-orbit resonance is widespread and it plays an important role in sculpting dynamical structures of rotation. The well-known examples are the Mercury and our Moon. In particular, the Mercury is locked inside the 3:2 spin-orbit resonance \citep{lemaitre20063,peale1965rotation} and our Moon is locked inside a synchronous (1:1) resonance with the Earth \citep{peale1969generalized}. It is known that most regular satellites around giant planets in the Solar System are locked in synchronous resonances, with stable orbital and spin evolution. However, some natural satellites are found to have chaotic rotations, e.g. Hyperion \citep{wisdom1984chaotic}, Nyx and Hydra \citep{correia2015spin}. Additionally, spin-orbit resonance is a common phenomenon existing in binary asteroid systems \citep{pravec2016binary,naidu2015near,scheeres2006dynamical,pravec2019asteroid}.

The classical model for describing spin-orbit resonances corresponds to a restricted case where the rotational angular momentum is negligible compared to the orbital angular momentum. Thus, it is assumed that the mutual orbits are decoupled from rotation, and such an approximation works well for the secondary bodies under Sun-planet and planet-satellite systems \citep{Goldreich1966Spin,Celletti1990AnalysisI,Celletti1990AnalysisII}. Under such a classical model, \citet{wisdom1984chaotic} investigated the onset of chaos through Chirikov's criterion \citep{chirikov1979universal} and they derived a critical value for asphericity parameter $\alpha_c$, above which the large-scale chaos may occur. \citet{celletti2000hamiltonian} studied the stability of periodic orbits under a nearly-integrable Hamiltonian model. \citet{flynn2005second} took advantage of Lie-series transformation to formulate a second-order Hamiltonian model for describing spin-orbit resonances. In particular, they derived an integral of motion for both resonant and non-resonant configurations \citep{flynn2005second}. According to pendulum approximations, \citet{jafari2015widespread,jafari2016chirikov} derived explicit expressions of resonant half-width in terms of eccentricity and asphericity parameter for spin-orbit resonances and they updated the Chirikov criterion given in \citet{wisdom1984chaotic}. Under a fourth-order Hamiltonian model, \citet{nadoushan2016geography} provided a genealogy for spin-orbit resonances and they provided an analytical estimation for amplitudes of a set of spin-orbit-spin resonances. In addition, secondary spin-orbit resonances inside the synchronous resonance have been studied under the classical spin-orbit model. In this respect, \citet{wisdom2004spin} took a perturbative method to study the secondary 3:1 resonance of Enceladus. In particular, they discussed tidal dissipation inside secondary resonance and found that forced libration due to secondary resonance may be large and thus would imply large tidal heating. \citet{gkolias2016theory,gkolias2019accurate} formulated high-order normal Hamiltonian model by means of Lie-series transformation theory \citep{1966Theory,1969Canonical} and their analytical theory can fully reproduce the dynamics of the 3:1, 2:1 and 1:1 secondary resonances inside the synchronous spin-orbit primary resonance.

In binary asteroid systems, rotation and orbit are strongly coupled in evolution and it is necessary to introduce the spin-orbit coupling model. \citet{naidu2015near} adopted the technique of Poincar\'e sections to explore coupled motions of rotation and translation for binary asteroid systems. Under the planar ellipsoid-ellipsoid and ellipsoid-sphere configurations, \citet{hou2017note} updated spin-orbit coupled Hamiltonian models and systematically studied spin-orbit, spin-spin and spin-orbit-spin resonances. They found that the resonance center may change with the mass ratio and the mutual distance between two asteroids. For the spin-orbit coupling problem, \citet{jafari2023surfing} formulated a fourth-order-and-degree Hamiltonian model, which determines a 2-DOF dynamical model depending on the total angular momentum. In particular, they provided a criterion for dynamical closeness of asteroid pairs based on the angular momentum ratio between rotation and translation.

Regarding the spin-orbit coupling problem, the dynamics are significantly different when the mass ratio of asteroid pairs and mutual distance are varied, as pointed out by \citet{hou2017note}. It has been widely investigated for the spin-orbit coupling for the secondary in a binary asteroid system and it is shown that there is no qualitative difference compared to the classical restricted spin-orbit resonances \citep{hou2017note, wang2020secondary, jafari2023surfing}. However, it is totally different for the primary. Because of primary's strong coupling between rotations and orbits, the assumptions of invariant orbit made in the traditional model are no longer appropriate in the binary asteroid system. Regarding this topic, spin-orbit resonances and their dynamical stability are studied by taking advantage of the approach of periodic orbits in \citet{wang2022stability}. In theory, we know less about the primary's spin-orbit coupling. To this end, in this work we focus on the dynamics of spin-orbit resonances for the primary body in a binary asteroid system.

The remaining part of this work is organised as follows. In Sect. \ref{Sect2}, two equivalent dynamical models for the spin-orbit coupling problem are formulated and numerical analysis are performed in Sect. \ref{Sect3}. In Sect. \ref{Sect4}, elliptic expansions of Hamiltonian function are made and Sect. \ref{Sect5} discusses the pendulum approximations for spin-orbit resonances. Perturbative treatments are adopted in Sect. \ref{Sect6} to formulate resonant Hamiltonian model and conclusions are summarised in Sect. \ref{Sect7}.

\section{Dynamical model}
\label{Sect2}

In this work, we concentrate on the spin-orbit coupling of the primary body under a binary asteroid system. The primary asteroid with mass $m_p$ is of triaxial ellipsoid, the secondary asteroid with mass $m_s$ is of spherical shape and they move around each other on perturbed Keplerian orbits under their mutual gravity field. The reduced mass of system is denoted by $m = \frac{m_p m_s}{m_p + m_s}$. Due to the spin-orbit coupling, angular momentum may exchange between orbit and rotation to conserve the total angular momentum, leading to the fact that their mutual orbit is changed \citep{hou2017note,jafari2023surfing}. It means that the semimajor axis, eccentricity and longitude of pericentre for describing orbits are varied in the spin-orbit coupling problem. This is distinctly different from the classical spin-orbit problem with an invariant orbit \citep{Goldreich1966Spin}.

\begin{figure*}
\centering
\includegraphics[width=0.5\columnwidth]{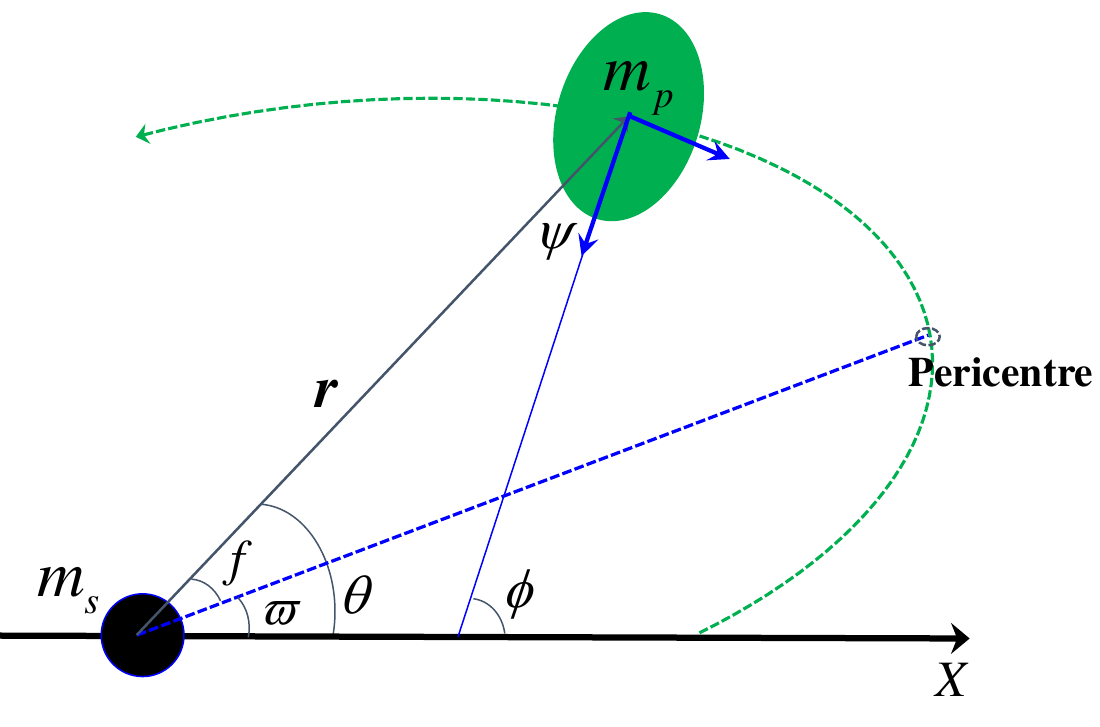}
\caption{Schematic diagram for the relative geometry of spin-orbit coupling in binary asteroid systems. The mass of the primary is denoted by $m_p$ and that of the secondary is denoted by $m_s$. The radius vector from the secondary to the primary is $\bm r$, the true anomaly is $f$ and the longitude of pericentre is $\varpi$, and the true longitude is $\theta = f + \varpi$. In the spin-orbit coupling problem, the longitude of pericentre $\varpi$ is varied. The rotational angle of the primary is denoted by $\phi$ and the relative angle between the line joining binary asteroids and the longest semi-major axis of the primary is $\psi = \phi - \theta = \left(\phi - \varpi\right) - f$. All the angles defined are measured under a reference frame centred at the secondary.}
\label{Fig1}
\end{figure*}

In the secondary-centred reference frame, Fig. \ref{Fig1} depicts the relative geometry of spin-orbit coupling problem considered in this work, where the true longitude $\theta = \varpi + f$ describes translation and the angle $\phi$ describes rotation. The mutual distance between the primary and secondary is denoted by $r$, and the physical semi-major axes of the primary are denoted by $a_p$, $b_p$ and $c_p$, satisfying $a_p \ge b_p \ge c_p$. Under the homogeneous assumption, the primary's moments of inertia along principal axes are computed by
\begin{equation*}
\left\{I_1,I_2,I_3\right\}= \frac{1}{5}m_p \left\{b_p^2 + c_p^2,a_p^2 + c_p^2,a_p^2 + b_p^2\right\}.
\end{equation*}
To simplify the spin-orbit coupling model, the following two assumptions are made \citep{jafari2023surfing, hou2017note}: (a) the primary asteroid spins about its principal axis of inertia, and (b) the principal axis of inertia is perpendicular to the mutual orbit plane. The mutual gravitational potential under the binary asteroid system truncated at the fourth order in $a_p/r$ is expressed as \citep{jafari2023surfing}\footnote{It should be noted that there is a typo for the term associated with $C_{42}$ in the expression of potential function given in \citet{jafari2023surfing}.}
\begin{equation}\label{Eq0}
\begin{aligned}
U (r,\theta,\phi) =&  - {{\cal G} m_p m_s} \left[ {\frac{1}{r} - \frac{{a_p^2{C_{20}}}}{{2{r^3}}} + \frac{{3a_p^4{C_{40}}}}{{8{r^5}}}} + {\left( {\frac{{3a_p^2{C_{22}}}}{{{r^3}}} - \frac{{15a_p^4{C_{42}}}}{{2{r^5}}}} \right)\cos \left( {2\phi  - 2\theta } \right)}\right.\\
&+ \left. \frac{{105a_p^4{C_{44}}}}{{{r^5}}}\cos \left( {4\phi  - 4\theta } \right) \right],
\end{aligned}
\end{equation}
where ${\cal G}$ is the universal gravitational constant. The dynamical model up to the eighth order in $a_p/r$ is discussed in Appendix. The spherical harmonics coefficients $C_{ij}$ are defined by \citep{balmino1994gravitational}
\begin{equation*}
\begin{aligned}
{C_{20}} &= \frac{1}{{5a_p^2}}\left( {c_p^2 - \frac{{a_p^2 + b_p^2}}{2}} \right),\quad {C_{22}} = \frac{1}{{20a_p^2}}\left( {a_p^2 - b_p^2} \right)\\
{C_{40}} &= \frac{{15}}{7}\left( {C_{20}^2 + 2C_{22}^2} \right),\;{C_{42}} = \frac{5}{7}{C_{20}}{C_{22}},\;{C_{44}} = \frac{5}{{28}}C_{22}^2.
\end{aligned}
\end{equation*}
As a result, the Hamiltonian function, governing the orbital and rotational evolution of the primary body, can be written as the sum of rotational and transnational kinetic energy and potential energy,
\begin{equation}\label{Eq1}
\begin{aligned}
{\cal H} =& T + U\\
=& \frac{1}{2} m \left({\dot r}^2 + r^2{\dot\theta}^2 \right) + \frac{1}{2}I_3{\dot\phi}^2+U\\
=& \frac{p_r^2}{2m} + \frac{p_\psi ^2}{2 m r^2} + \frac{1}{2I_3}\left(G_{\rm tot} + {p_\psi}\right)^2\\
&- {{\cal G}m_p m_s} \left[ { \frac{1}{r}- \frac{{a_p^2{C_{20}}}}{{2{r^3}}} + \frac{{3a_p^4{C_{40}}}}{{8{r^5}}} + \frac{{105a_p^4{C_{44}}}}{{{r^5}}}\cos 4\psi } \right.\\
&\left. { + \left( {\frac{{3a_p^2{C_{22}}}}{{{r^3}}} - \frac{{15a_p^4{C_{42}}}}{{2{r^5}}}} \right)\cos 2\psi } \right]
\end{aligned}
\end{equation}
where $p_r = m \dot r$ is the conjugate moment of $r$, the angle $\psi$ is $\psi = \phi - \theta$, its conjugate moment is $p_{\psi} = - p_{\theta} = - m {r^2}\dot \theta$, and the total angular moment is a conserved quantity, given by
\begin{equation*}
{G_{\rm tot}} = {p_\theta} + {p_\phi},
\end{equation*}
with ${p_\theta} = m r^2 \dot\theta$ and ${p_\phi} = I_3 \dot\phi$ as the orbital and rotational angular momentum.

The Hamiltonian (\ref{Eq1}) determines a 2-DOF dynamical model, depending on the total angular momentum $G_{\rm tot}$. The Hamiltonian canonical relations \citep{morbidelli2002modern} leads to the equations of motion, given by
\begin{equation}\label{Eq2}
\begin{aligned}
\dot r &= \frac{{\partial {\cal H}}}{{\partial {p_r}}},\quad {{\dot p}_r} =  - \frac{{\partial {\cal H}}}{{\partial r}},\\
\dot \psi  &= \frac{{\partial {\cal H}}}{{\partial {p_\psi }}},\quad {{\dot p}_\psi } =  - \frac{{\partial {\cal H}}}{{\partial \psi }}.
\end{aligned}
\end{equation}
In the Hamiltonian model, the state is denoted by ${\bm X} = \left(r,p_r,\psi,p_{\psi}\right)^T$. Usually, it is convenient for us to describe the spin-orbit state by means of ${\bm Z} = \left( {a, e, f ,\phi-\varpi ,\dot \phi} \right)^T$. It is not difficult to realise mutual transformations between $\bm X$ and $\bm Z$.

For convenience of computation, we adopt the following units of length, mass and time:
\begin{equation*}
[L] = {a_p},\;[M] = \frac{{{m_p}{m_s}}}{{{m_p} + {m_s}}},\;[T] = \sqrt {{{{\cal G}\left( {{m_p} + {m_s}} \right)} \mathord{\left/{\vphantom {{{\cal G}\left( {{m_p} + {m_s}} \right)} {a_p^3}}} \right.\kern-\nulldelimiterspace} {a_p^3}}}. 
\end{equation*}
Under the normalised system of units, the gravitational parameter $\mu = {\cal G}(m_p+m_s)$ is equal to unity and ${\cal G} m_p m_s = 1$. In the entire work, we use LU to stand for unit of length and TU to stand for unit of time. It is noted that the system of normalised units adopted here is in agreement with that in \citet{jafari2023surfing} (see the statement before Eq. (14) of their work), but it is different from that shown in \citet{hou2017note}.

Let us define the primary-to-secondary mass ratio as $\alpha_m = m_p/m_s > 1$. As a consequence, the normalised mass of the primary and secondary are specified by
\begin{equation*}
{m_s} = 1 + \frac{1}{{{\alpha _m}}} \in \left( {1,2} \right),\quad
{m_p} = 1 + {\alpha _m} \in \left( {2,\infty } \right),
\end{equation*}
and the total angular momentum can be expressed as
\begin{equation}\label{Eq5}
\begin{aligned}
{G_{\rm tot}} &= {p_\phi } + {p_\theta }\\
&= {I_3}\dot \phi  + m r^2\dot\theta\\
&= \frac{1}{5} m_p \left( {a_p^2 + b_p^2} \right)\dot \phi + m\sqrt{\mu a (1-e^2)}\\
&= \frac{1}{5}\left( {1 + {\alpha _m}} \right)\left( {1 + b_p^2} \right)\dot \phi  + \sqrt {a\left( {1 - {e^2}} \right)}. 
\end{aligned}
\end{equation}
In practice, we take the reference semimajor axis $a_{\rm ref}$ and the reference eccentricity $e_{\rm ref}$ to characterise the total angular momentum by
\begin{equation}\label{Eq6}
{G_{\rm tot}} = \frac{1}{5}\left( {1 + {\alpha _m}} \right)\left( {1 + b_p^2} \right) {\dot\phi}_{\rm ref}  + \sqrt {a_{\rm ref}\left( {1 - {e_{\rm ref}^2}} \right)}
\end{equation}
where the reference spin frequency is taken as ${\dot\phi}_{\rm ref}=n_{\rm ref}$ for the synchronous (1:1) spin-orbit resonance, ${\dot\phi}_{\rm ref}=1.5 n_{\rm ref}$ for the 2:3 resonance and ${\dot\phi}_{\rm ref}= 0.5 n_{\rm ref}$ for the 2:1 resonance (i.e., at the nominal location of resonance centre). The reference mean motion is computed from $n_{\rm ref}^2 a_{\rm ref}^3 = \mu$. Please refer to Table \ref{Tab1} for physical parameters of those known doubly synchronous binary asteroid systems. It is observed that the mass ratio of the primary and secondary is not far from unity except the binary system (624) Hektor. It should be noted that $a_{\rm ref}$ and $e_{\rm ref}$ are just two parameters to characterise the total angular momentum. If we change $a_{\rm ref}$ (fixing $e_{\rm ref}$), both the semi-major axis and the spin frequency of the primary are varied.

Unless otherwise stated, we take the following normalised parameters of system to perform all simulations in this work:
\begin{equation*}
a_p = 1.0,\quad b_p = 0.95,\quad c_p = 0.85,\quad \alpha_m = 10.
\end{equation*}
Please refer to Table (\ref{TabA1}) for the harmonics coefficients of the ellipsoidal primary. The moment of inertia along the shortest axis is equal to $I_3=4.1855$. The normalised mass of the primary is $m_p=11$ and the normalised mass of the secondary is $m_s=1.1$.

\begin{figure*}
\centering
\includegraphics[width=0.49\columnwidth]{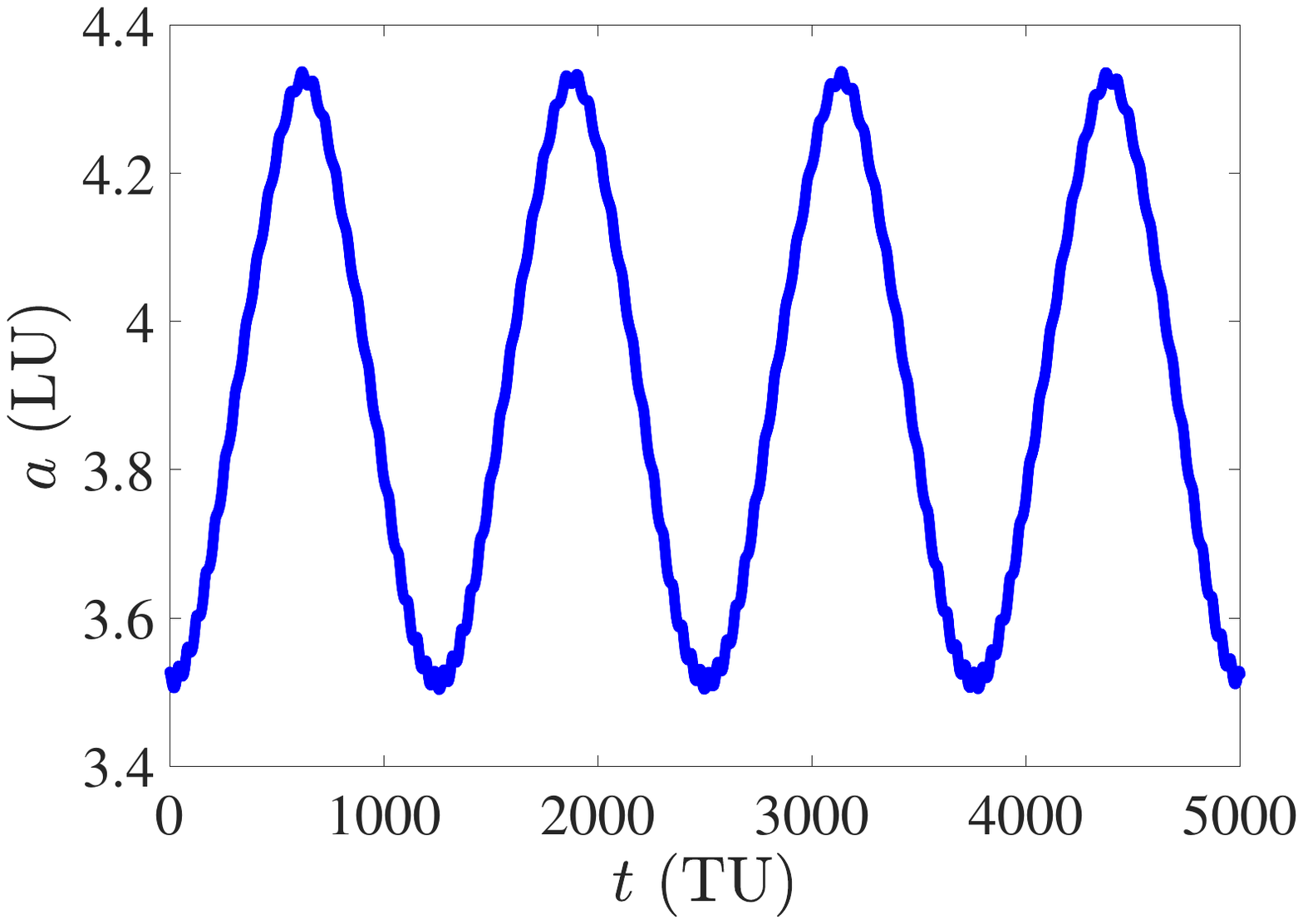}
\includegraphics[width=0.49\columnwidth]{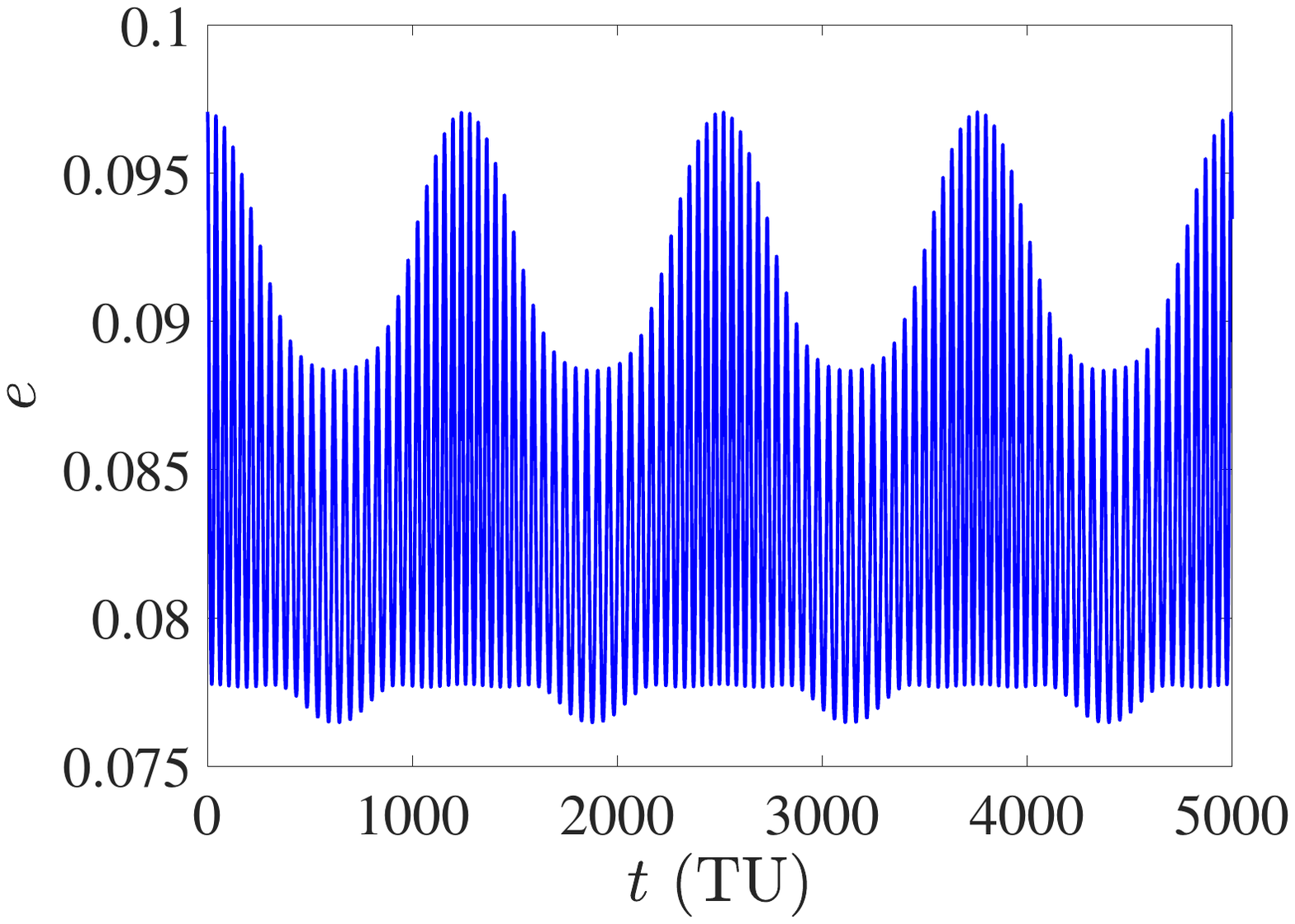}\\
\includegraphics[width=0.49\columnwidth]{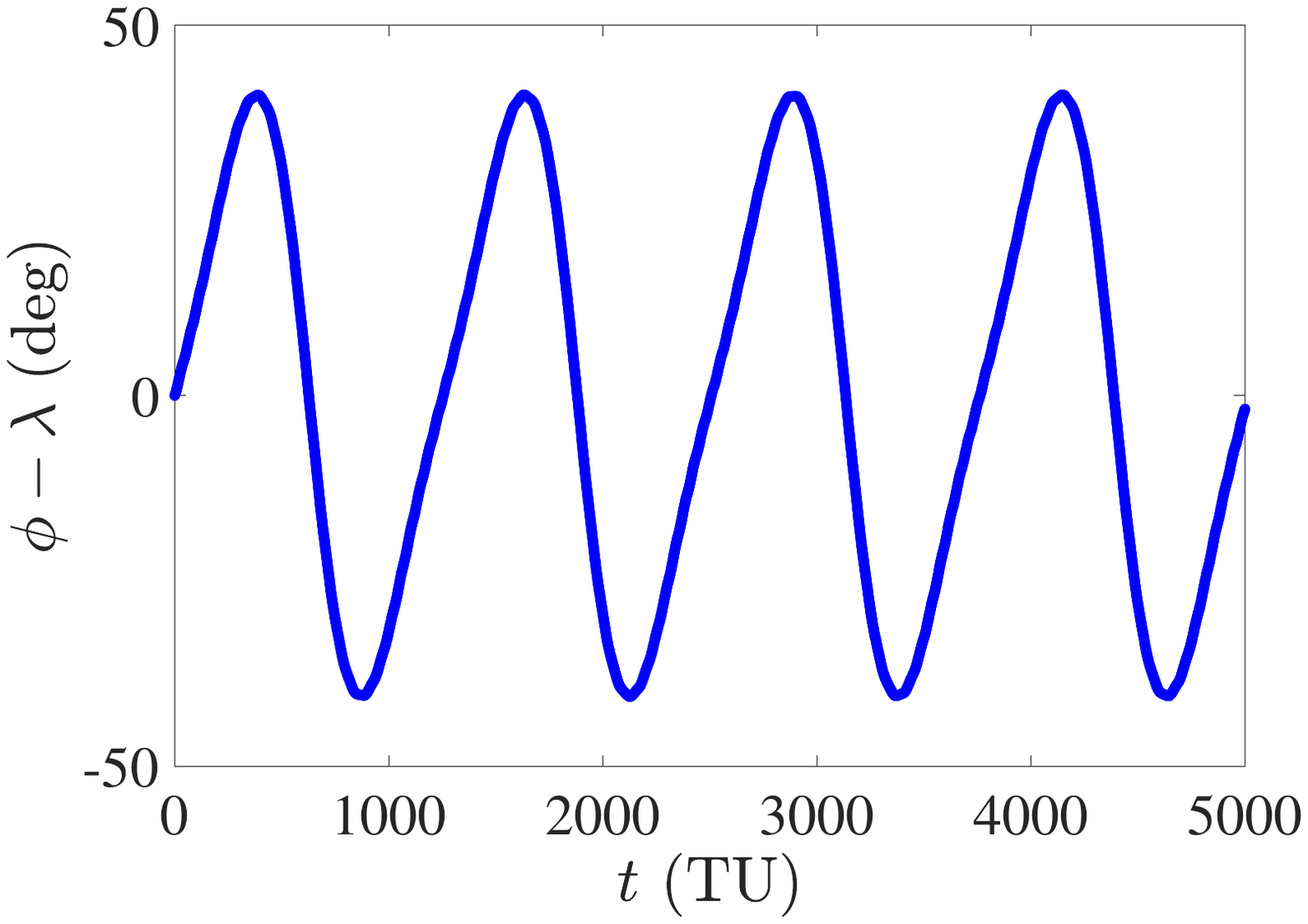}
\includegraphics[width=0.49\columnwidth]{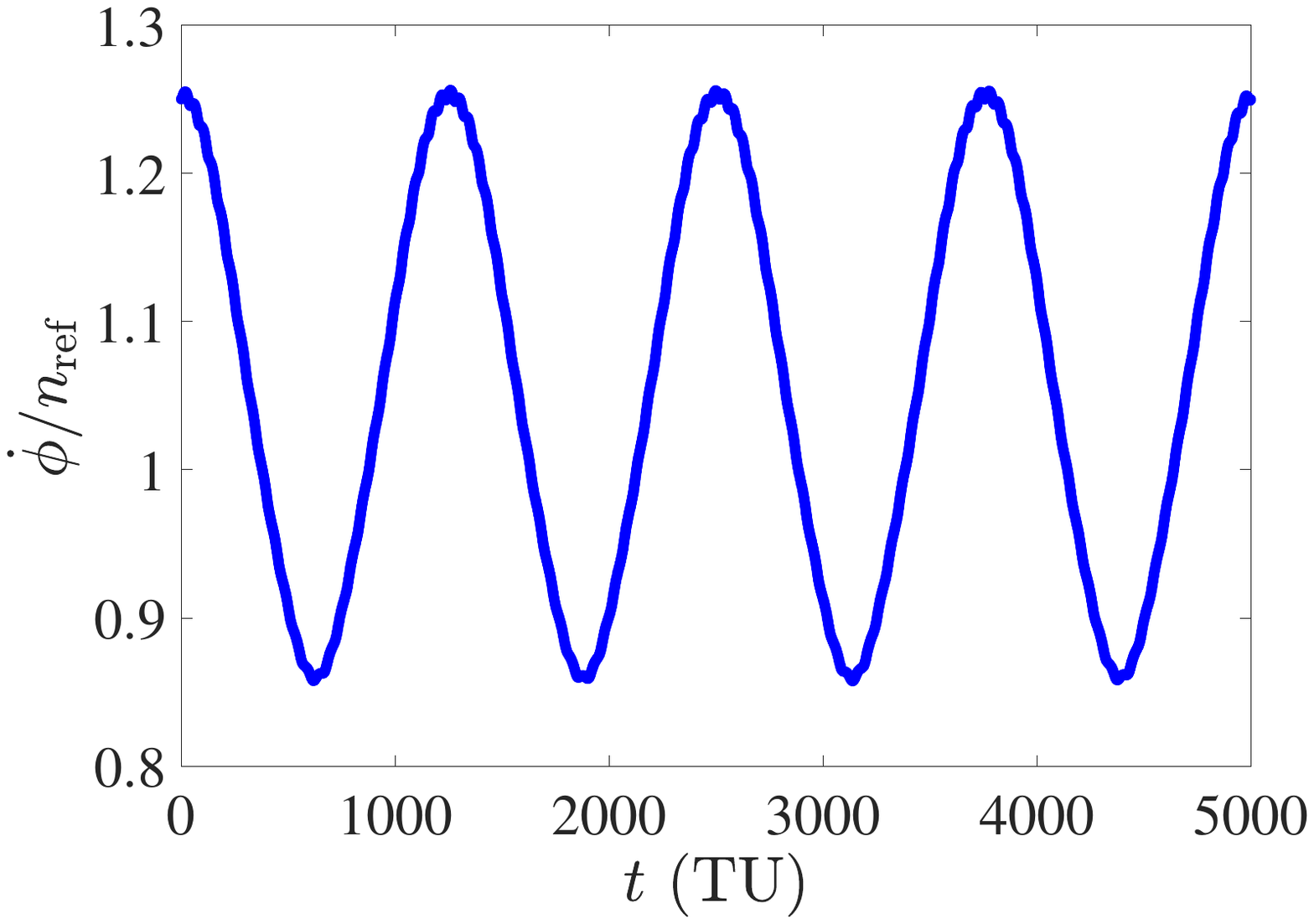}
\caption{Time evolution of the semimajor axis $a$, eccentricity $e$, critical argument of the 1:1 resonance $\sigma = \phi-\lambda$ ($\lambda = \varpi + M$ is the mean longitude) and rotational frequency of the primary $\dot\phi$ for an example orbit propagated under the spin-orbit coupling model. The reference mean motion $n_{\rm ref}$ is calculated by $n_{\rm ref}^2 a_{\rm ref}^3 = \mu$, where the reference semimajor axis is taken as $a_{\rm ref} = 4.0$ in this example.}
\label{Fig2}
\end{figure*}

\begin{table*}
\small
\centering
\caption{Physical parameters of doubly synchronous binary asteroid systems (please refer to http://www.asu.cas.cz/$\sim$asteroid/binastdata.htm for detailed data ). The subscript `p' stands for the primary and the subscript `s' stands for the secondary. The symbol `$-$' shows that the corresponding parameter is not determinant. $D_p$ and $D_s$ are, respectively, the diameters of the primary and secondary and $T_{\rm orb}$ is the orbital period. $a_{c,1}$ is the critical reference semimajor axis of the 1:1 spin-orbit resonance for the ellipsoidal primary. In this table, we take the current semimajor axis as the reference semimajor axis.}
\resizebox{0.95\textwidth}{!}{
\begin{tabular*}{\textwidth}{@{\extracolsep{\fill}}lcccccccccc@{\extracolsep{\fill}}}
\hline
{binary system}&{$D_p$ ($\rm km$)}&{$D_s$ ($\rm km$)}&{$T_{\rm orb}$ ($\rm h$)}&{$a_p/b_p$}&{$b_p/c_p$}&{$\alpha_m$}&{$a_{c,1}$}&{$a_{\rm ref}$ (LU)}\\
\hline
(90) Antiope&86.7&82.8&16.50&1.07&1.04&1.15&1.55&4.18\\
(809)Lundia&6.9&6.1&15.42&1.2&1.17&1.45&1.58&5.18\\
(854) Frostia&6.0&5.7&37.71&1.37&$-$&1.17&1.41&9.86\\
(1089)Tama&9.1&8.0&16.45&1.29&$-$&1.47&1.54&5.46\\
(1139) Atami&5.0&4.0&27.45&1.24&$-$&1.95&1.71&6.92\\
(1313) Berna&9.5&9.2&25.46&1.14&$-$&1.1&1.49&6.90\\
(4492) Debussy&12.6&12.0&26.61&1.32&$-$&1.16&1.43&7.00\\
(2478) Tokai&7.6&6.6&25.90&1.31&$-$&1.53&1.55&7.24\\
(3905) Doppler&7.0&6.0&50.80&1.33&$-$&1.59&1.56&11.74\\
(4951) Iwamoto&4.2&3.7&118&1.24&$-$&1.46&1.56&19.24\\
(5674) Wolff&4.6&3.9&93.7&1.37&$-$&1.64&1.56&17.70\\
(7369) Gavrilin&4.6&3.2&49.12&1.21&$-$&2.97&2.00&9.88\\
(8474) Rettig&4.2&3.8&30.54&1.35&$-$&1.35&1.52&7.78\\
(624) Hektor&220&11&6.92&2.4&$-$&8000&75.06&27.70\\
\hline
\end{tabular*}
}
\label{Tab1}
\end{table*}

\section{Numerical explorations}
\label{Sect3}

The equations of motion represented by Eq. (\ref{Eq2}) are numerically integrated over a certain period and the resulting trajectories are plotted in Fig. \ref{Fig2}. It is observed that there is a strong coupling between translation and rotation. The argument $\sigma = \phi-\lambda$ ($\lambda$ is the mean longitude) is librating around $0^{\circ}$, indicating that this example trajectory is located inside the spin-orbit synchronous resonance. From the evolution of eccentricity (see the top-right panel), we can see that the example trajectory is composed of long-period and short-period contributions. However, the short-period amplitudes of $a$, $\sigma = \phi-\lambda$ and $\dot\phi$ are much smaller than the corresponding long-period amplitudes. In addition, the evolution of $a$, $e$, $\sigma = \phi-\lambda$ and $\dot\phi$ shows that they have almost the same frequencies about the long-period components.

\begin{figure*}
\centering
\includegraphics[width=0.49\columnwidth]{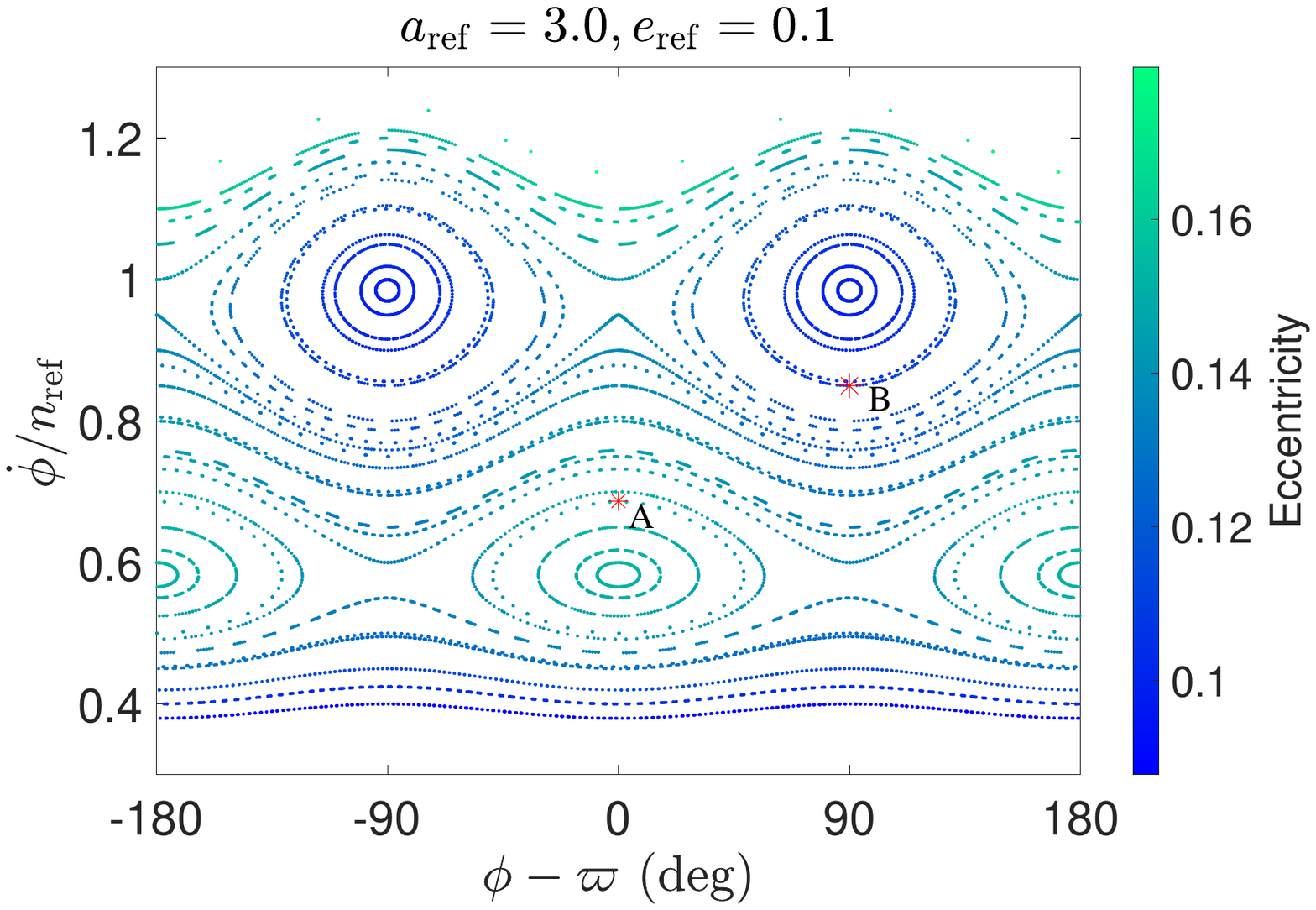}
\includegraphics[width=0.49\columnwidth]{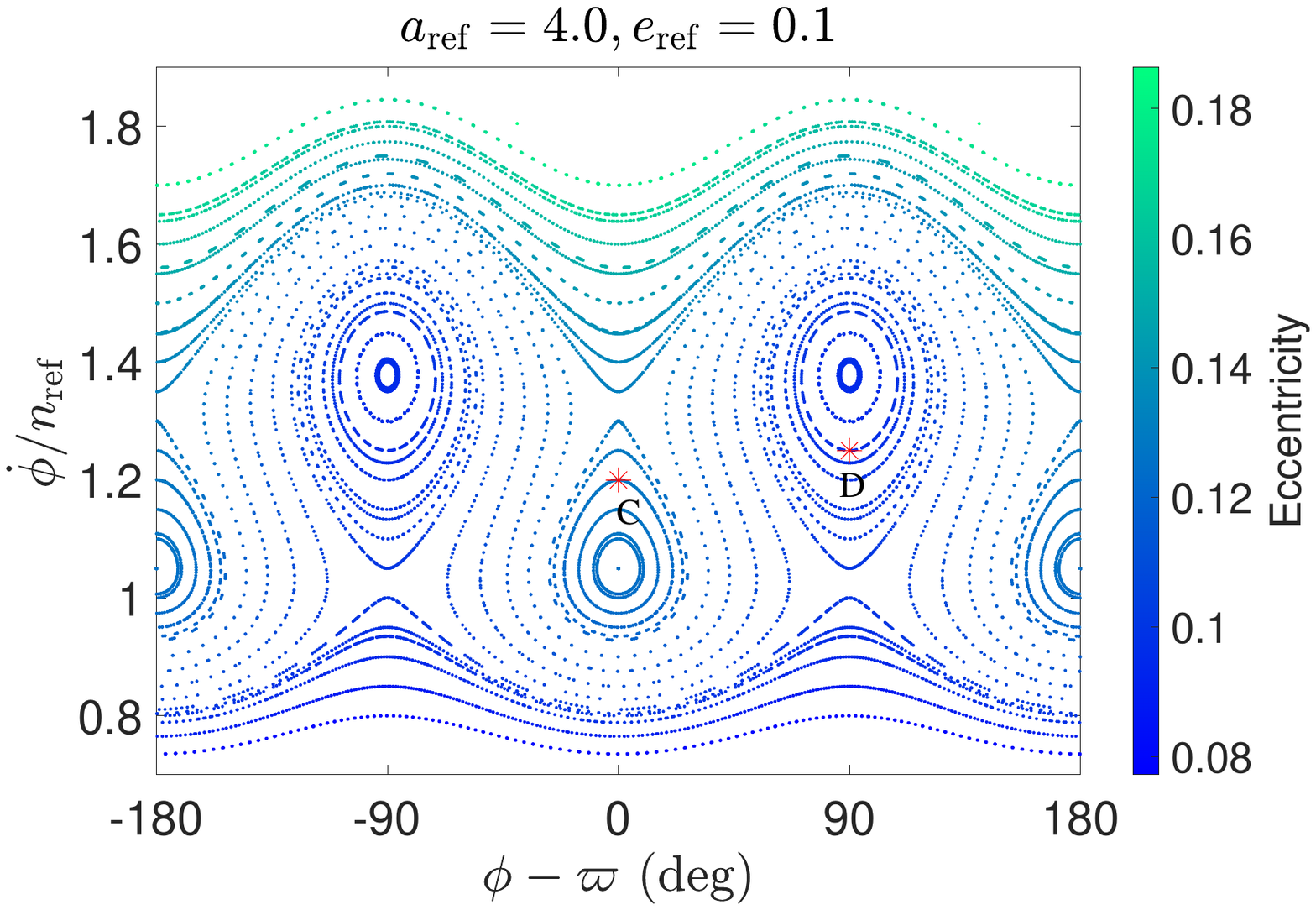}\\
\includegraphics[width=0.49\columnwidth]{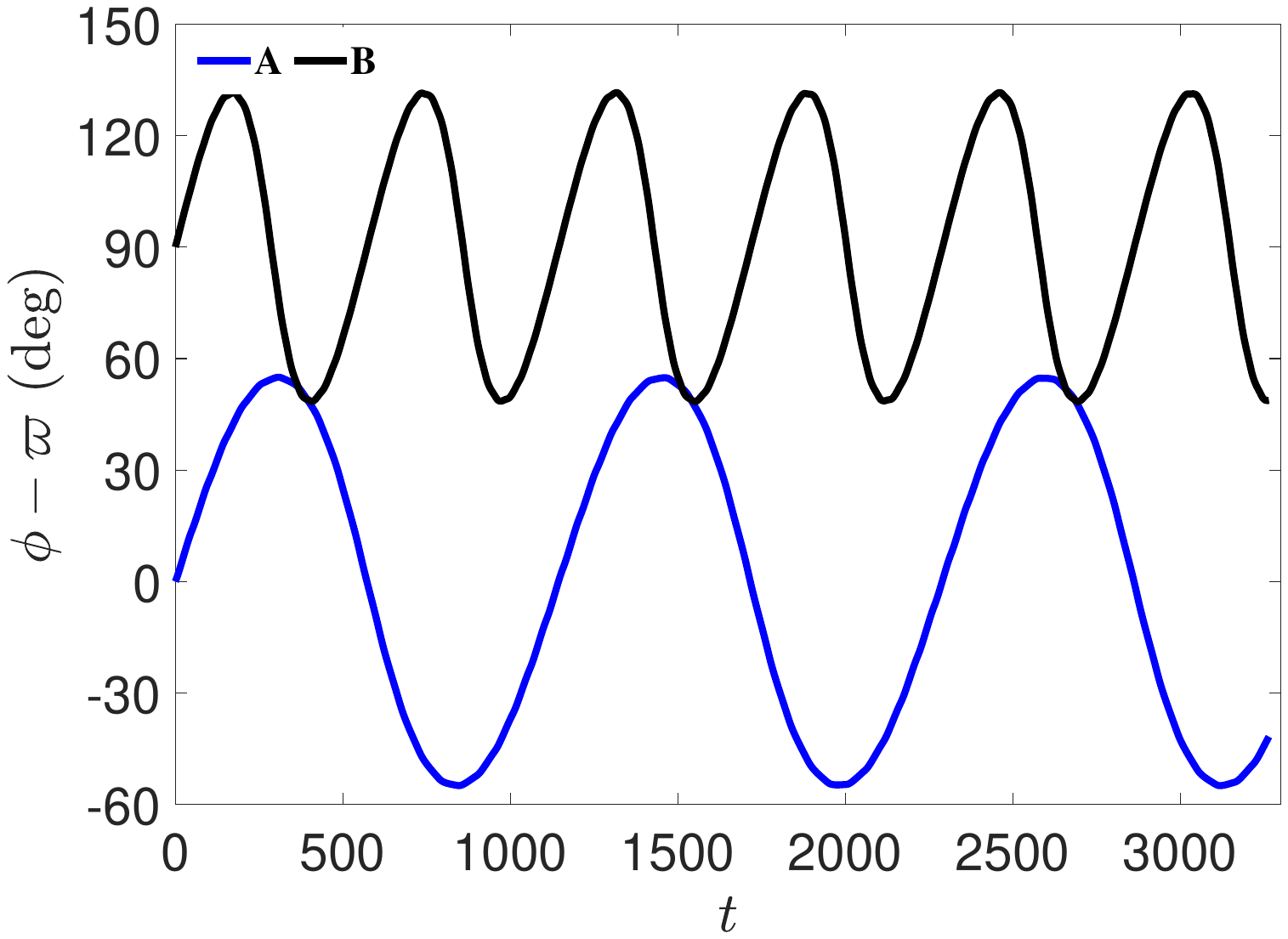}
\includegraphics[width=0.49\columnwidth]{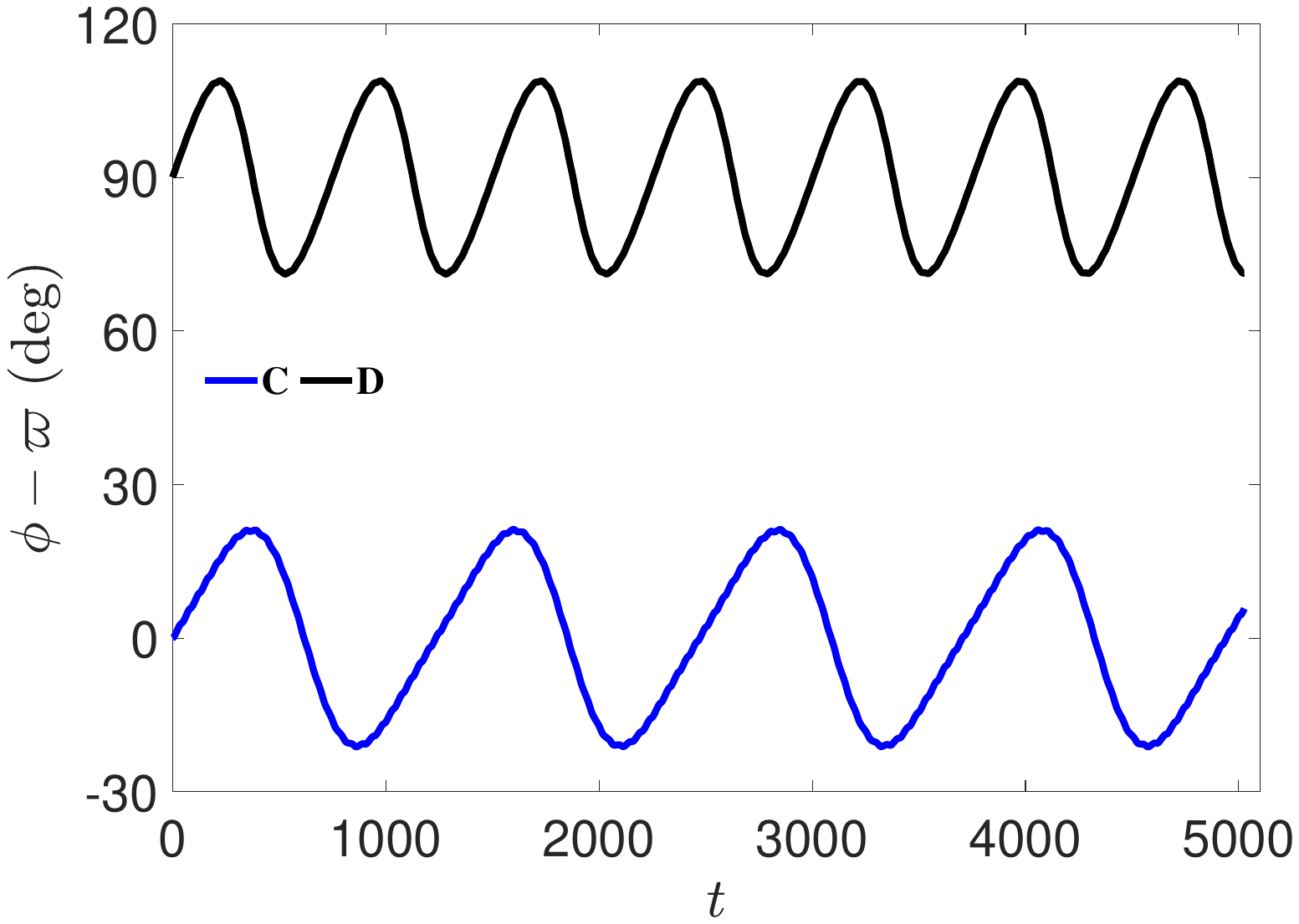}
\caption{Poincar\'e sections (\textit{top panels}) and time evolution of the argument of 1:1 resonance (\textit{bottom panels}). The total angular momentum is specified by $a_{\rm ref}$ and $e_{\rm ref}$. In the top panels, the index shown in the colour bar stands for the magnitude of eccentricity. The starting points of orbits A, B, C and D are marked by red stars in Poincar\'e sections.}
\label{Fig3}
\end{figure*}

To see the global structure of spin-orbit coupling, we choose to produce Poincar\'e sections, defined by 
\begin{equation*}
{\rm mod} (f,2\pi)=0.
\end{equation*}
To generate Poincar\'e sections, the total angular momentum and the Hamiltonian should be given. As for the 1:1 resonance, the total angular momentum is parameterised by equation (\ref{Eq6}) in terms of $a_{\rm ref}$ and $e_{\rm ref}$, and the Hamiltonian is also parameterised in terms of $a_{\rm ref}$ and $e_{\rm ref}$ in the following form,
\begin{equation}\label{Eq7}
{\cal H} = {\left. {\cal H} \right|_{a = a_{\rm ref}, e = e_{\rm ref}, f = 0, \phi-\varpi = \pi/2, \dot\phi = n_{\rm ref}}}.
\end{equation}
It means that a given set of parameters $(a_{\rm ref},e_{\rm ref})$ is used to uniquely characterise the total angular momentum $G_{\rm tot}$ and the Hamiltonian ${\cal H}$.

Poincar\'e sections are produced and presented in the top two panels of Fig. \ref{Fig3}, where the index shown in the colour bar stands for the magnitude of eccentricity $e$. The structures arising in Poincar\'e sections are completely different from the classical pendulum structures and also different from the ones associated with ellipsoidal secondary. It is observed that almost the entire phase space is filled with regular motions (chaotic layers are too thin to observe). In addition, numerical trajectories inside libration islands are shown in the bottom panels of Fig. \ref{Fig3}.

In the case of $a_{\rm ref} = 3.0$ and $e_{\rm ref} = 0.1$ (see the top-left panel of Fig. \ref{Fig3}), there are four islands of libration as well as four saddle points arising in the Poincar\'e section. In particular, two islands centred at $\pm 90^{\circ}$ hold higher values of $\dot\phi$ (lower values of eccentricity) than the other two centred at $0^{\circ}$ and $180^{\circ}$. An evident feature is that their dynamical separatrices are completely separable. Two trajectories marked by A and B inside libration islands are numerically integrated and the results show that both of them are locked inside the synchronous spin-orbit resonance (see the bottom-left panel of Fig. \ref{Fig3}). In particular, the trajectory A is librating around $\sigma = 0$ and the trajectory B is librating around $\sigma = 90^{\circ}$.

In the case of $a_{\rm ref} = 4.0$ and $e_{\rm ref} = 0.1$ (see the top-right panel of Fig. \ref{Fig3}), there are also four islands of libration and four saddle points arising in the Poincar\'e section. However, the phase-space structures are distinctly different from the former case. In particular, the two islands centred at $\pm 90^{\circ}$ move up to $\dot\phi/n_{\rm ref}\sim 1.5$ and the other two centred at $0^{\circ}$ and $180^{\circ}$ move to $\dot\phi/n_{\rm ref}\sim 1.0$. Two trajectories marked by C and D inside libration islands are numerically integrated and the numerical results show that both of them are located inside the synchronous spin-orbit resonance (see the bottom-right panel of Fig. \ref{Fig3}). In particular, the trajectory C is librating around $\sigma = 0$ and the trajectory D is librating around $\sigma = 90^{\circ}$.

\section{Expansion of Hamiltonian function}
\label{Sect4}

The elliptic expansion takes the form \citep{hughes1981computation},
\begin{equation*}
{\left( {\frac{a}{r}} \right)^l}\cos \left( {mf} \right) = \sum\limits_{s =  - \infty }^\infty  {X_s^{ - l,m}\left( e \right)\cos \left( {sM} \right)}
\end{equation*}
where $X_c^{a,b} (e)$ is Hansen coefficient dependent on eccentricity \citep{murray1999solar} and its lowest order in eccentricity is equal to $|b-c|$. After performing elliptic expansion, the Hamiltonian (\ref{Eq1}) can be written as
\begin{small}
\begin{equation}\label{Eq8}
\begin{aligned}
{\cal H} =&  - \frac{1}{{2a}} + \frac{{p_\phi ^2}}{{2{I_3}}}\\
& - \sum\limits_{s =  - \infty }^\infty  {\left\{ {\left[ { - \frac{{{C_{20}}}}{{2{a^3}}}X_s^{ - 3,0}\left( e \right) + \frac{{3{C_{40}}}}{{8{a^5}}}X_s^{ - 5,0}\left( e \right)} \right]\cos \left( {sM} \right)} \right.} \\
& + \left[ {\frac{{3{C_{22}}}}{{{a^3}}}X_s^{ - 3,2}\left( e \right)  - \frac{{15{C_{42}}}}{{2{a^5}}}X_s^{ - 5,2}\left( e \right)} \right]\cos \left( {2\phi  - 2\varpi  - sM} \right)\\
&\left. { + \frac{{105{C_{44}}}}{{{a^5}}}X_s^{ - 5,4}\left( e \right)\cos \left( {4\phi  - 4\varpi  - sM} \right)} \right\}
\end{aligned}
\end{equation}
\end{small}
where the normalised system of units has been used to simplify the expression (i.e., $m = 1.0$, $a_p =1.0$ and $\mu=1.0$). For convenience, we introduce a set of canonical variables as follows:
\begin{equation*}
\begin{aligned}
l &= M,\quad L = \sqrt a,\\
g &= \omega ,\quad G = \sqrt {a\left( {1 - {e^2}} \right)},\\
h &= \Omega, \quad H= G\cos{i},\\
\phi &, \quad p_\phi,
\end{aligned}
\end{equation*}
and perform the following canonical transformation,
\begin{equation}\label{Eq9}
\begin{aligned}
{\psi _1} &= \phi  - g - h = \phi - \varpi,\quad {\Psi _1} = {p_\phi },\\
{\psi _2} &= l,\quad {\Psi _2} = L,\\
{\psi _3} &= g,\quad {\Psi _3} = G + {p_\phi},\\
{\psi _4} &= h,\quad {\Psi _4} =H + {p_\phi}.
\end{aligned}
\end{equation}
In terms of canonical variables $(\psi _{1,2,3,4}, \Psi _{1,2,3,4})$, the Hamiltonian (\ref{Eq8}) truncated up to the fourth order in eccentricity can be organised by
\begin{equation}\label{Eq10}
\begin{aligned}
{\cal H} =&  - \frac{1}{{2\Psi _2^2}} + \frac{{\Psi _1^2}}{{2{I_3}}}\\
& - \sum\limits_{n =  - 4}^4 {\left[ { - \frac{{{C_{20}}}}{{2\Psi _2^6}}X_n^{ - 3,0} + \frac{{3{C_{40}}}}{{8\Psi _2^{10}}}X_n^{ - 5,0}} \right]\cos \left( {n{\psi _2}} \right)} \\
& - \sum\limits_{n =  - 2}^6 {\left[ {\frac{{3{C_{22}}}}{{\Psi _2^6}}X_n^{ - 3,2} - \frac{{15{C_{42}}}}{{2\Psi _2^{10}}}X_n^{ - 5,2}} \right]\cos \left( {2{\psi _1} - n{\psi _2}} \right)} \\
& - \sum\limits_{n = 0}^8 {\frac{{105{C_{44}}}}{{\Psi _2^{10}}}X_n^{ - 5,4}\cos \left( {4{\psi _1} - n{\psi _2}} \right)} 
\end{aligned}
\end{equation}
We need to remember that, in equation (\ref{Eq10}), Hansen coefficients $X_c^{a,b} (e)$ are related to the eccentricity $e$, which can be expressed in terms of the canonical variables by $e = \sqrt {1 - \frac{{{G^2}}}{{{L^2}}}}  = \sqrt {1 - \frac{{{{\left( {{G_{\rm tot}} - {\Psi _1}} \right)}^2}}}{{\Psi _2^2}}}$.

It is observed that the angles ${\psi _3}$ and ${\psi _4}$ are cyclic variables under the dynamical model determined by the Hamiltonian (\ref{Eq10}). Thus, their conjugate momenta ${\Psi _3} \equiv G_{\rm tot}$ (the total angular momentum) and ${\Psi _4}$ are motion integral. We can see that the dynamical model specified by the Hamiltonian (\ref{Eq10}) is of 2 degrees of freedom, depending on the total angular momentum. When the total angular momentum is given (please recall that $a_{\rm ref}$ and $e_{\rm ref}$ are used to specify $G_{\rm tot}$), the first degree of freedom $(\psi_1,\Psi_1)$ stands for rotation and the second degree of freedom $(\psi_2,\Psi_2)$ stands for translation.

In particular, spin-orbit resonances may take place if the rotational period and orbital period of the primary are commensurable. In order to describe resonant dynamics, it is necessary to introduce the resonant argument. To this end, the following canonical transformation is introduced for the $k_1$:$k_2$ resonance,
\begin{equation}\label{Eq11}
\begin{aligned}
{\sigma _1} &= {\psi _1} - \frac{{{k_2}}}{{{k_1}}}{\psi _2},\quad {\Sigma _1} = {\Psi _1},\\
{\sigma _2} &= {\psi _2},\quad {\Sigma _2} = {\Psi _2} + \frac{{{k_2}}}{{{k_1}}}{\Psi _1}.
\end{aligned}
\end{equation}
The angle $\sigma_1$ stands for the resonant argument. In particular, it is $\sigma_1 = (\phi-\varpi) -M$ for the synchronous spin-orbit resonance and, it is $\sigma_1 = (\phi-\varpi) - 3M/2$ for the 2:3 spin-orbit resonance and it is $\sigma_1 = (\phi-\varpi) - M/2$ for the 2:1 spin-orbit resonance. As a result, the Hamiltonian (\ref{Eq10}) becomes
\begin{small}
\begin{equation}\label{Eq12}
\begin{aligned}
{\cal H} =&  - \frac{1}{{2{{\left( {{\Sigma _2} - \frac{{{k_2}}}{{{k_1}}}{\Sigma _1}} \right)}^2}}} + \frac{{\Sigma _1^2}}{{2{I_3}}}\; - \sum\limits_{n =  - 4}^4 {\left[ { - \frac{{{C_{20}}X_n^{ - 3,0}}}{{2{{\left( {{\Sigma _2} - \frac{{{k_2}}}{{{k_1}}}{\Sigma _1}} \right)}^6}}}} \right.} \\
&\left. { + \frac{{3{C_{40}}X_n^{ - 5,0}}}{{8{{\left( {{\Sigma _2} - \frac{{{k_2}}}{{{k_1}}}{\Sigma _1}} \right)}^{10}}}}} \right]\cos \left( {n{\sigma _2}} \right) - \sum\limits_{n =  - 2}^6 {\left[ {\frac{{3{C_{22}}X_n^{ - 3,2}}}{{{{\left( {{\Sigma _2} - \frac{{{k_2}}}{{{k_1}}}{\Sigma _1}} \right)}^6}}}} \right.} \\
&\left. { - \frac{{15{C_{42}}X_n^{ - 5,2}}}{{2{{\left( {{\Sigma _2} - \frac{{{k_2}}}{{{k_1}}}{\Sigma _1}} \right)}^{10}}}}} \right]\cos \left[ {2{\sigma _1} + \left( {\frac{{2{k_2}}}{{{k_1}}} - n} \right){\sigma _2}} \right]\\
& - \sum\limits_{n = 0}^8 {\frac{{105{C_{44}}X_n^{ - 5,4}}}{{{{\left( {{\Sigma _2} - \frac{{{k_2}}}{{{k_1}}}{\Sigma _1}} \right)}^{10}}}}\cos \left[ {4{\sigma _1} + \left( {\frac{{4{k_2}}}{{{k_1}}} - n} \right){\sigma _2}} \right]} 
\end{aligned}
\end{equation}
\end{small}
For convenience, we denote the Hamiltonian (\ref{Eq12}) by the following compact form,
\begin{equation}\label{Eq13}
{\cal H}= \sum\limits_{{k_1},{k_2}} {{{{\cal C}_{{k_1},{k_2}}}\left( {{G_{\rm tot}};{\Sigma _1},{\Sigma _2}} \right)\cos \left( {{k_1}{\sigma _1} + {k_2}{\sigma _2}} \right)} } 
\end{equation}
which specifies a 2-DOF Hamiltonian model, depending on the total angular momentum $G_{\rm tot}$.

Based on the Hamiltonian formulation given by Eq. (\ref{Eq12}), in the following two sections we will take advantage of pendulum approximation as well as adiabatic approximation to analytically study spin-orbit resonances.

\section{Pendulum approximations}
\label{Sect5}

In this section, we make some discussions about pendulum approximations to the spin-orbit coupling problem \citep{murray1999solar,hou2017note,jafari2023surfing}.

For the synchronous (1:1) spin-orbit resonance (i.e., $k_1 =k_2 =1$), performing average for Hamiltonian (\ref{Eq12}) over one period of $\sigma_2$ leads to the resonant Hamiltonian,
\begin{equation}\label{Eq14}
\begin{aligned}
{{\cal H}_{1:1}} =&  - \frac{1}{{2{{\left( {{\Sigma _2} - {\Sigma _1}} \right)}^2}}} + \frac{{\Sigma _1^2}}{{2{I_3}}} + \frac{{{C_{20}}}}{{2{{\left( {{\Sigma _2} - {\Sigma _1}} \right)}^6}}} - \frac{{3{C_{40}}}}{{8{{\left( {{\Sigma _2} - {\Sigma _1}} \right)}^{10}}}}\\
& - \left[ {\frac{{3{C_{22}}}}{{{{\left( {{\Sigma _2} - {\Sigma _1}} \right)}^6}}} - \frac{{15{C_{42}}}}{{2{{\left( {{\Sigma _2} - {\Sigma _1}} \right)}^{10}}}}} \right]\cos 2{\sigma _1}\\
& - \frac{{105{C_{44}}}}{{{{\left( {{\Sigma _2} - {\Sigma _1}} \right)}^{10}}}}\cos 4{\sigma _1}
\end{aligned}
\end{equation}
which is truncated at the first order in eccentricity. The resonant Hamiltonian (\ref{Eq14}) is coincident with that provided in \citet{jafari2023surfing} (see equation 23 in their study). Ignoring the influence of $C_{20}$ and $C_{40}$ upon the nominal resonance, we can obtain the nominal location of 1:1 resonance at $\Sigma_{1,\rm res} = I_3 \sqrt{\mu/a_{\rm ref}^3}$. By expanding resonant Hamiltonian (\ref{Eq14}) around the nominal resonance in Taylor series and removing high-order terms \citep{morbidelli2002modern}, the approximate resonant Hamiltonian takes the form,
\begin{equation}\label{Eq15}
\begin{aligned}
\Delta {{\cal H}_{1:1}} =& \frac{1}{2}\left( {\frac{1}{{{I_3}}} - \frac{3}{{\left( {{\Sigma _2} - {\Sigma _1}} \right)_{\rm res}^4}}} \right)\Delta \Sigma _1^2\\
&- \left[ {\frac{{3{C_{22}}}}{{\left( {{\Sigma _2} - {\Sigma _1}} \right)_{\rm res}^6}} - \frac{{15{C_{42}}}}{{2\left( {{\Sigma _2} - {\Sigma _1}} \right)_{\rm res}^{10}}}} \right]\cos 2{\sigma _1}\\
&- \frac{{105{C_{44}}}}{{\left( {{\Sigma _2} - {\Sigma _1}} \right)_{\rm res}^{10}}}\cos 4{\sigma _1}
\end{aligned}
\end{equation}
where the constant terms have been removed, the term $(\Sigma_2 - \Sigma_1)_{\rm res}$ is evaluated at the nominal location of resonance centre and it is equal to $(\Sigma_2 - \Sigma_1)_{\rm res} = \sqrt{a_{\rm ref}}$ for the synchronous resonance. According to \citet{nadoushan2016geography}, the fourth-order harmonic term associated with $C_{44}$ has no contribution to the width of the synchronous resonance. Thus, it can be removed from the Hamiltonian. As a consequence, we can reach the pendulum approximation for the resonant Hamiltonian (\ref{Eq15}), given by
\begin{equation}\label{Eq16}
\Delta {{\cal H}_{1:1}} = \frac{1}{2}\left( {\frac{1}{{{I_3}}} - \frac{3}{{{a_{\rm ref}^2}}}} \right)\Delta \Sigma _1^2 - \left[ {\frac{{3{C_{22}}}}{{{a_{\rm ref}^3}}} - \frac{{15{C_{42}}}}{{2{a_{\rm ref}^5}}}} \right]\cos 2{\sigma _1}.
\end{equation}
Similarly, the pendulum approximation of resonant Hamiltonian for the 2:3 resonance ($k_1=2$ and $k_2=3$) can be written as
\begin{equation}\label{Eq17}
\Delta {{\cal H}_{2:3}} = \frac{1}{2}\left( {\frac{1}{{{I_3}}} - \frac{{27}}{{4{a_{\rm ref}^2}}}} \right)\Delta \Sigma _1^2 - \left[ {\frac{{21{C_{22}}e_{\rm ref}}}{{2{a_{\rm ref}^3}}} - \frac{{135{C_{42}}e_{\rm ref}}}{{4{a_{\rm ref}^5}}}} \right]\cos 2{\sigma _1}, 
\end{equation}
and the pendulum approximation for the 2:1 resonance ($k_1=2$ and $k_2=1$) can be expressed as
\begin{equation}\label{Eq18}
\Delta {{\cal H}_{2:1}} = \frac{1}{2}\left( {\frac{1}{{{I_3}}} - \frac{3}{{4{a_{\rm ref}^2}}}} \right)\Delta \Sigma _1^2 + \left[ {\frac{{3{C_{22}}e_{\rm ref}}}{{2{a_{\rm ref}^3}}} + \frac{{15{C_{42}}e_{\rm ref}}}{{4{a_{\rm ref}^5}}}} \right]\cos 2{\sigma _1}.
\end{equation}
From the pendulum approximations, the location of libration centre is dependent on the sign of $S_{1:1}$ for synchronous (1:1) resonance, of $S_{2:3}$ for the 2:3 resonance and of $S_{2:1}$ for the 2:1 resonance,
\begin{equation}\label{Eq19}
S_{1:1} = {\frac{1}{{{I_3}}} - \frac{3}{{{a_{\rm ref}^2}}}},\quad
S_{2:3} = {\frac{1}{{{I_3}}} - \frac{{27}}{{4{a_{\rm ref}^2}}}},\quad
S_{2:1} = \frac{1}{{{I_3}}} - \frac{3}{{4{a_{\rm ref}^2}}}.  
\end{equation}
In particular, the libration centre is located at
\begin{equation*}
2{\sigma _{1,c}} = \left\{ \begin{array}{l}
0,\quad {\rm if}\;{S_{1:1}} > 0\\
\pi,\quad {\rm if}\;{S_{1:1}} < 0
\end{array} \right.
\end{equation*}
for the synchronous (1:1) resonance, it is located at
\begin{equation*}
2{\sigma _{1,c}} = \left\{ \begin{array}{l}
0,\quad {\rm if}\;{S_{2:3}} > 0\\
\pi,\quad {\rm if}\;{S_{2:3}} < 0
\end{array} \right.
\end{equation*}
for the 2:3 resonance and it is located at
\begin{equation*}
2{\sigma _{1,c}} = \left\{ \begin{array}{l}
\pi,\quad {\rm if}\;{S_{2:1}} > 0\\
0,\quad {\rm if}\;{S_{2:1}} < 0
\end{array} \right.
\end{equation*}
for the 2:1 resonance.

When $I_3$ is given, the sign of $S_{1:1}$ (or $S_{2:3}$ and $S_{2:1}$) is determined by the value of $a_{\rm ref}$. It is observed that the libration centre may be changed if the reference semimajor axis is different. This is consistent with the conclusion made by \citet{hou2017note}.

In particular, for the synchronous resonance, the critical reference semimajor axis $a_{c,1}$ is
\begin{equation}\label{Eq20}
{S_{1:1}} = 0 \Rightarrow {a_{c,1}} = \sqrt {\frac{3}{5}\left( {1 + {\alpha _m}} \right)\left( {a_p^2 + b_p^2} \right)}
\end{equation}
for the 2:3 resonance, the critical semimajor axis $a_{c,2}$ is 
\begin{equation}\label{Eq21}
{S_{2:3}} = 0 \Rightarrow {a_{c,2}} = \frac{3}{2} {a_{c,1}},
\end{equation}
and, for the 2:1 resonance, the critical semimajor axis $a_{c,3}$ is
\begin{equation}\label{Eq22}
{S_{2:1}} = 0 \Rightarrow {a_{c,3}} = \frac{1}{2} {a_{c,1}}.
\end{equation}
As a result, the libration centre $\sigma_{1,c}$ is dependent on the reference semimajor axis,
\begin{equation*}
2{\sigma _{1,c}} = \left\{ \begin{array}{l}
0,\quad {\rm if}\;a_{\rm ref} > {a_c}\\
\pi,\quad {\rm if}\;a_{\rm ref}  < {a_c}
\end{array} \right.
\end{equation*}
for the synchronous (1:1) resonance and the 2:3 resonance, and it is 
\begin{equation*}
2{\sigma _{1,c}} = \left\{ \begin{array}{l}
\pi,\quad {\rm if}\;a_{\rm ref} > {a_c}\\
0,\quad {\rm if}\;a_{\rm ref}  < {a_c}
\end{array} \right.
\end{equation*}
for the 2:1 resonance. It takes $a_c = a_{c,1}$ for the synchronous resonance, $a_c = a_{c,2}$ for the 2:3 resonance and $a_c = a_{c,3}$ for the 2:1 resonance. Please refer to Table \ref{Tab1} for the critical value of reference semimajor axis $a_{c,1}$ for those known doubly synchronous binary asteroid systems. It is observed that it satisfies $a_{\rm ref} < a_{c,1}$ for (624) Hektor meaning that its resonance centre is at $2{\sigma _{1,c}} = \pi$, and it satisfies $a_{\rm ref} > a_{c,1}$ for other binary asteroid systems, meaning that their resonance centres are at $2{\sigma _{1,c}} = 0$. This is in agreement with the statement made by \citet{hou2017note}.

\begin{figure*}
\centering
\includegraphics[width=0.49\columnwidth]{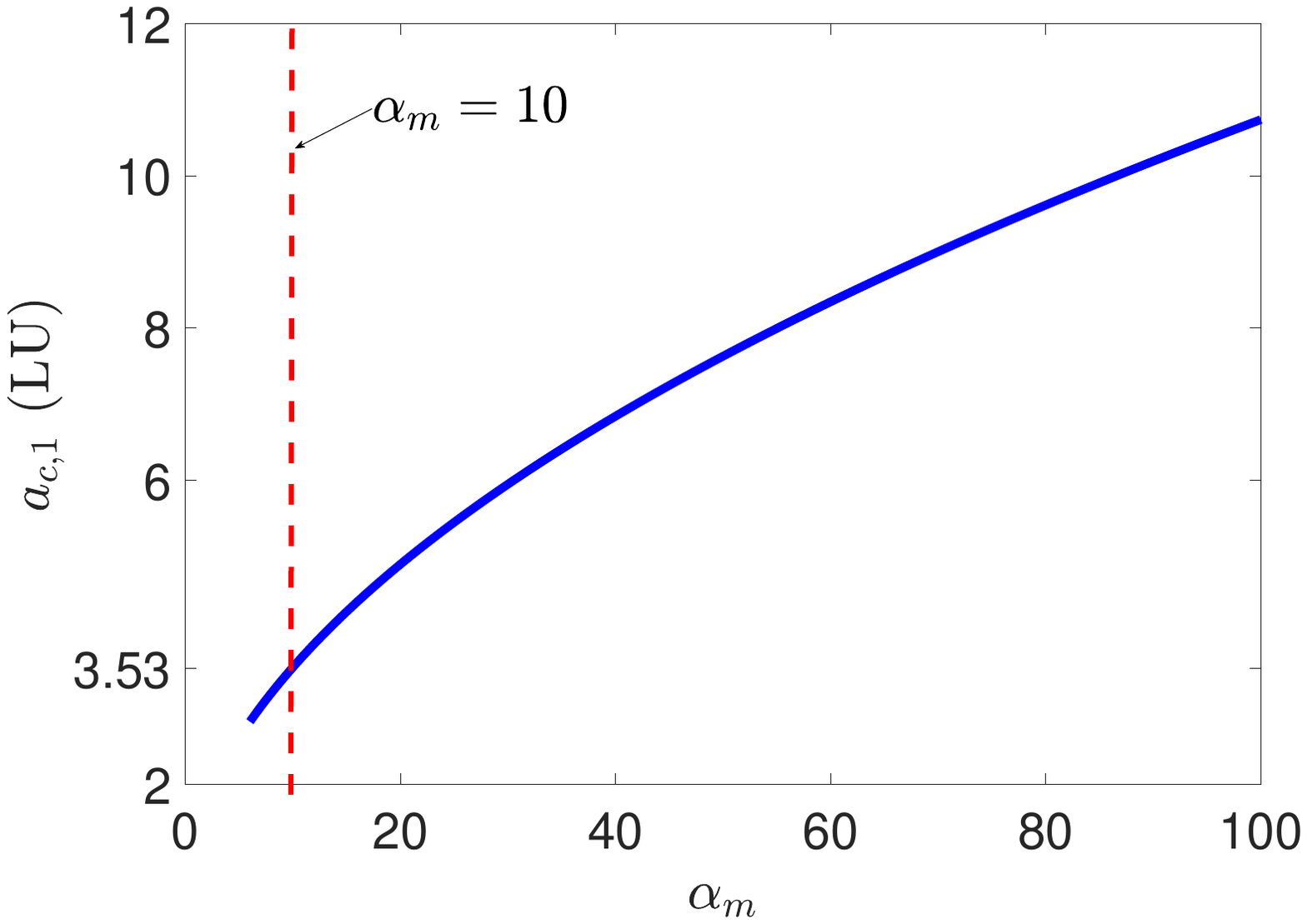}
\includegraphics[width=0.49\columnwidth]{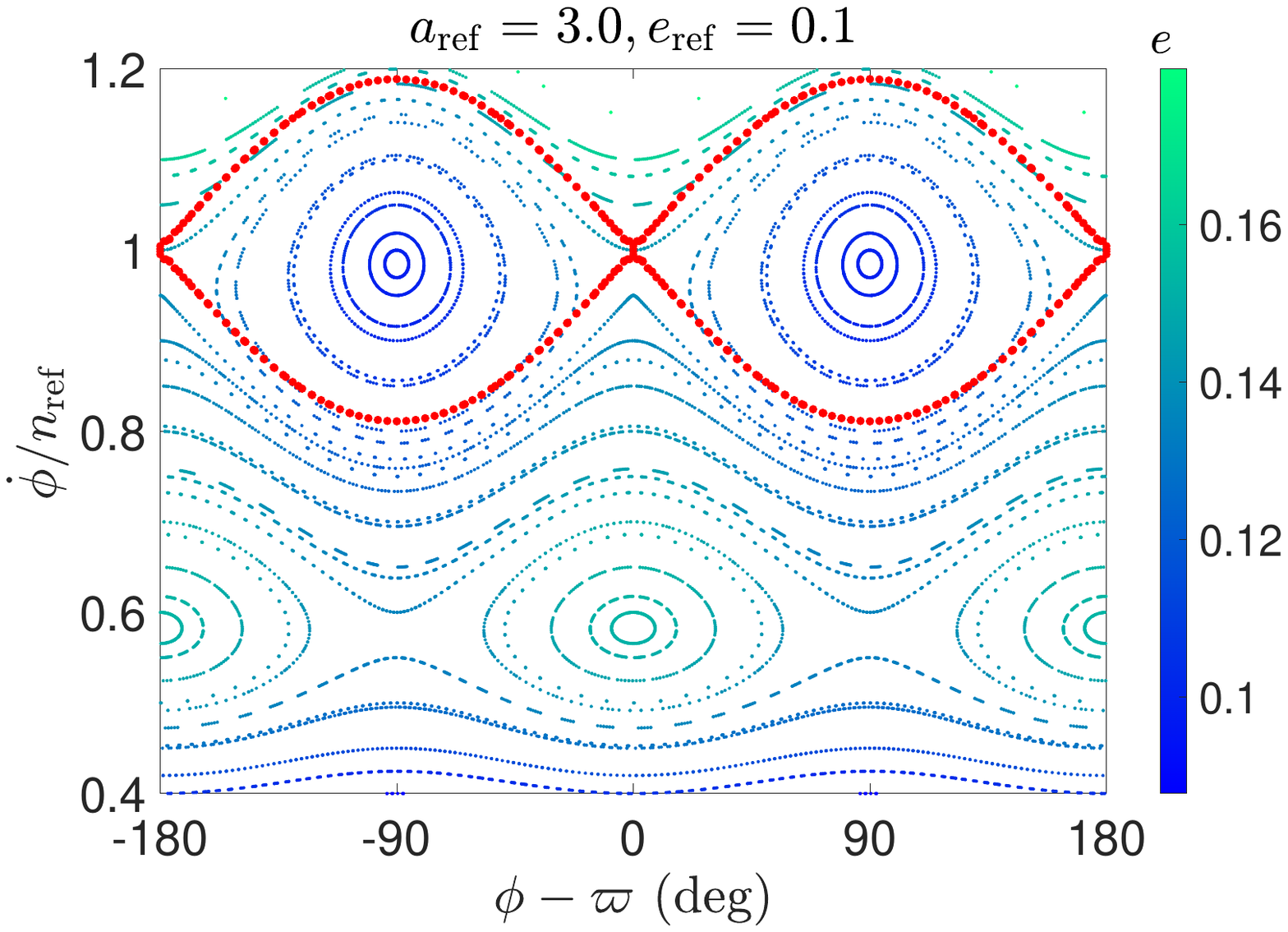}\\
\includegraphics[width=0.49\columnwidth]{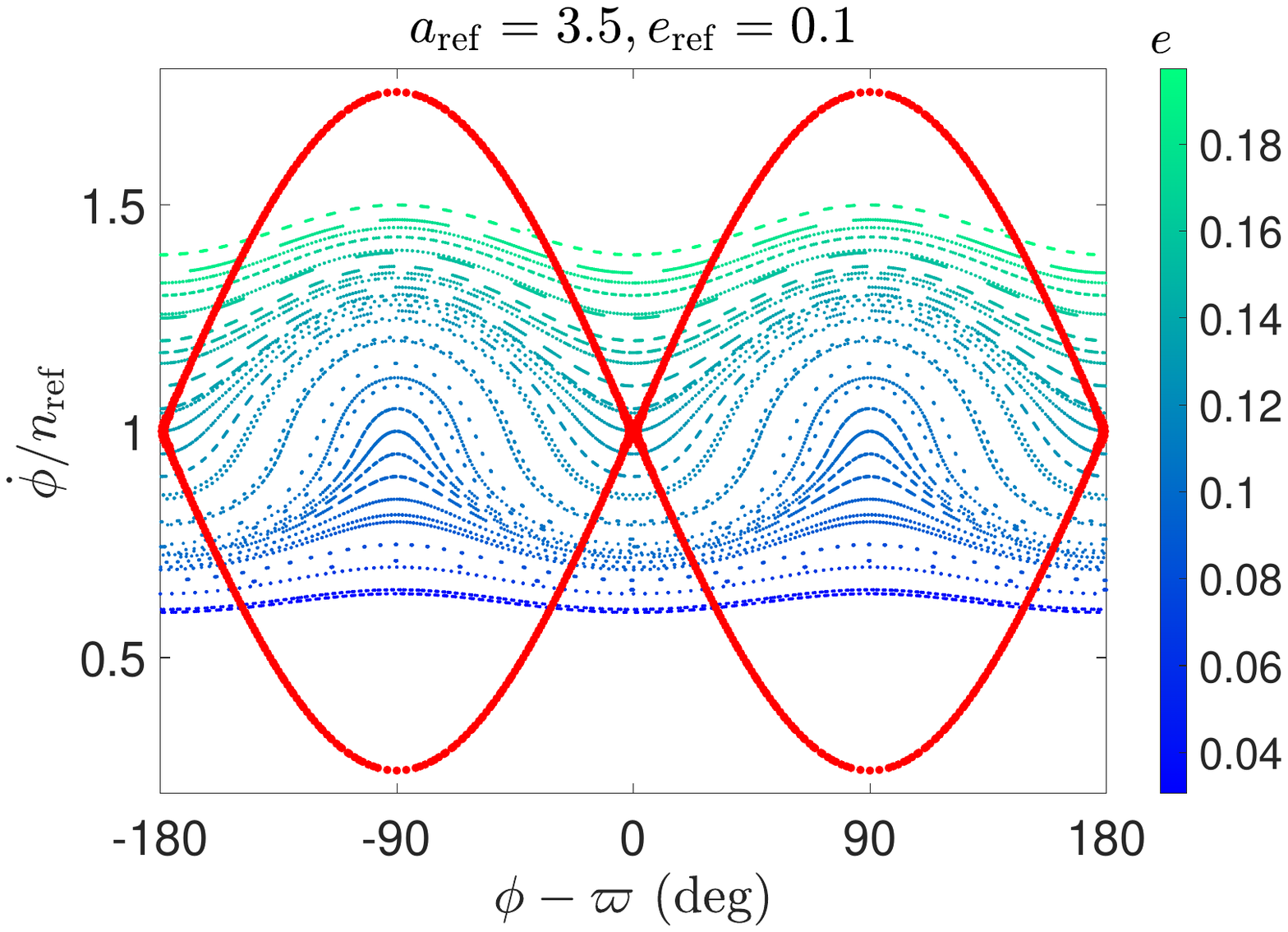}
\includegraphics[width=0.49\columnwidth]{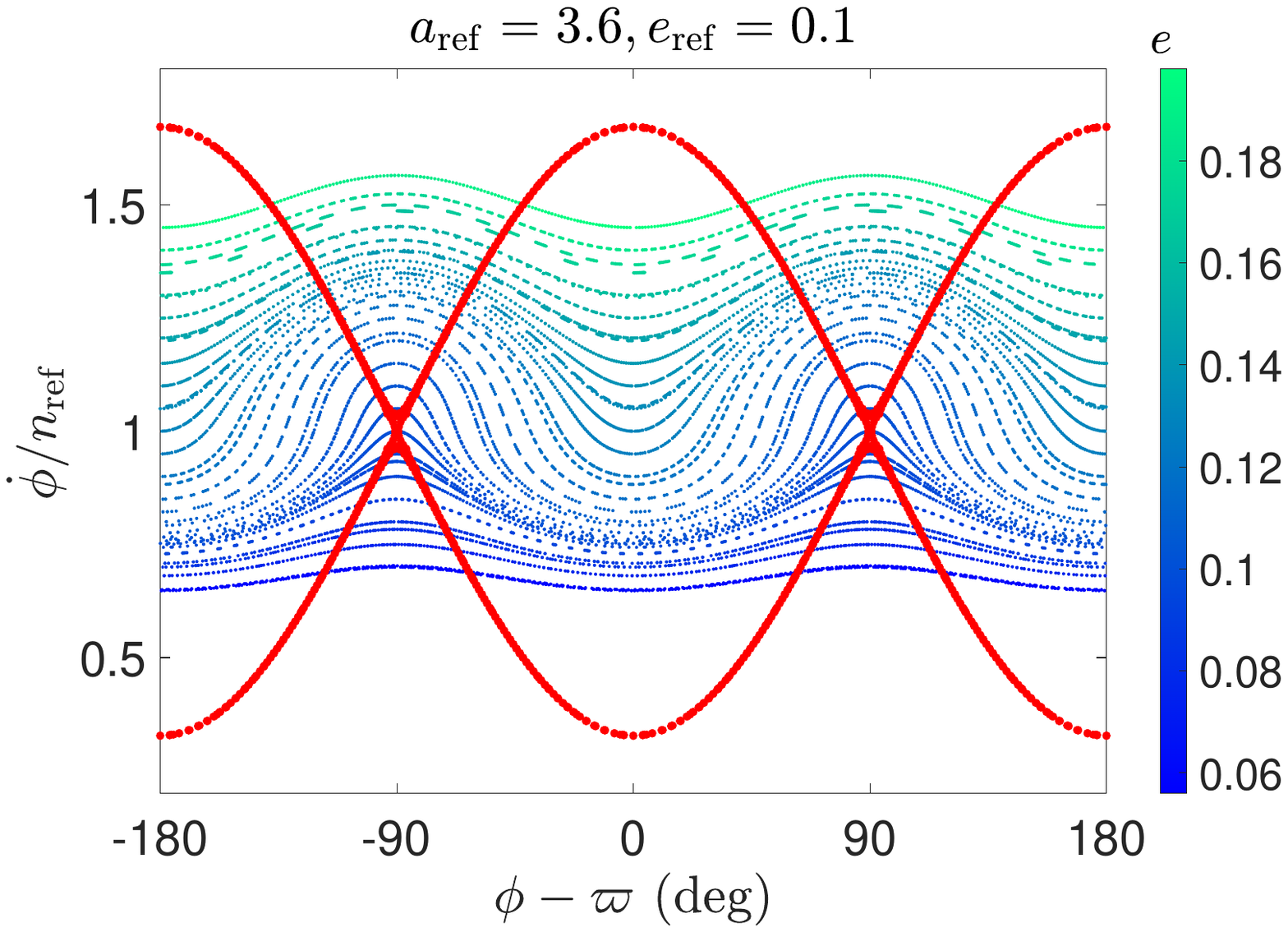}\\
\includegraphics[width=0.49\columnwidth]{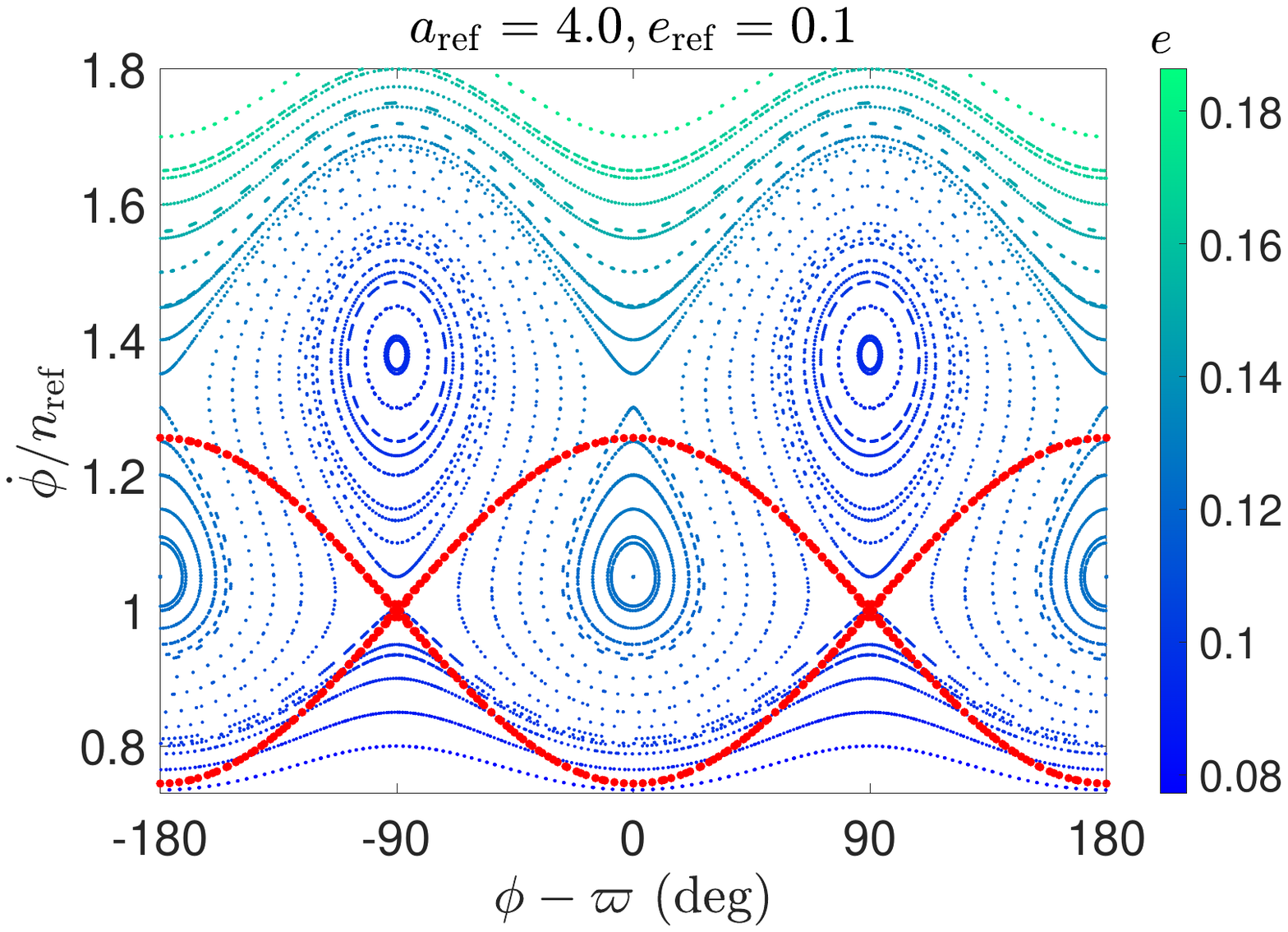}
\includegraphics[width=0.49\columnwidth]{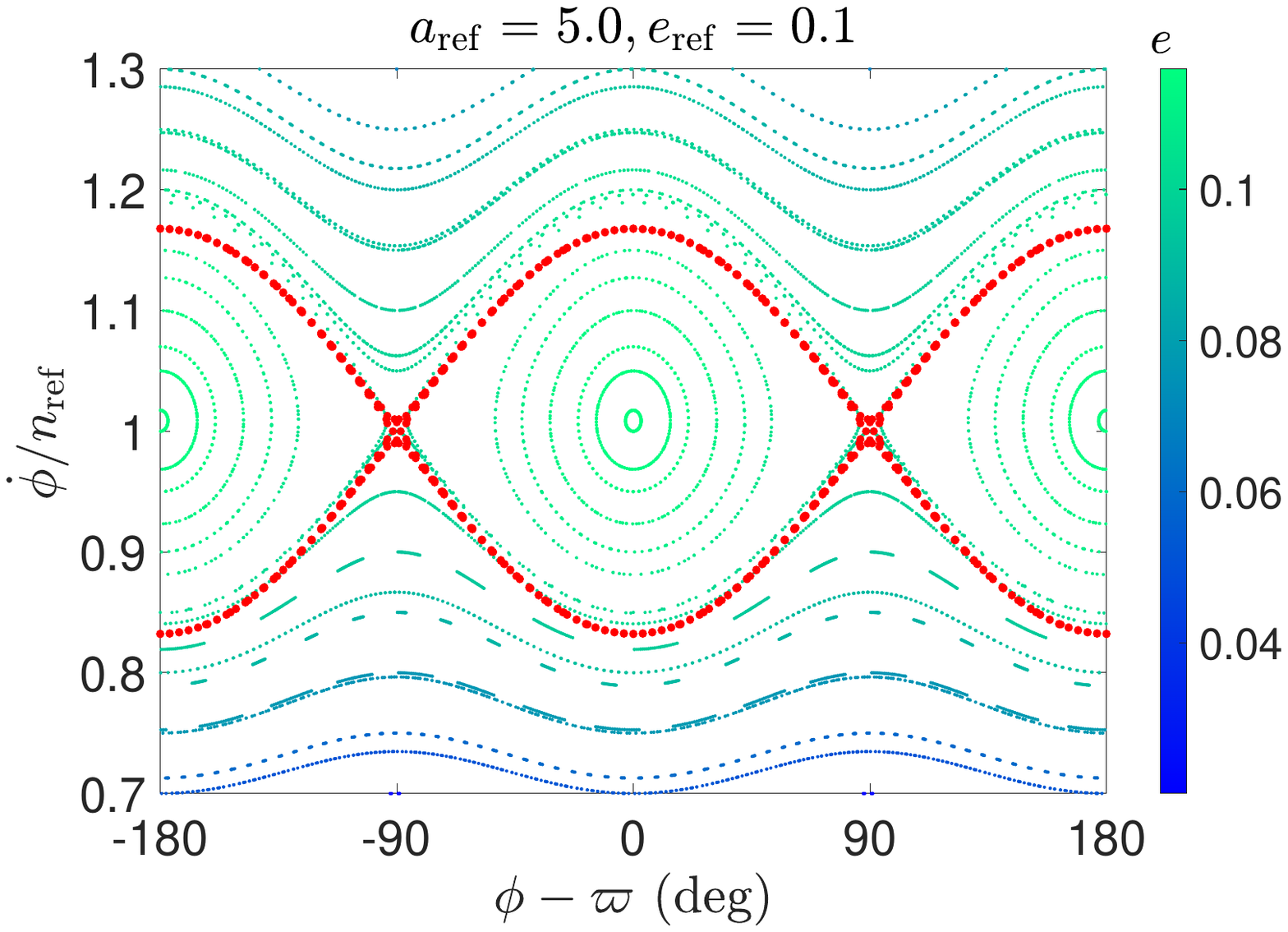}
\caption{Critical semimajor axis as a function of the primary-to-secondary mass ratio $\alpha_m$ for the synchronous (1:1) resonance (\textit{top-left panel}) and Poincar\'e section for the synchronous resonance together with the associated dynamical separatrices under the pendulum approximations for the cases of $a_{\rm ref} = 3.0$, $a_{\rm ref} = 3.5$, $a_{\rm ref} = 3.6$, $a_{\rm ref} = 4.0$ and $a_{\rm ref} = 5.0$ (\textit{the last five panels}). Separatrices under the pendulum approximations are marked by red dots. In simulations, the primary-to-secondary mass ratio is taken as $\alpha_m = 10$, at which the critical semimajor axis is $a_{c,1}=3.53$ for the synchronous (1:1) resonance.}
\label{Fig4}
\end{figure*}

If the equatorial elongation is fixed at $a_p/b_p = 1/0.95$ (the elongation adopted in the entire work), the critical semimajor axis is an increasing function of the primary-to-secondary mass ratio $\alpha_m$, as shown in the top-left panel of Fig. \ref{Fig4}. In particular, when $\alpha_m = 10$, the critical semimajor axes are $a_{c,1}=3.53$, $a_{c,2}=5.31$ and $a_{c,3}=1.76$ in normalised unit of length. As the critical semimajor axis $a_{c,3}=1.76$ is close to contact, it shows that the exchange of libration centre is physically impossible for the 2:1 resonance in the case of $\alpha_m = 10$. Thus, in the following simulations, we will not present simulation results for the 2:1 resonance.

To see the influence of reference semimajor axis (with fixed reference eccentricity at $e_{\rm ref}=0.1$) upon dynamical structures, we take the synchronous (1:1) resonance as an example and adopt three representative reference semimajor axes at $a_{\rm ref} = 3.0$ (below the critical value), $a_{\rm ref}=3.5$ and $3.6$ (near the critical value), and $a_{\rm ref}=4.0$ and $5.0$ (above the critical value) to produce Poincar\'e sections. With a given set of $(a_{\rm ref}, e_{\rm ref})$, the total angular momentum and the Hamiltonian are determined. Please refer to Fig. \ref{Fig4} for Poincar\'e sections. The cases of $a_{\rm ref} = 3$ and $a_{\rm ref}=4$ show that there are four libration islands and four saddle points (but their structures are different). Dynamical structures are distinctly different for the cases of $a_{\rm ref}=3.5$ and $3.6$ which are near the critical semimajor axis. Please refer to the middle-row panels of Fig. \ref{Fig4}, which show that there are no libration islands in the phase space. In the case of $a_{\rm ref}=5$, there are two islands of libration and two saddle points in the Poincar\'e section.

From the pendulum approximations, we can further derive the expression of resonant half-width. The resonant half-width in terms of $\Delta\dot\phi$ is given by
\begin{equation}\label{Eq23}
{\left( {\Delta \dot \phi /n} \right)_{1:1}} = \sqrt {\frac{{12{C_{22}}{a_{\rm ref}^2} - 30{C_{42}}}}{{{I_3}\left( {{a_{\rm ref}^2} - 3{I_3}} \right)}}},
\end{equation}
for the synchronous resonance, it is
\begin{equation}\label{Eq24}
{\left( {\Delta \dot \phi /n} \right)_{2:3}} = \sqrt {\frac{\left(42C_{22}a_{\rm ref}^2 - 135C_{42}\right)e_{\rm ref}}{{{I_3}\left( {{a_{\rm ref}^2} - \frac{{27}}{4}{I_3}} \right)}}},
\end{equation}
for the 2:3 spin-orbit resonance and it is
\begin{equation}\label{Eq25}
{\left( {\Delta \dot \phi /n} \right)_{2:1}} = \sqrt {\frac{{\left( {6{C_{22}}{a_{\rm ref}^2} + 15{C_{42}}} \right)e_{\rm ref}}}{{{I_3}\left( {{a_{\rm ref}^2} - \frac{3}{4}{I_3}} \right)}}}
\end{equation}
for the 2:1 spin-orbit resonance. Resonant half-width of the synchronous resonance is independent on the eccentricity while resonant width of the 2:3 (and 2:1) resonance is an increasing function of the eccentricity. Comparing to Table 1 of \citet{jafari2023surfing}, we can see that the expressions of resonant half-width are different\footnote{The author of \citet{jafari2023surfing} confirmed that there exist typos in their expressions (private communications).}.

Under the pendulum approximations (\ref{Eq16}-\ref{Eq18}), dynamical separatrix corresponds to the level curve of Hamiltonian passing through saddle points and it plays an important role in dividing the entire phase space into regions of libration and circulation. As a result, the dynamical separatrix is expressed by
\begin{equation*}
\frac{1}{2}\left( {\frac{1}{{{I_3}}} - \frac{3}{{{a_{\rm ref}^2}}}} \right)\Delta \Sigma _1^2 - \left[ {\frac{{3{C_{22}}}}{{{a_{\rm ref}^3}}} - \frac{{15{C_{42}}}}{{2{a_{\rm ref}^5}}}} \right]\left( {\cos 2{\sigma _1} \pm 1} \right) = 0
\end{equation*}
for the synchronous (1:1) resonance, it is expressed by
\begin{equation*}
\frac{1}{2}\left( {\frac{1}{{{I_3}}} - \frac{{27}}{{4{a_{\rm ref}^2}}}} \right)\Delta \Sigma _1^2 - \left[ {\frac{{21{C_{22}}}}{{2{a_{\rm ref}^3}}} - \frac{{135{C_{42}}}}{{4{a_{\rm ref}^5}}}} \right]e_{\rm ref}\left( {\cos 2{\sigma _1} \pm 1} \right) = 0
\end{equation*}
for the 2:3 resonance and it is expressed by
\begin{equation*}
\frac{1}{2}\left( {\frac{1}{{{I_3}}} - \frac{3}{{4{a_{\rm ref}^2}}}} \right)\Delta \Sigma _1^2 + \left[ {\frac{{3{C_{22}}}}{{2{a_{\rm ref}^3}}} + \frac{{15{C_{42}}}}{{4{a_{\rm ref}^5}}}} \right]e_{\rm ref}\left( {\cos 2{\sigma _1} \mp 1} \right){\rm{ = }}0
\end{equation*}
for the 2:1 resonance. In the expressions of separatrix, the upper sign is for the case of $a_{\rm ref}>a_{c}$ and the lower sign is for the case of $a_{\rm ref}<a_{c}$.

Dynamical separatrices of synchronous resonance with reference semimajor axes at $a_{\rm ref}=3.0$, $a_{\rm ref}=3.5$, $a_{\rm ref}=3.6$ $a_{\rm ref}=4.0$ and $a_{\rm ref}=5.0$ are compared with the associated Poincar\'e sections, as shown in Fig. \ref{Fig4}. It should be noted that, at the section defined by $M=0$, it holds $\sigma_1 = \phi-\varpi$. We can see that, under the pendulum approximations, the libration centre is located at $2(\phi-\varpi) = \pi$ for the cases of $a_{\rm ref} < 3.53$ and it is located at $2 (\phi-\varpi) = 0$ for the cases of $a_{\rm ref}>3.53$. This is in consistent with our previous discussion.

In the case of $a_{\rm ref}=3.0$, the dynamical separatrix under the pendulum approximation can qualitatively describe the high-$\dot\phi$ libration islands but it cannot describe those low-$\dot\phi$ islands of libration. For the cases of $a_{\rm ref}=3.5$ and $3.6$ (near the critical value), there are no islands of libration in the Poincar\'e section. Evidently, the dynamical separatrix under the pendulum approximation cannot describe the associated structures arising in the Poincar\'e section. In the case of $a_{\rm ref}=4.0$, there are four islands of libration. The separatrix under the pendulum approximation cannot describe the phase-space structures. In the case of $a_{\rm ref}=5.0$, the pendulum approximation can well describe the numerical structures. 

To conclude, we can see that the pendulum approximations can only describe numerical structures for those cases with reference semimajor axes far away from the critical value. For those cases with reference semimajor axes which are not far from the critical value, pendulum approximations fail to describe the numerical structures. The failure of pendulum approximations near the critical case is due to the strong coupling between rotation and translation under the considered problem. Thus, we may ask: can we formulate a general resonant Hamiltonian which is applicable for describing spin-orbit coupling in the entire range of reference semimajor axes? The purpose of Section \ref{Sect6} is to address this problem.

\section{Adiabatic approximation}
\label{Sect6}

In this section, we intend to formulate resonant Hamiltonian to describe resonant structures under the spin-orbit coupling problem by taking advantage of perturbative treatments developed by \citet{wisdom1985perturbative}.

\subsection{Resonant Hamiltonian model}
\label{Sect6-1}

When the primary body is locked inside a certain $k_1$:$k_2$ spin-orbit resonance, the associated argument $\sigma_1$ becomes a long-period variable and the other angle $\sigma_2$ is a short-period variable. With a given total angular momentum, $(\sigma_1,\Sigma_1)$ corresponds to the slow DOF and $(\sigma_2,\Sigma_2)$ corresponds to the fast DOF. As a result, the Hamiltonian (\ref{Eq13}) determines a typical separable dynamical model \citep{henrard1990semi}. During one period of the fast DOF, the variables associated with the slow DOF have negligible changes. According to the concept of perturbative treatment proposed by \citet{wisdom1985perturbative}, the variables associated with the slow DOF can be treated as parameters when considering the evolution of the fast DOF over its timescale. Such an approximation is usually known as adiabatic approximation, which has been widely applied in varieties of problems \citep{henrard1987perturbative,henrard1989motion,yokoyama1996simple,saillenfest2020long,lei2022systematic,lei2022zeipel,saillenfest2016long}. 

In practice, the variables $(\sigma_1,\Sigma_1)$ are considered as parameters when considering the evolution of the fast DOF $(\sigma_2,\Sigma_2)$. Under the assumption, the subsystem associated with the fast DOF becomes integrable and the solutions are level curves of the Hamiltonian in the $(\sigma_2,\Sigma_2)$ space \citep{saillenfest2020long}. As a result, the action-angle variables can be introduced as follows \citep{morbidelli2002modern}:
\begin{equation}\label{Eq26}
\sigma _2^{\rm{*}}{\rm{ = }}\frac{{2\pi }}{T}t,\quad \Sigma _2^* = \frac{1}{{2\pi }}\int\limits_0^{2\pi } {{\Sigma _2}{\rm d}{\sigma _2}}
\end{equation}
where $T$ is the period and the Arnold action $\Sigma_2$ stands for the signed area bounded by the solution curve during one period of the fast DOF. Performing the canonical transformation given by equation (\ref{Eq26}), we can write the Hamiltonian (\ref{Eq13}) as an integrable Hamiltonian form,
\begin{equation}\label{Eq27}
{\cal H}\left( {{G_{\rm tot}};{\sigma _1},{\Sigma _1},{\sigma _2},{\Sigma _2}} \right) \Rightarrow {\cal H}\left( {{G_{\rm tot}};{\sigma _1},{\Sigma _1},\Sigma _2^*} \right).
\end{equation}
The angle $\sigma_2^*$ is absent from the Hamiltonian ${\cal H}\left( {{G_{\rm tot}};{\sigma _1},{\Sigma _1},\Sigma _2^*} \right)$, resulting in that the Arnold action $\Sigma_2^*$ becomes a motion integral. For convenience, we denote the motion integral as the adiabatic invariant $J = \Sigma_2^*$.

During the long-term evolution, there are two conserved quantities including $J$ and ${\cal H}$ besides the total angular momentum. Thus, the original 2-DOF Hamiltonian model is integrable. Naturally, global structures can be explored by analysing phase portraits. In practice, phase portraits are produced by plotting level curves of adiabatic invariant $J$ with given Hamiltonian ${\cal H}$ and total angular momentum $G_{\rm tot}$ \citep{lei2022zeipel}. 

\subsection{Analytical and numerical structures}
\label{Sect6-2}

According to the discussions made in Section \ref{Sect3}, ${\cal H}$ and $G_{\rm tot}$ can be characterised by reference semimajor axis $a_{\rm ref}$ and reference eccentricity $e_{\rm ref}$. Please refer to equation (\ref{Eq6}) for the total angular momentum and refer to equation (\ref{Eq7}) for the Hamiltonian ${\cal H}$ in the case of synchronous resonance. For the 2:3 resonance, the Hamiltonian is characterised by
\begin{equation}\label{Eq28}
{\cal H} = {\left. {\cal H} \right|_{a = a_{\rm ref}, e = e_{\rm ref}, f = 0, \phi-\varpi = \pi/2, \dot\phi = 1.5 n_{\rm ref}}}.
\end{equation}
A pair of $(a_{\rm ref},e_{\rm ref})$ determines the set of conserved quantities $(G_{\rm tot},{\cal H})$. It should be noted that the 2:1 resonance is not considered in practical simulations because, in the case of $\alpha_m = 10$, the critical semimajor axis $a_{c,3} = 1.76$ is smaller than the allowed minimum distance.

\begin{figure*}
\centering
\includegraphics[width=0.49\columnwidth]{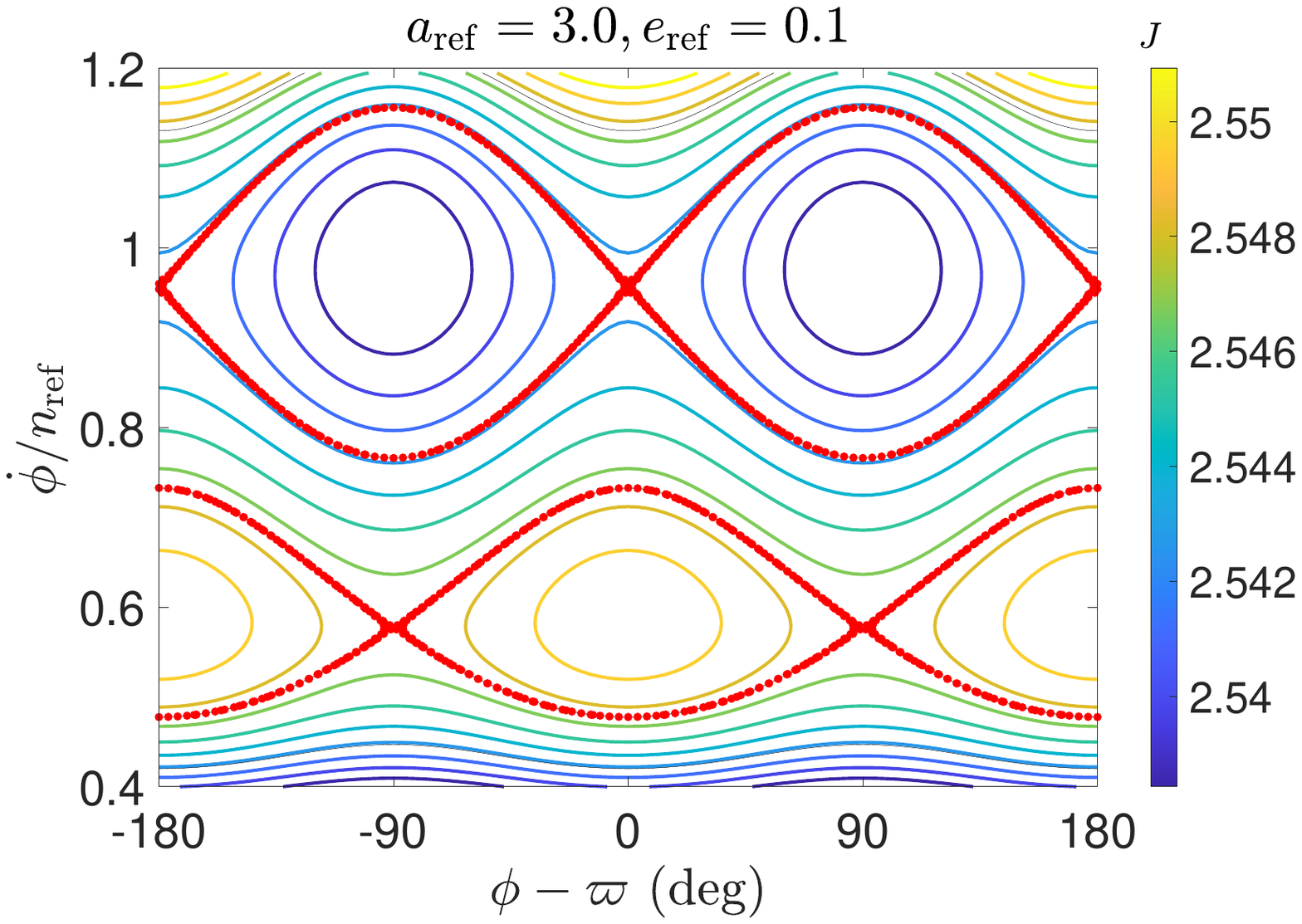}
\includegraphics[width=0.49\columnwidth]{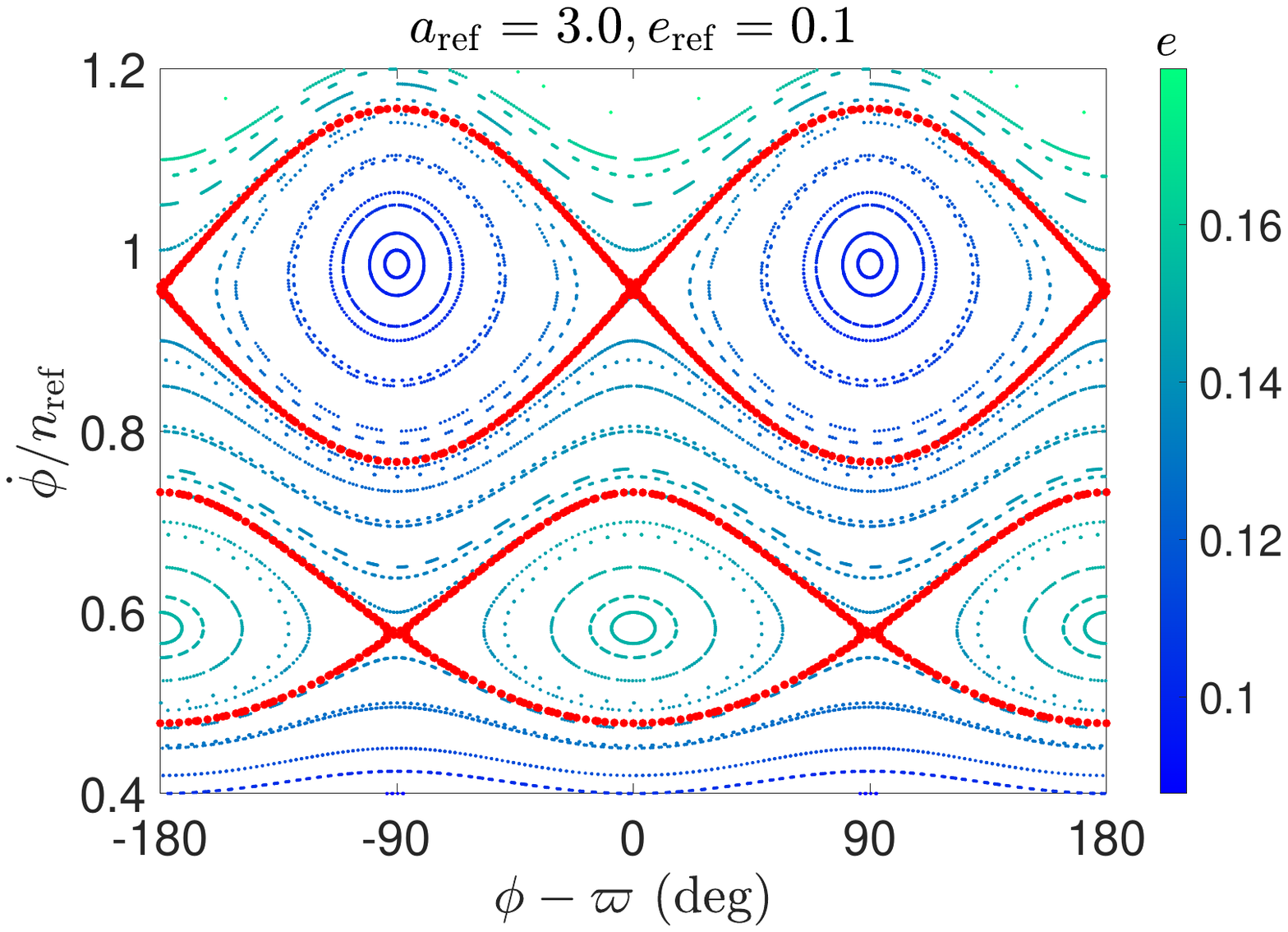}\\
\includegraphics[width=0.49\columnwidth]{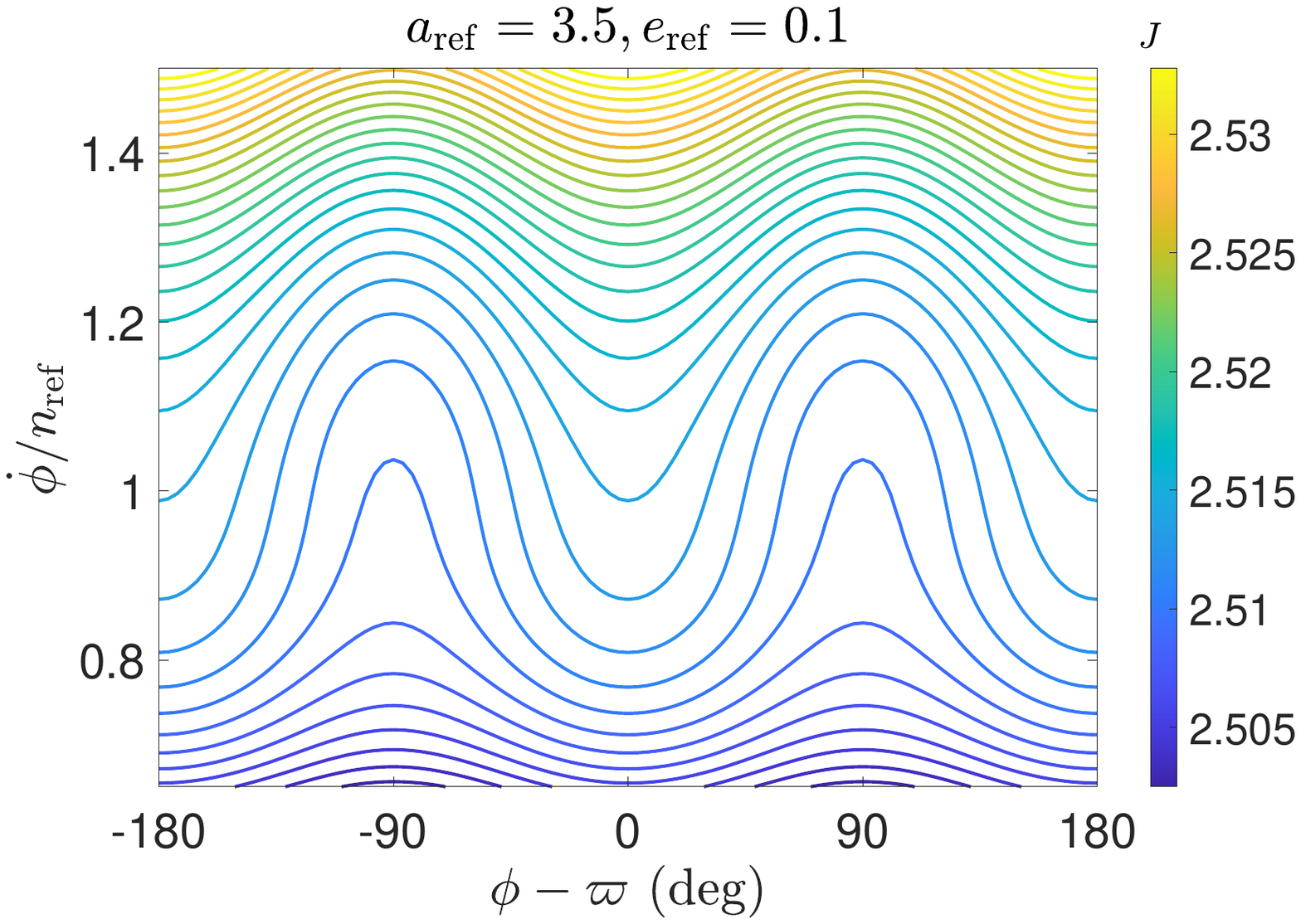}
\includegraphics[width=0.49\columnwidth]{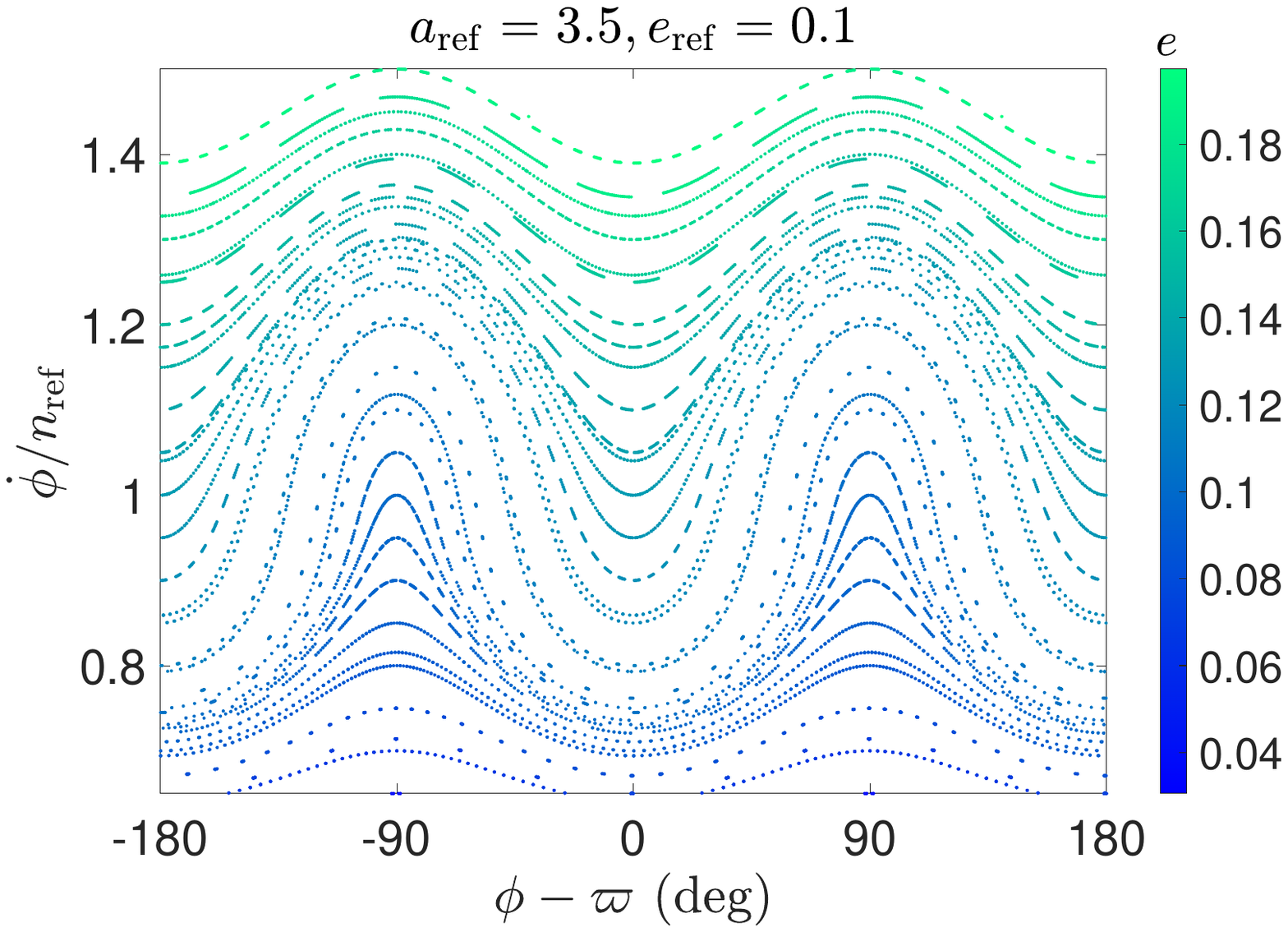}\\
\includegraphics[width=0.49\columnwidth]{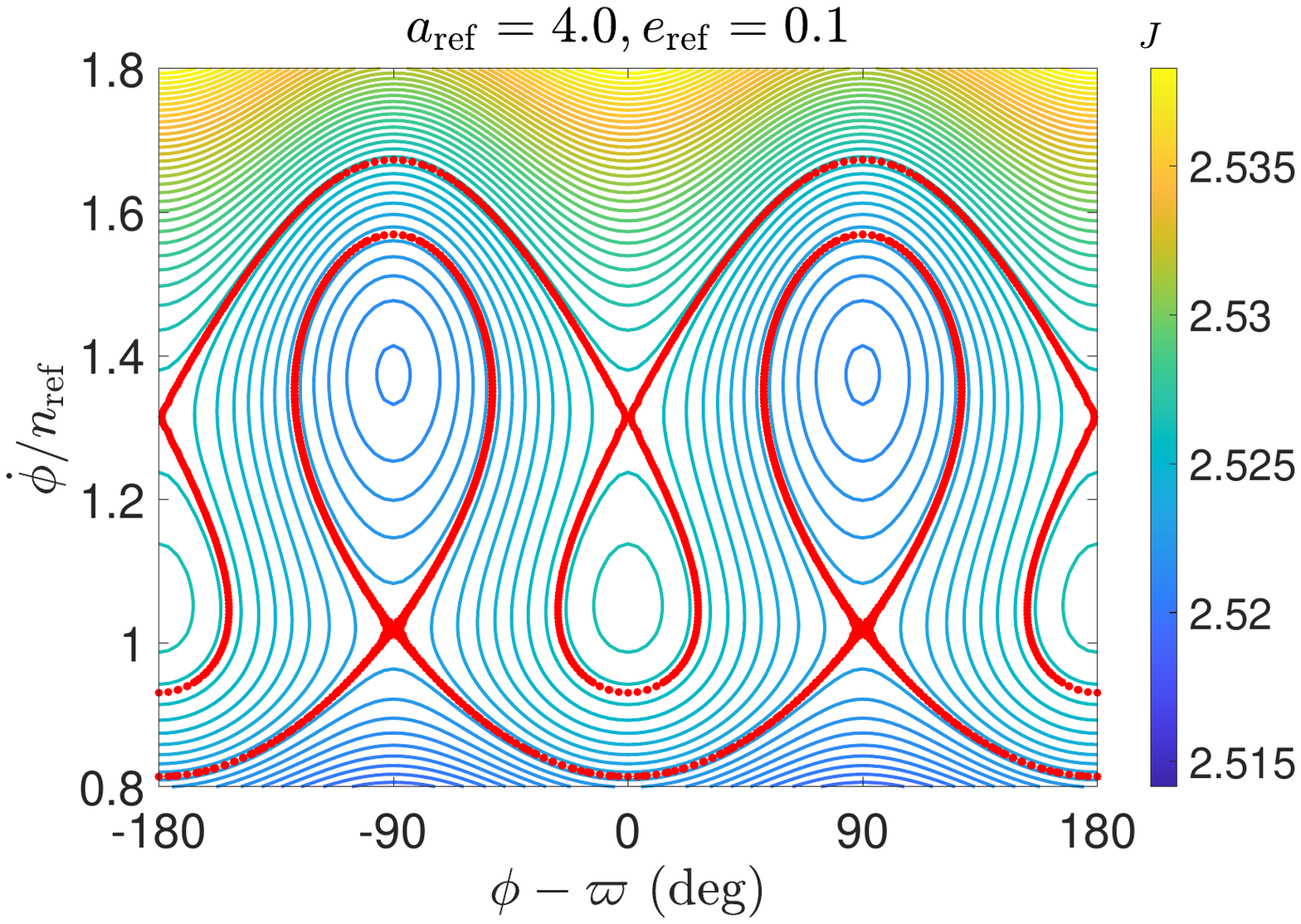}
\includegraphics[width=0.49\columnwidth]{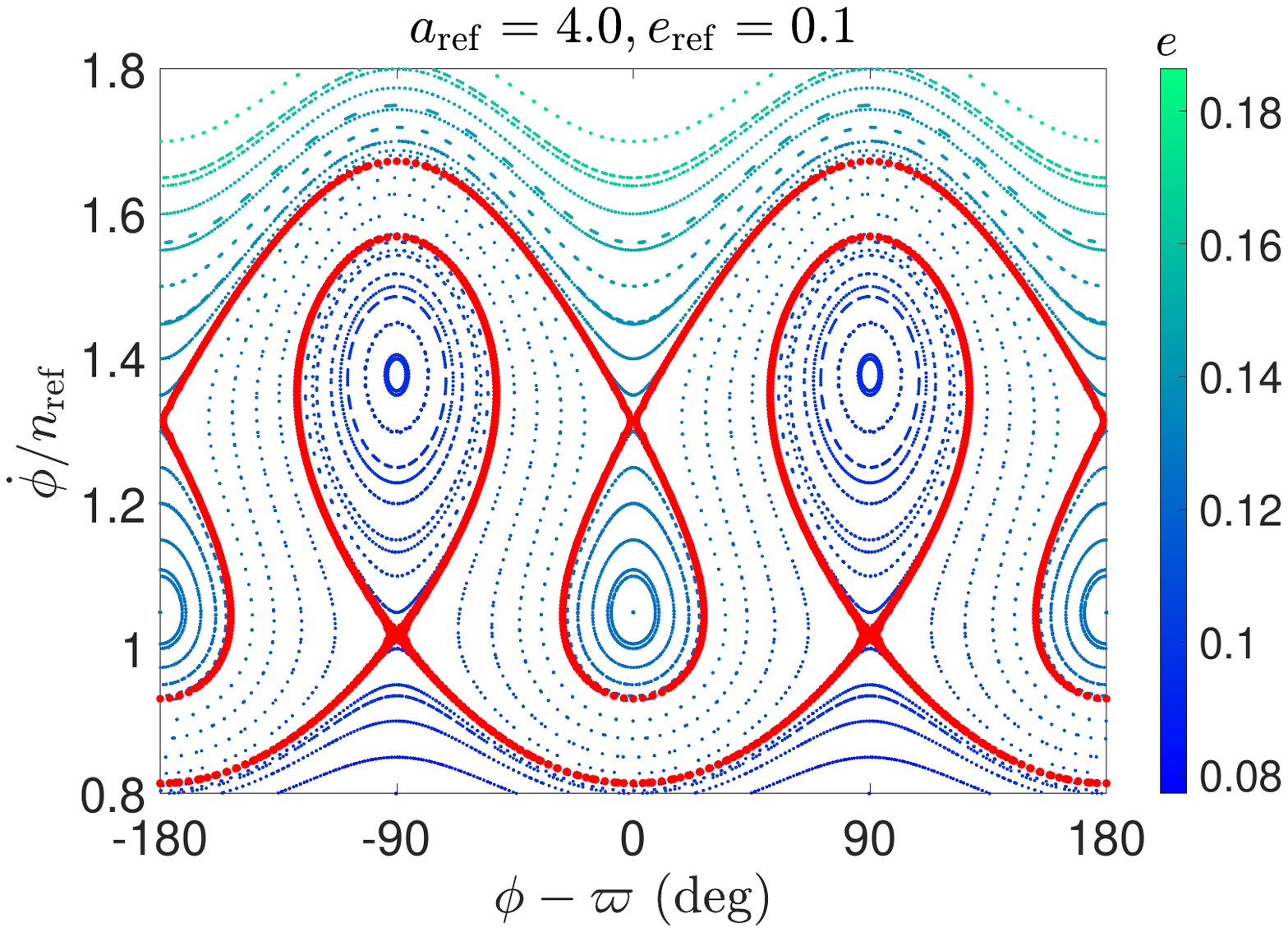}\\
\includegraphics[width=0.49\columnwidth]{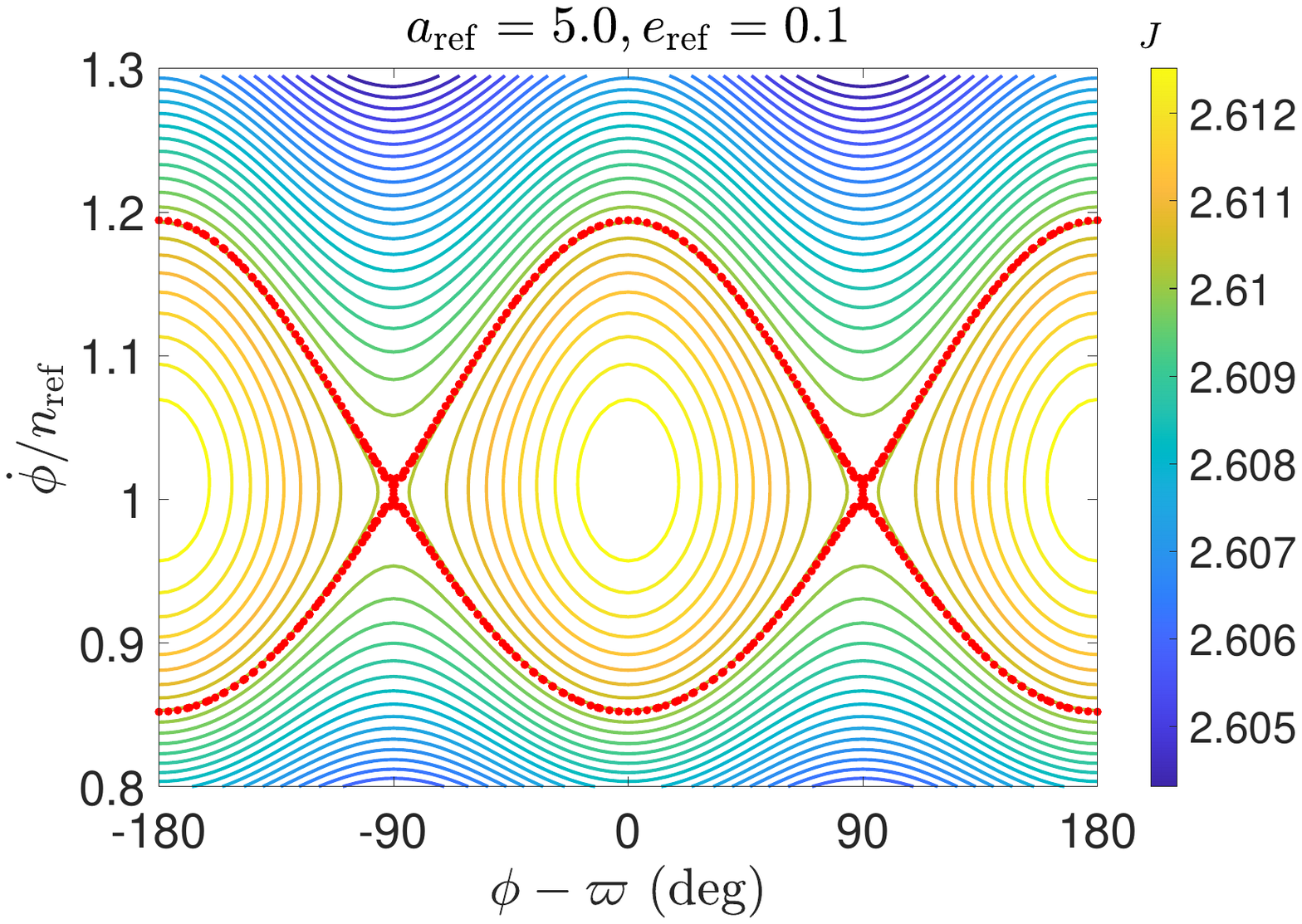}
\includegraphics[width=0.49\columnwidth]{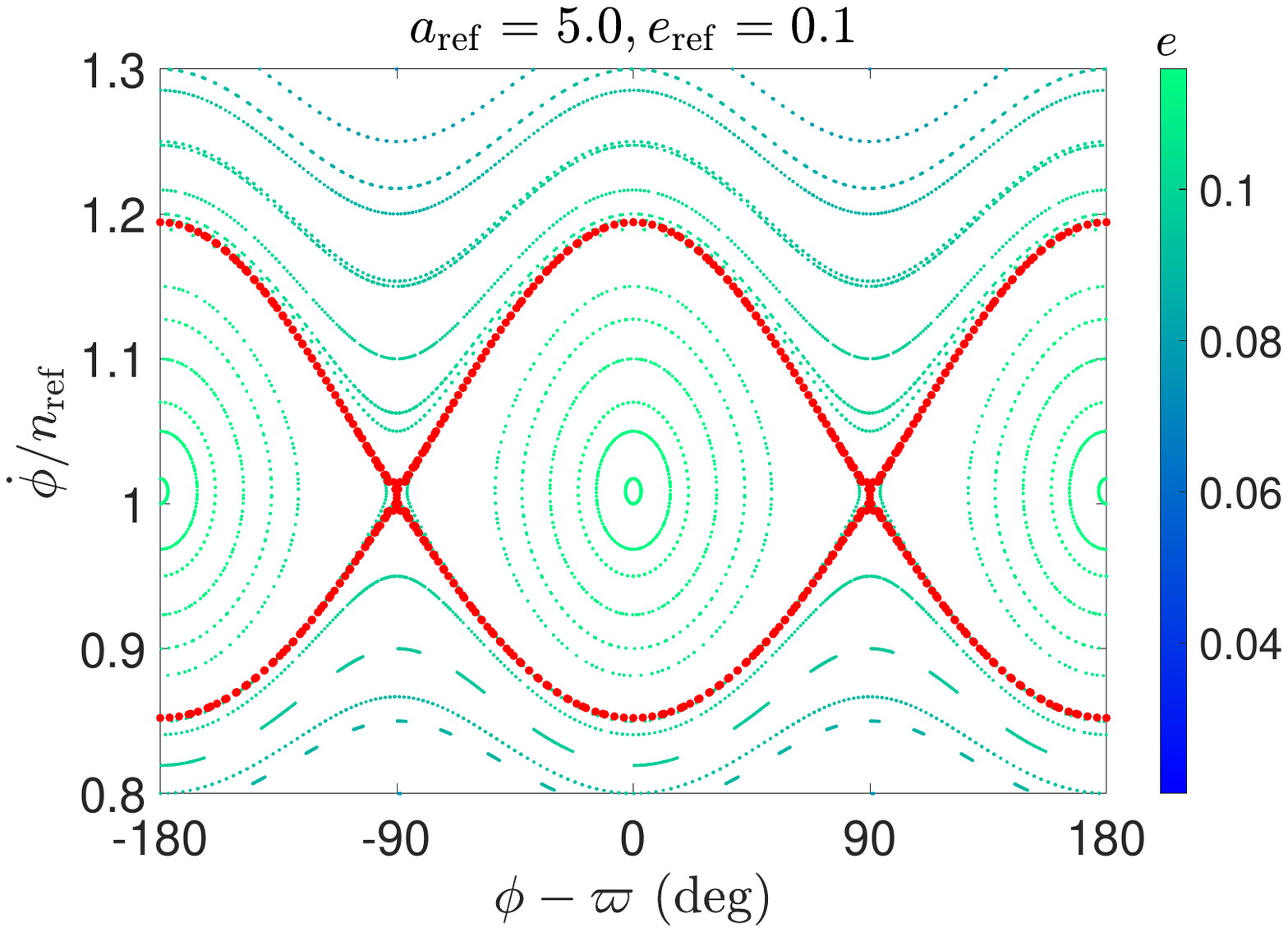}
\caption{Phase portraits (\textit{left panels}) and Poincar\'e sections (\textit{right panels}) for synchronous (1:1) resonances of the primary body. In all panels (except for the cases of $a_{\rm ref}=3.5$), the red dots stand for analytical dynamical separatrices, which divide the entire phase space into the regions of circulation and libration. In the case of $a_{\rm ref}=3.5$, there are no saddle points and thus no separatrices.}
\label{Fig5}
\end{figure*}

\begin{figure*}
\centering
\includegraphics[width=0.49\columnwidth]{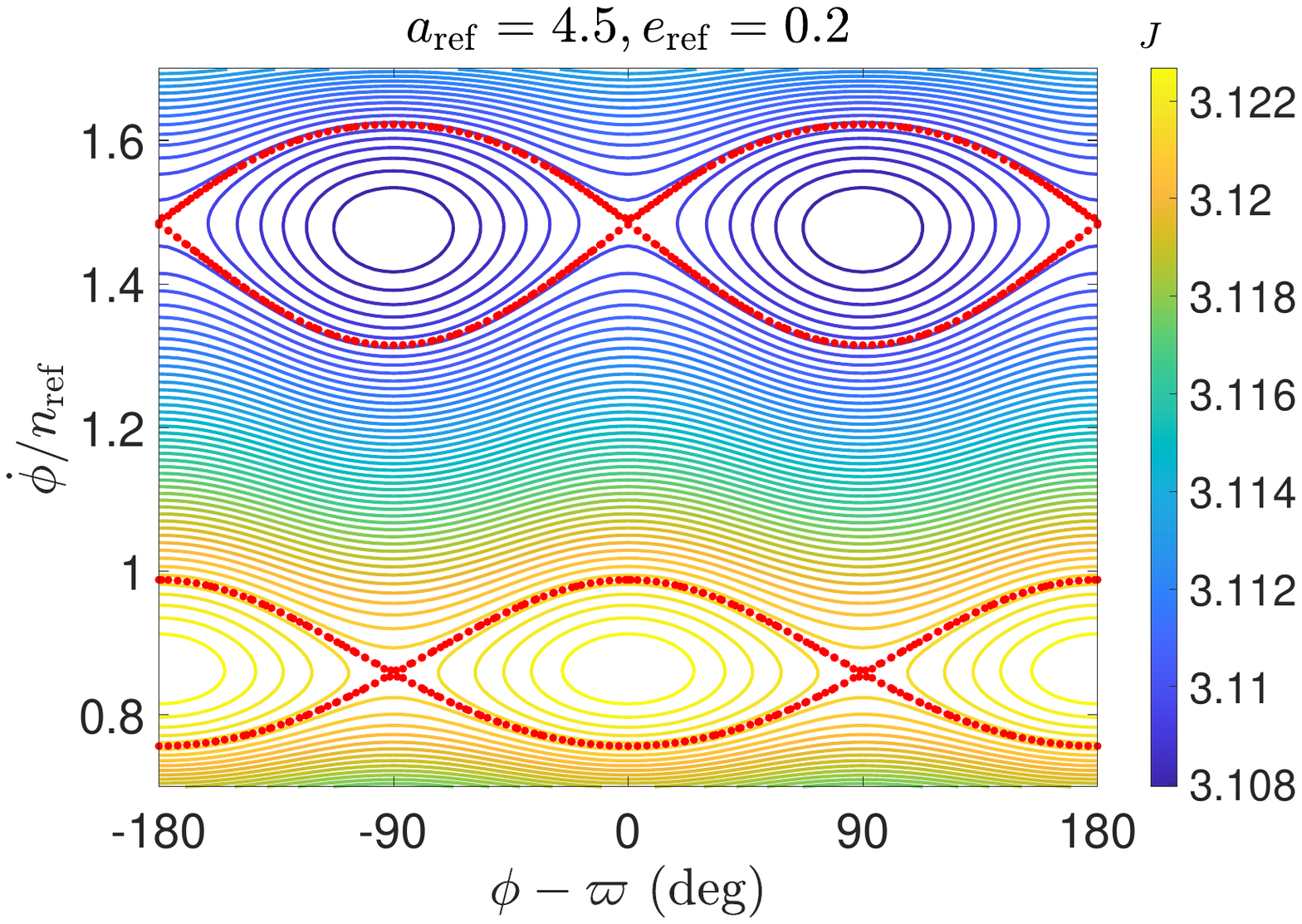}
\includegraphics[width=0.49\columnwidth]{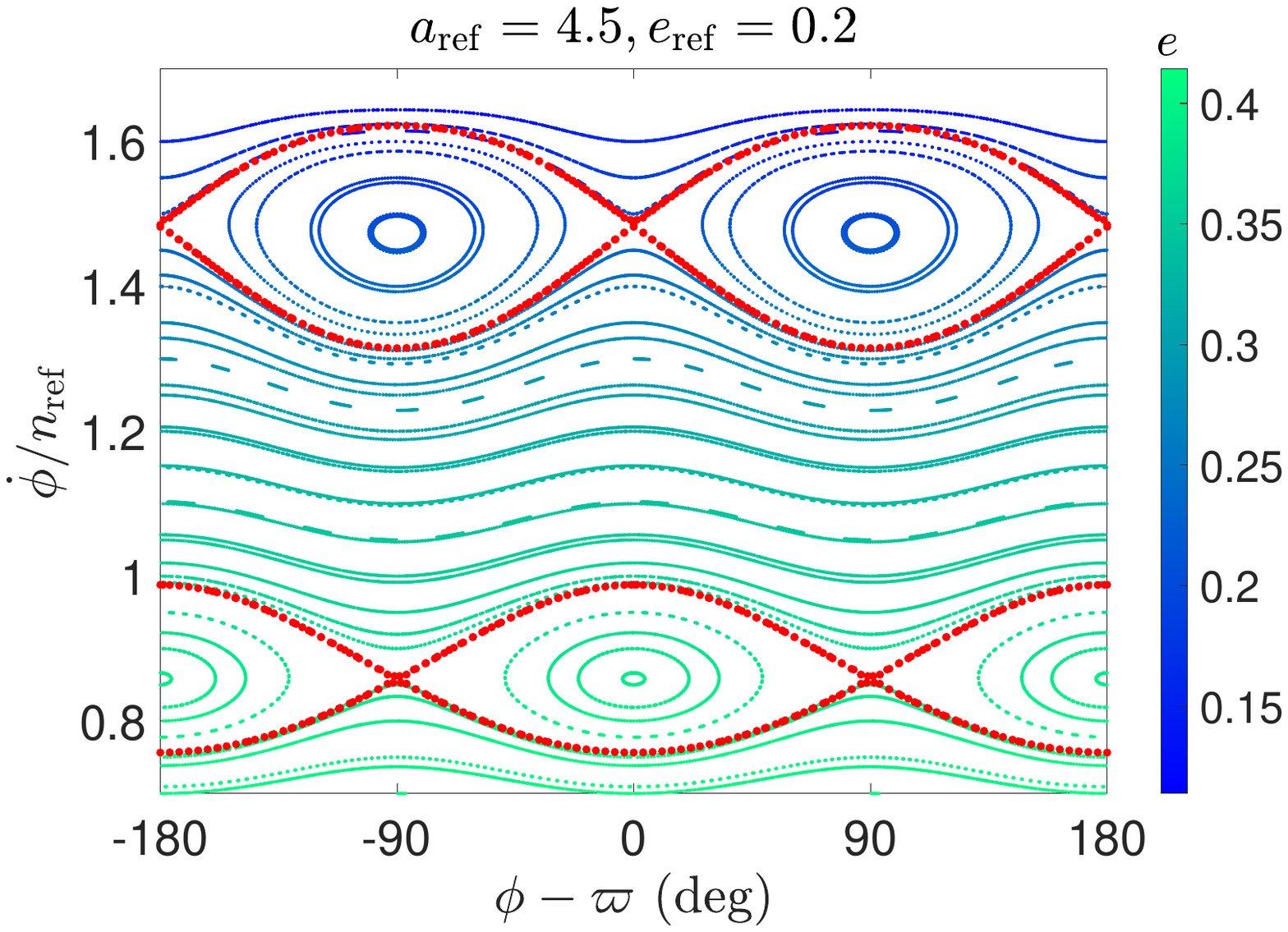}\\
\includegraphics[width=0.49\columnwidth]{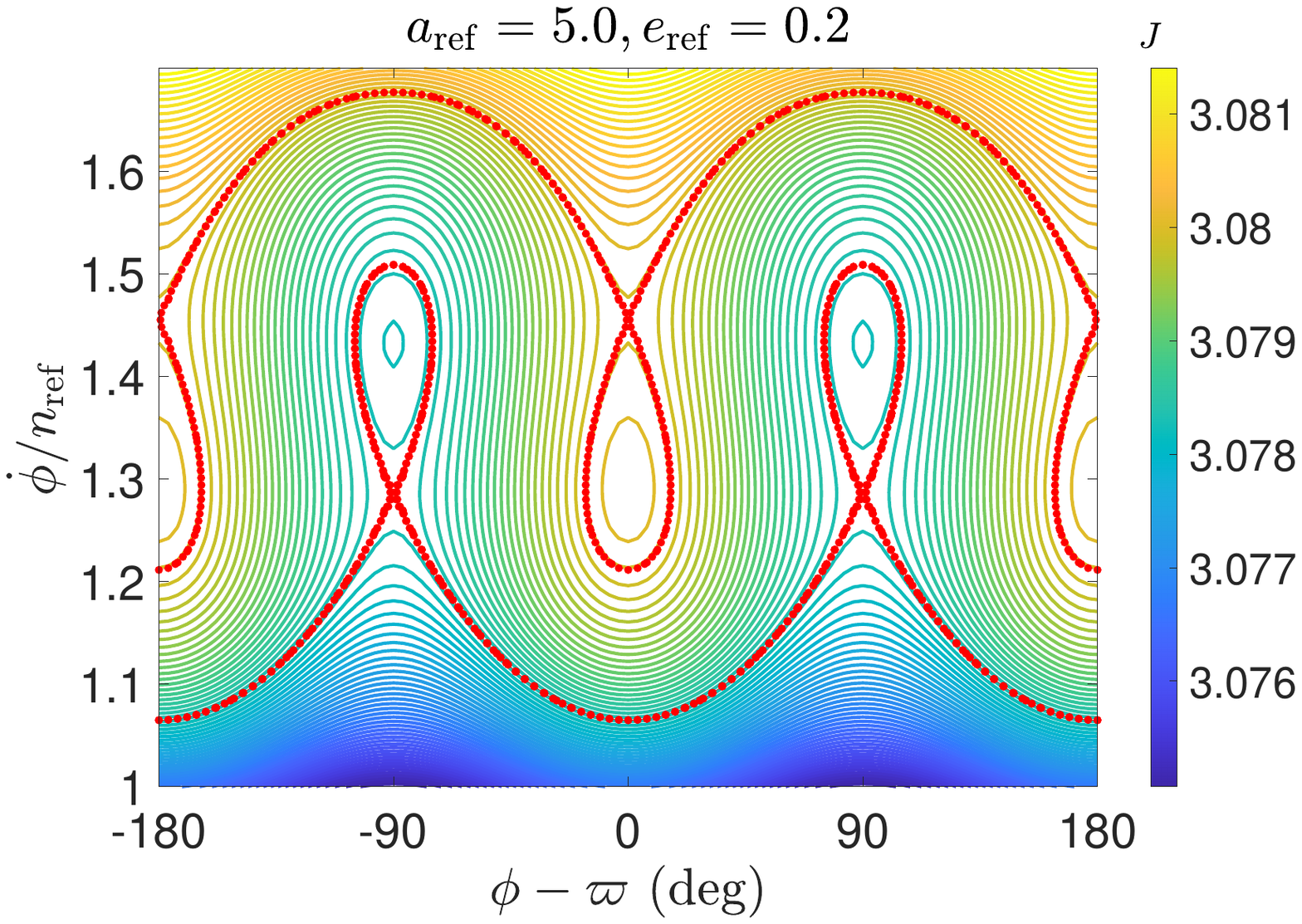}
\includegraphics[width=0.49\columnwidth]{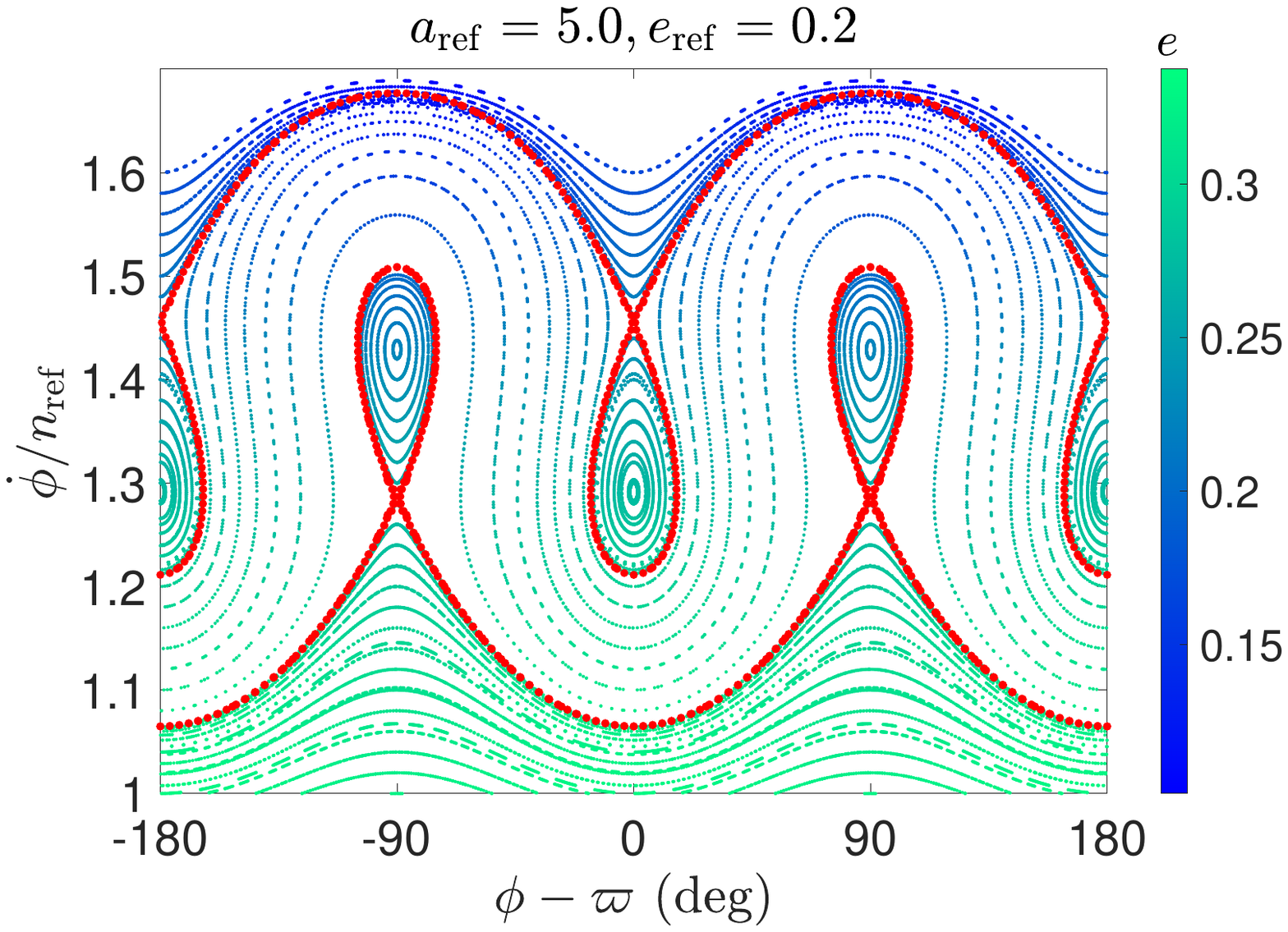}\\
\includegraphics[width=0.49\columnwidth]{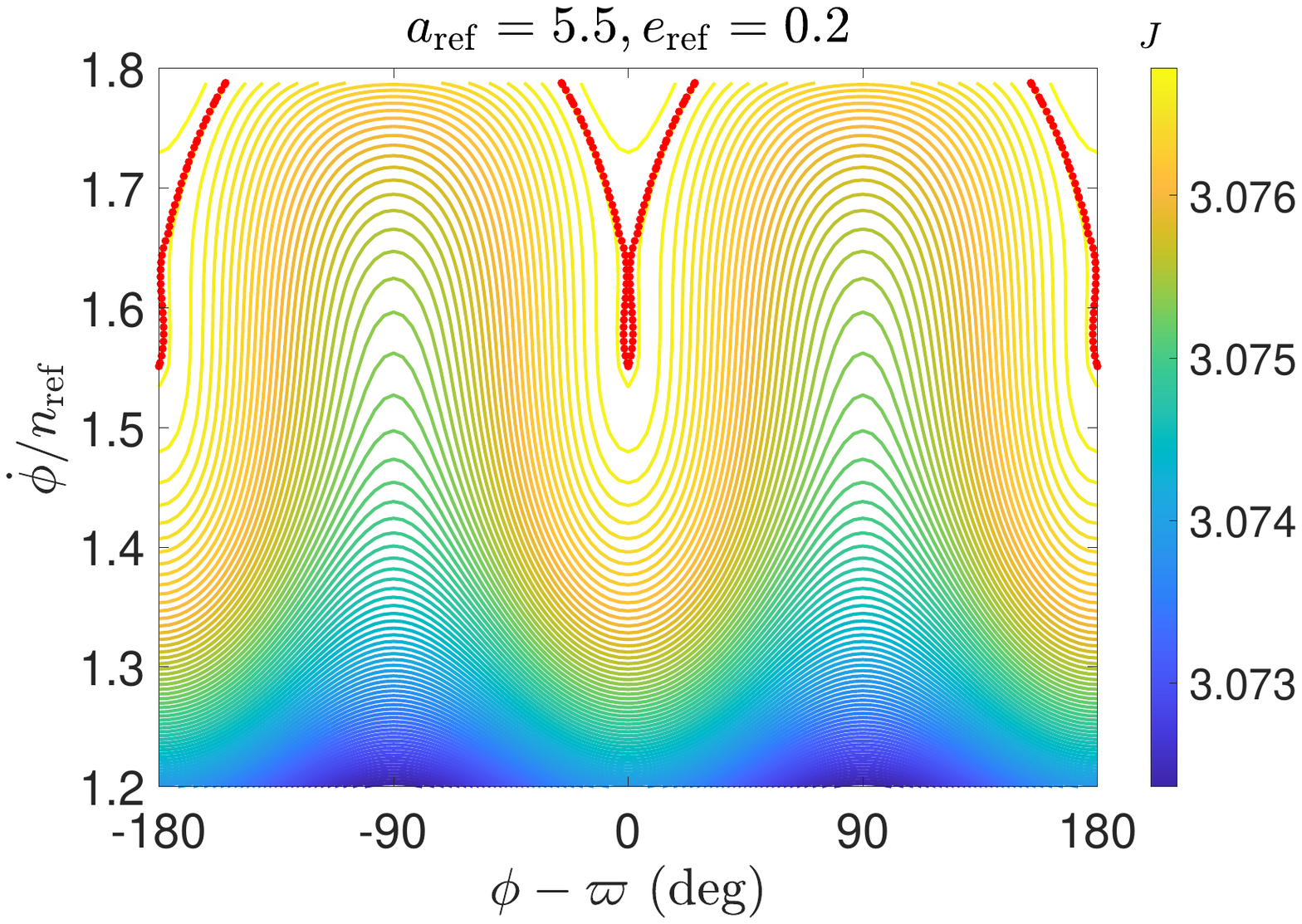}
\includegraphics[width=0.49\columnwidth]{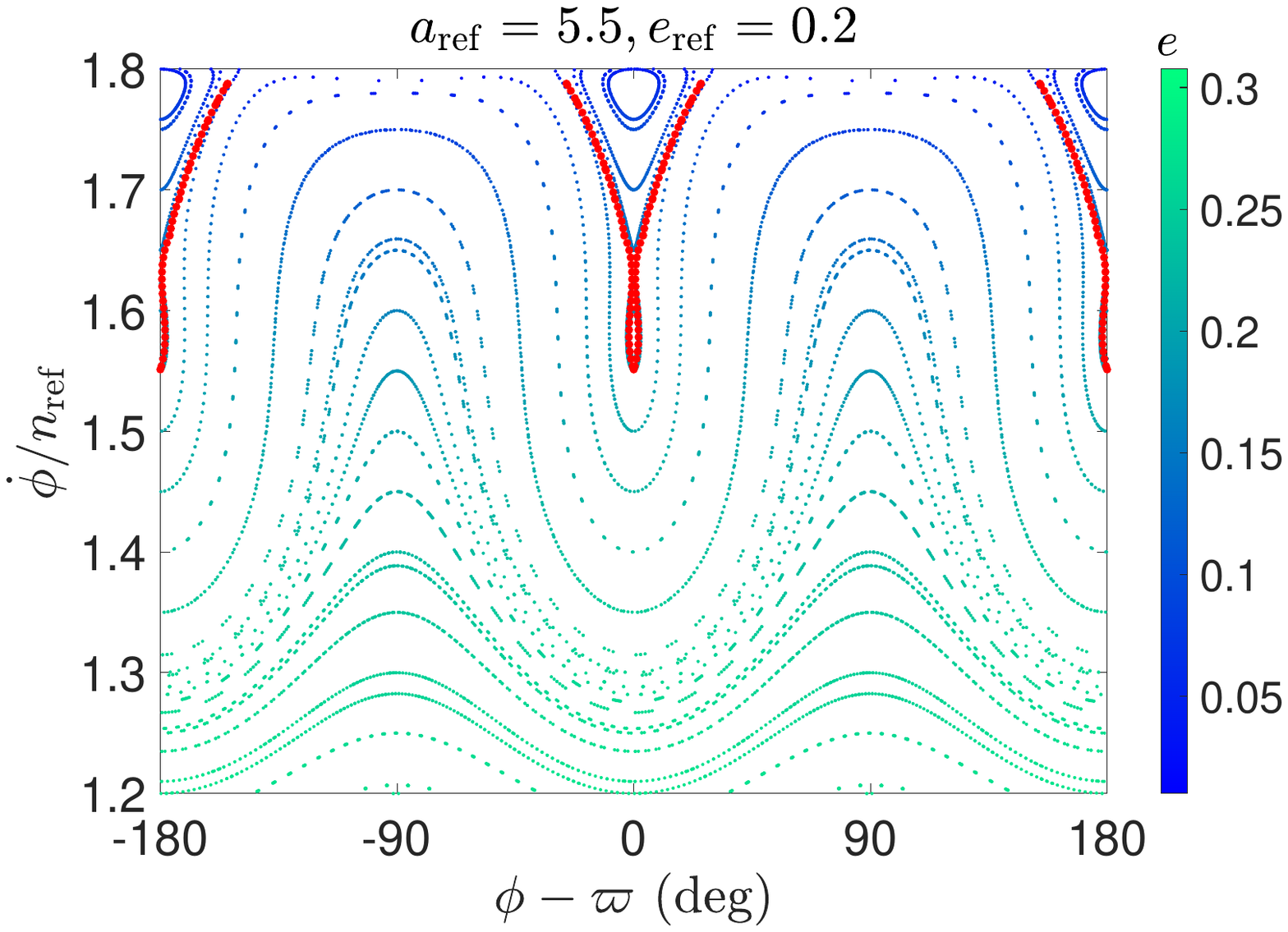}
\caption{Phase portraits (\textit{left panels}) and Poincar\'e sections (\textit{right panels}) for the 2:3 spin-orbit resonances of the primary body. In all panels, the red dots stand for analytical separatrices.}
\label{Fig6}
\end{figure*}

For the synchronous (1:1) resonance, phase portraits and Poincar\'e sections are produced and compared in Fig. \ref{Fig5} for the cases of $(a_{\rm ref}=3,e_{\rm ref}=0.1)$, $(a_{\rm ref}=3.5,e_{\rm ref}=0.1)$, $(a_{\rm ref}=4,e_{\rm ref}=0.1)$ and $(a_{\rm ref}=5,e_{\rm ref}=0.1)$. Analytical and numerical results for the 2:3 resonance are shown in Fig. \ref{Fig6} where the cases of $(a_{\rm ref}=4.5,e_{\rm ref}=0.2)$, $(a_{\rm ref}=5,e_{\rm ref}=0.2)$ and $(a_{\rm ref}=5.5,e_{\rm ref}=0.2)$ are considered. In both figures, left-column panels represent phase portraits and the right-column panels stand for Poincar\'e sections. The colour index shown in phase portraits stands for the magnitude of the adiabatic invariant $J$ and the colour index shown in Poincar\'e sections correspond to the orbital eccentricity. The dynamical separatrices corresponding to the level curves passing through saddle points are marked by red dots and they divide the entire phase space into regions of libration and circulation.

From Figs. \ref{Fig5} and \ref{Fig6}, we can observe an excellent correspondence between analytical structures arising in phase portraits and numerical structures arising in Poincar\'e sections. It validates the resonant Hamiltonian model for describing spin-orbit resonances for the primary body under strong coupling environments.

For the synchronous (1:1) resonance (see Fig. \ref{Fig5}), dynamical structures are different when the reference semimajor axis is changed. In the case of $a_{\rm ref} = 3$, the high-$\dot\phi$ islands of libration are bounded by the separatrix stemming from those high-$\dot\phi$ saddle points and those low-$\dot\phi$ islands of libration are bounded by the separatrix stemming from those low-$\dot\phi$ saddle points. In the case of $a_{\rm ref} = 3.5$, there are no islands of libration and saddle points in both the phase portraits and Poincar\'e sections. In the case of $a_{\rm ref} = 4$, those high-$\dot\phi$ islands of libration are bounded by the separatrix stemming from those low-$\dot\phi$ saddle points and vice versa. In the case of $a_{\rm ref} = 5$, the islands of libration in the high-$\dot\phi$ branch disappear and only those islands centred at the nominal resonance remain.

For the 2:3 resonance (see Fig. \ref{Fig6}), the dynamical structures are changed at different reference semimajor axes. In the case of $a_{\rm ref}=4.5$, the high-$\dot\phi$ branch and the low-$\dot\phi$ branch are distinctly separable. The high-$\dot\phi$ islands are bounded by the separatrix stemming from high-$\dot\phi$ saddle points and vice versa. It is different for the case of $a_{\rm ref}=5$. The high-$\dot\phi$ islands are bounded by the separatrix stemming from low-$\dot\phi$ saddle points and those low-$\dot\phi$ islands are bounded by the separatrix stemming from high-$\dot\phi$ saddle points. In the case of $a_{\rm ref}=5.5$, the islands of libration centred at $\phi-\varpi = \pm \pi/2$ disappear. However, the islands centred at $\phi-\varpi = 0$ become very small and additional islands centred at $\phi-\varpi = 0$ bifurcate in the high-$\dot\phi$ space.

\begin{figure*}
\centering
\includegraphics[width=0.49\columnwidth]{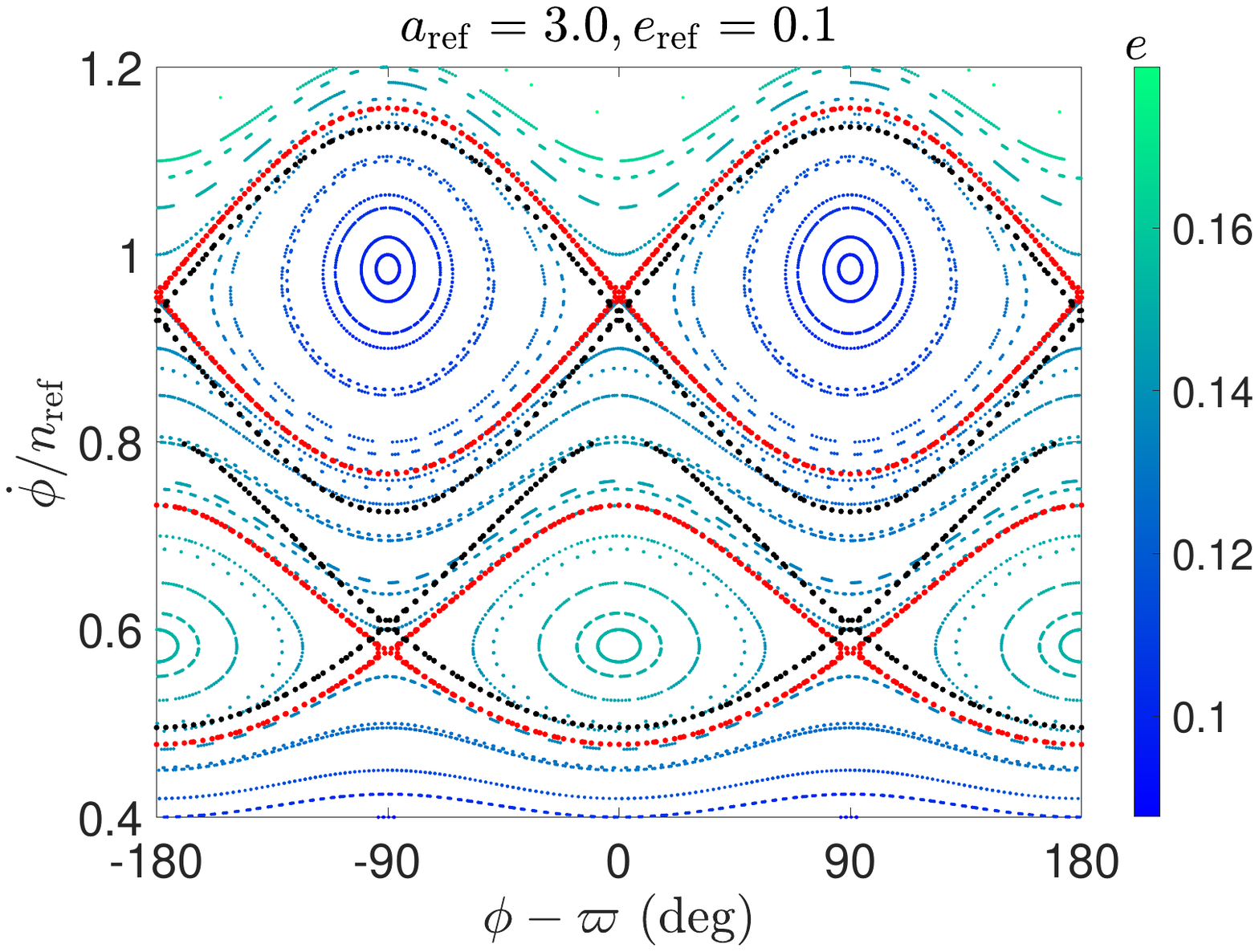}
\includegraphics[width=0.49\columnwidth]{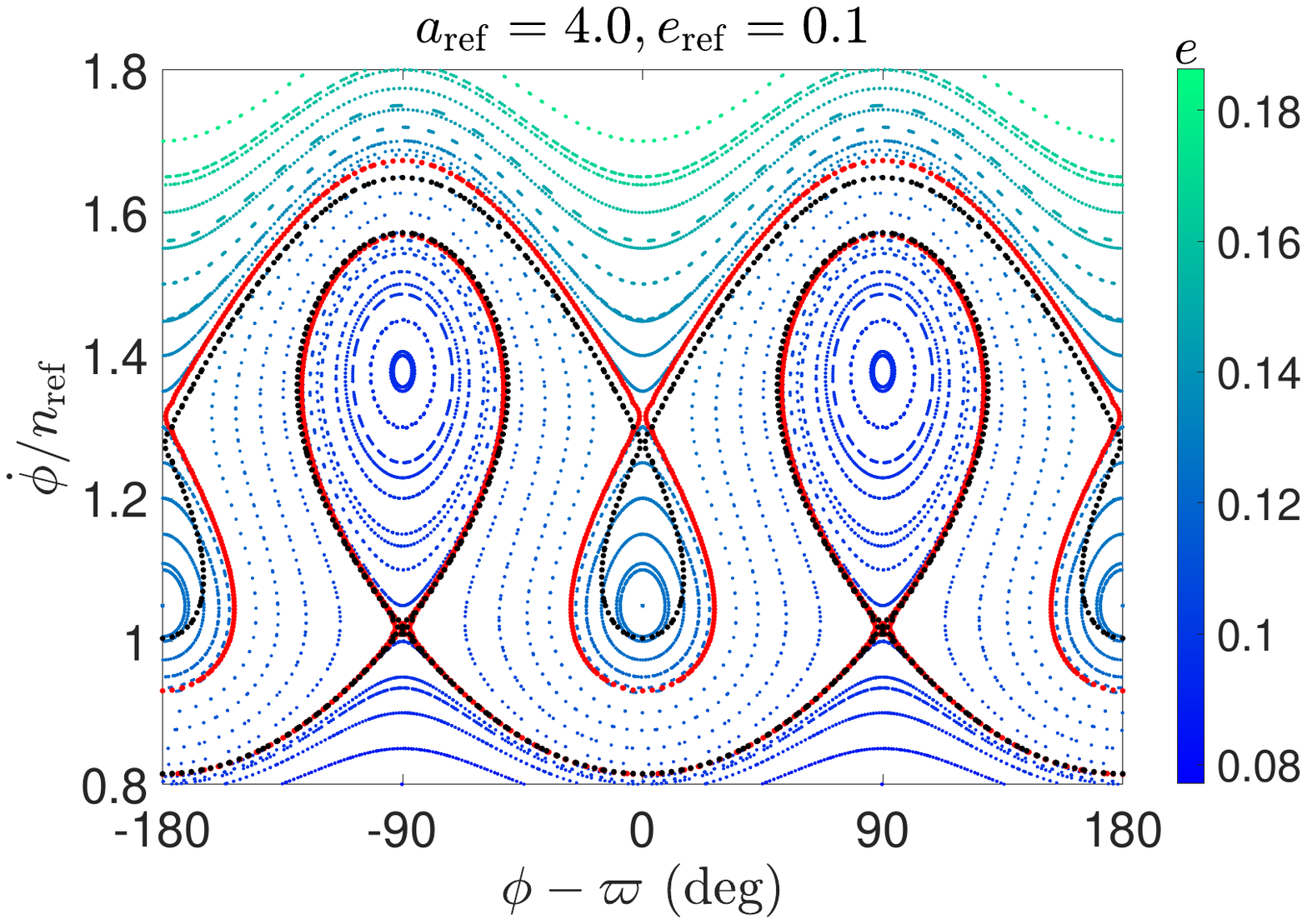}
\caption{Comparison of dynamical structures produced from the resonant Hamiltonian represented by Eq. (\ref{Eq14}) (black dots) and adiabatic approximation (red dots) for synchronous spin-orbit resonance. The structures shown in the background correspond to Poincar\'e surfaces of section.}
\label{Fig6_1}
\end{figure*}

In fact, it is still possible to discuss phase-space structures based on the resonant Hamiltonian by Eq. (\ref{Eq14}) for the synchronous 1:1 spin-orbit resonance. In Fig. \ref{Fig6_1}, we produce dynamical structures from the resonant Hamiltonian and from adiabatic approximation and compare analytical structures to the corresponding Poincar\'e sections. As desired, dynamical structures produced from adiabatic approximation have better agreement with numerical structures. 

\subsection{Distribution of spin equilibrium}
\label{Sect6-3}

\begin{figure*}
\centering
\includegraphics[width=0.49\columnwidth]{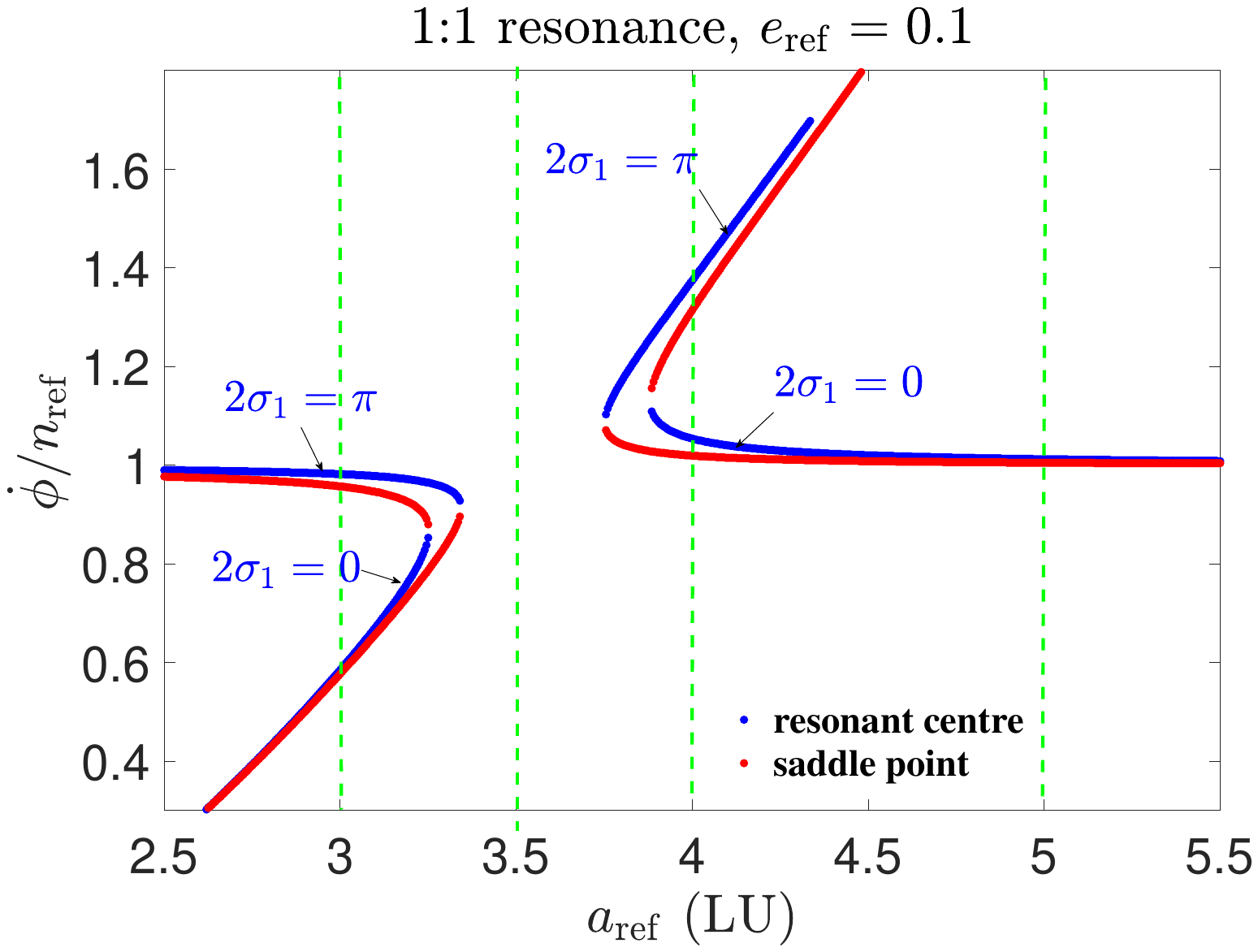}
\includegraphics[width=0.49\columnwidth]{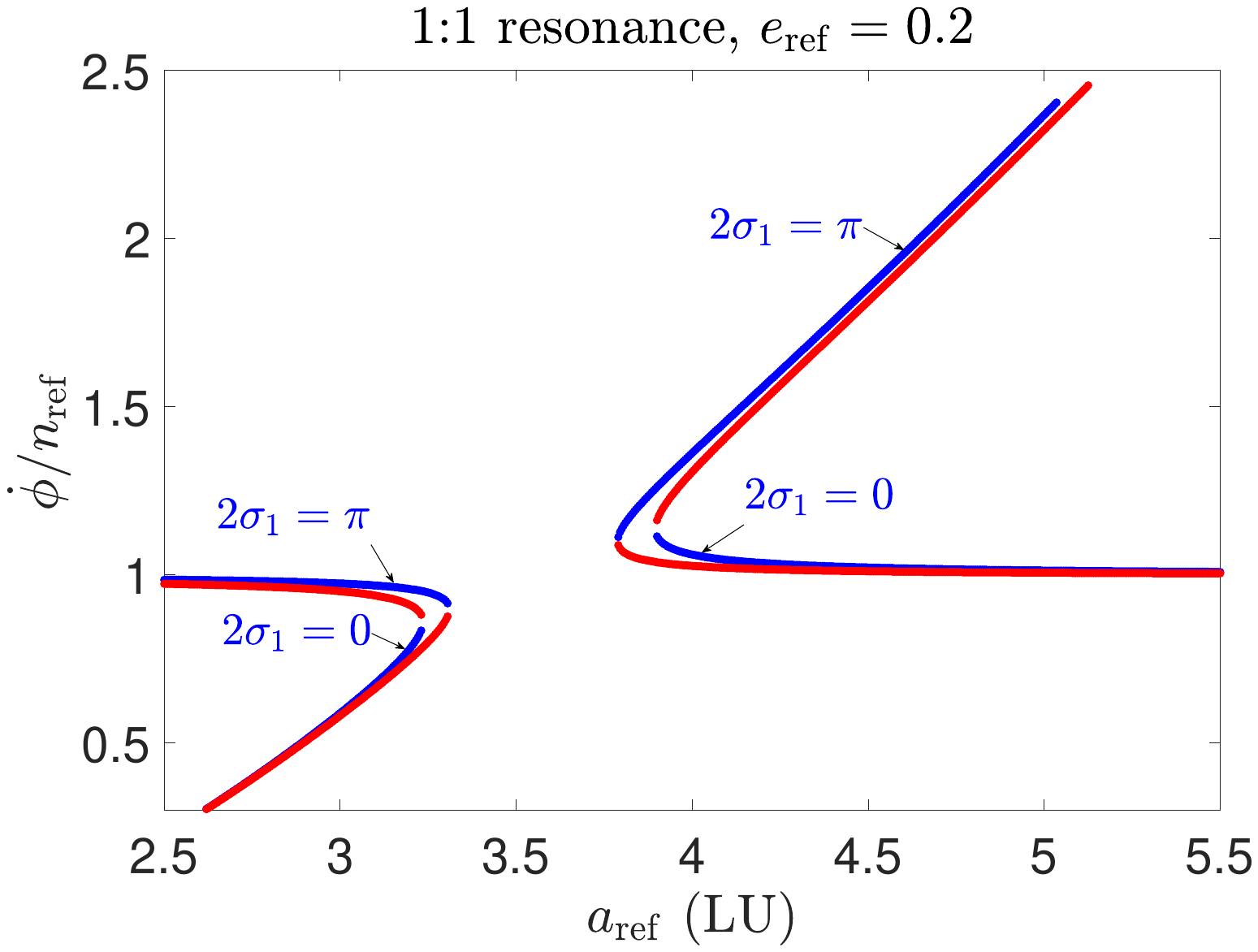}\\
\includegraphics[width=0.49\columnwidth]{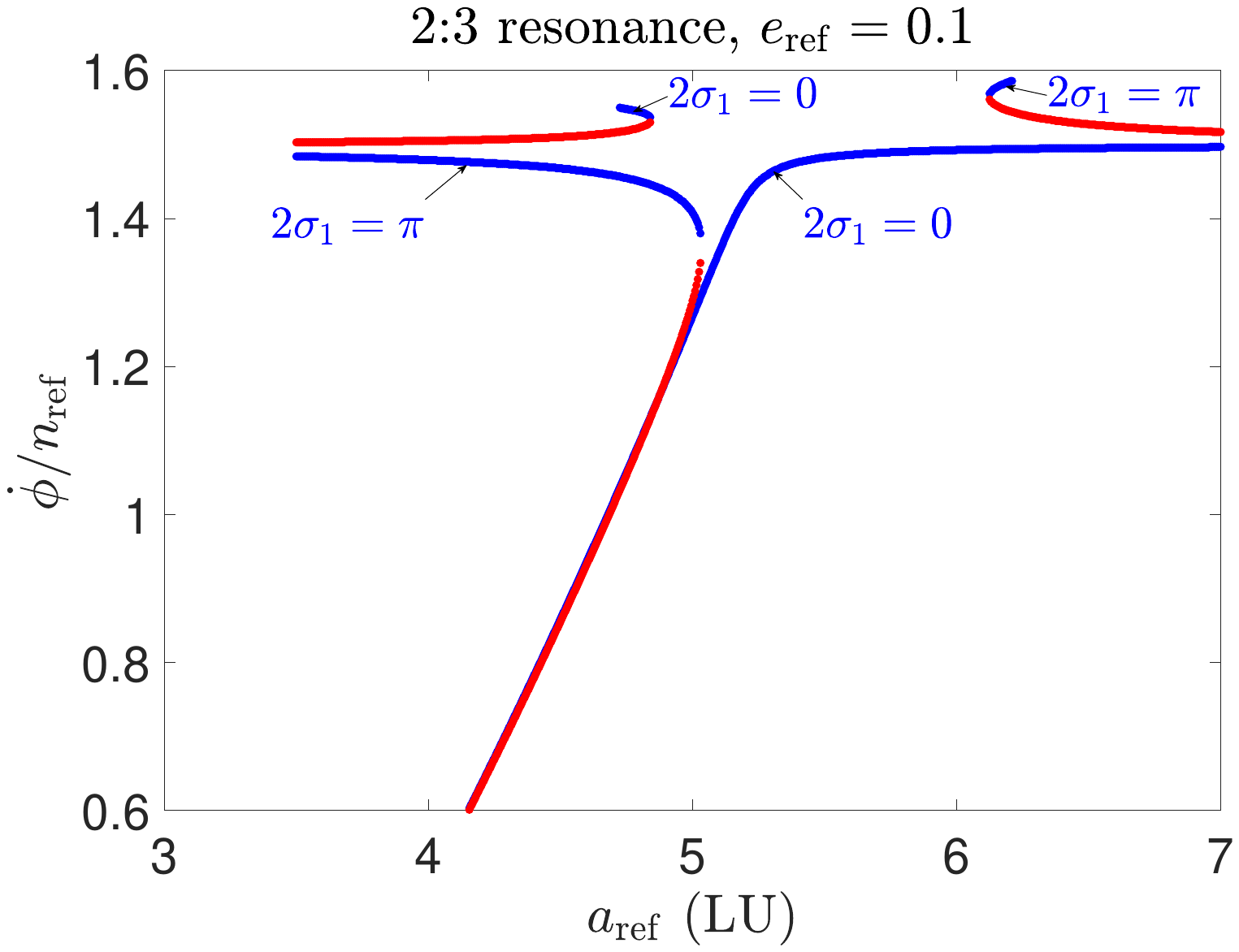}
\includegraphics[width=0.49\columnwidth]{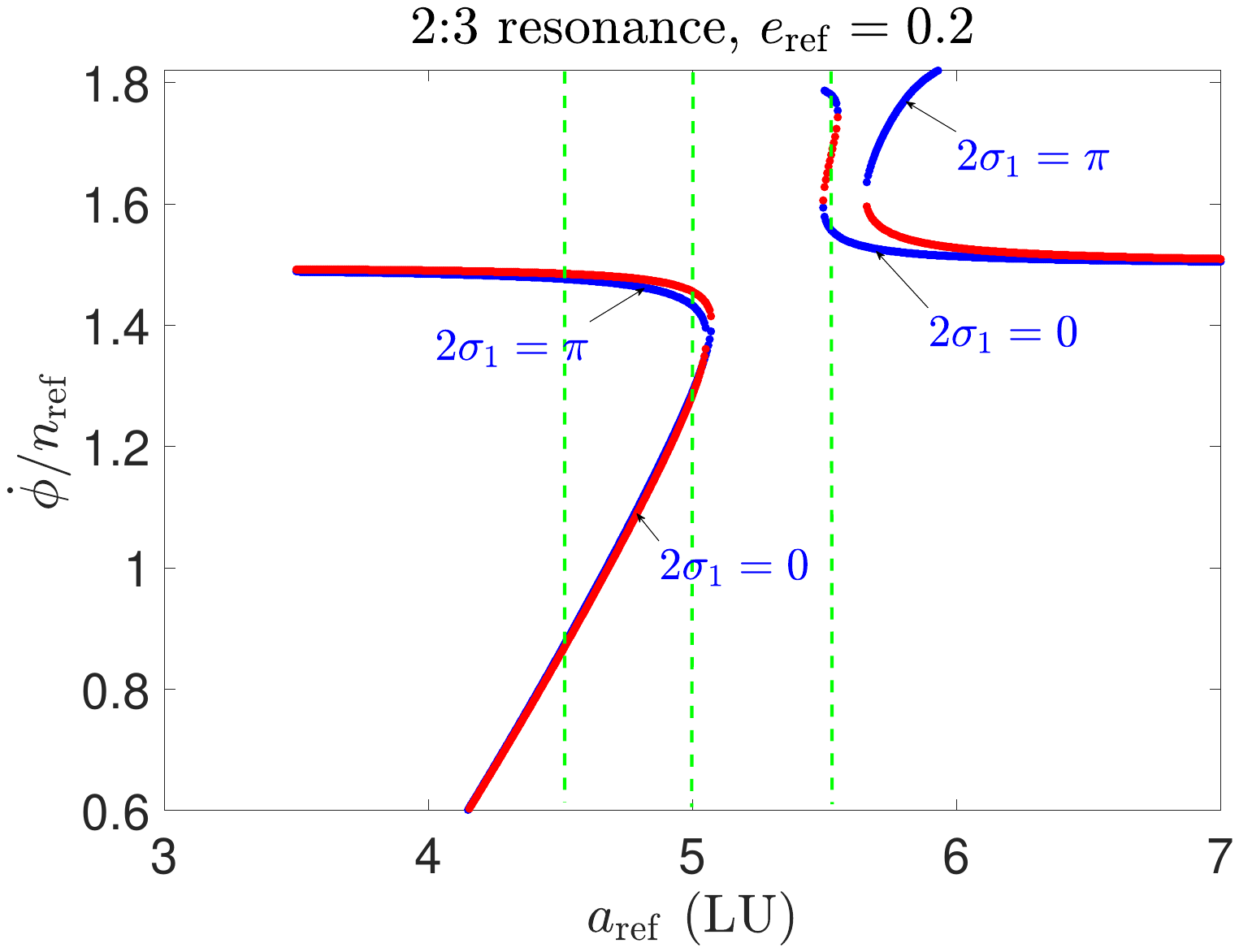}
\caption{Characteristic curves of resonant centres (blue dots) and saddle points (red dots) in the $(a_{\rm ref},\dot\phi)$ space for the synchronous resonance (\textit{top panels}) and the 2:3 resonance (\textit{bottom panels}). The vertical green dashed lines shown in the upper-left and bottom-right panels stand for the examples considered in Figs. \ref{Fig5} and \ref{Fig6}.}
\label{Fig7}
\end{figure*}

By analysing phase portraits produced from adiabatic approximation, we can further determine the location of libration centres and saddle points, as shown in Fig. \ref{Fig7}. Two cases of reference eccentricities at $e_{\rm ref} = 0.1$ and $e_{\rm ref} = 0.2$ are considered for the synchronous (1:1) resonance and 2:3 resonances. The resonant centres are shown in blue dots and saddle points are marked in red dots. The vertical dashed lines stand for the cases considered in Figs. \ref{Fig5} and \ref{Fig6}. According to the discussion made in previous section, the critical semimajor axis is $a_{c,1} = 3.53$ for the synchronous (1:1) resonance and $a_{c,2} = 5.31$ for the 2:3 resonance. 

For the synchronous resonance (see the top panels of Fig. \ref{Fig7}), the characteristic curves show that there is a gap of $a_{\rm ref}$ in the vicinity of $a_{c,1}$. In the space with $a_{\rm ref} < a_{c,1}$, there are two branches of libration centres: one is close to the nominal synchronous resonance with argument at $2\sigma_1 = \pi$ and the other branch with argument at $2\sigma_1 = 0$ occupies a lower-$\dot\phi$ space. In the space with $a_{\rm ref} > a_{c,1}$, there are also two branches of libration centres: one is close to the nominal synchronous resonance with argument at $2\sigma_1 = 0$ and the other branch with argument at $2\sigma_1 = \pi$ occupies a higher-$\dot\phi$ space. When the reference semimajor axis $a_{\rm ref}$ is higher (or lower) than a certain value, the high-$\dot\phi$ (or low-$\dot\phi$) branch will disappear. 

For the 2:3 resonance (see the bottom panels of Fig. \ref{Fig7}), there is no gap for the characteristic curves in the case of $e_{\rm ref}=0.1$ and there is a gap in the vicinity of $a_{c,2}$ for the case of $e_{\rm ref}=0.2$. Similar to the case of synchronous resonance, the high-$\dot\phi$ (or low-$\dot\phi$) branch may disappear if the reference semimajor axis $a_{\rm ref}$ is higher (or smaller) than a certain value.

\section{Conclusions}
\label{Sect7}

In this work, spin-orbit resonances of the primary body are studied in a binary asteroid system, composed of a triaxial ellipsoidal primary and a spherical secondary. The Hamiltonian is expanded up to a high order in eccentricity based on elliptic expansions. The resulting Hamiltonian determines a 2-DOF dynamical model, depending on the total angular momentum. We introduce two parameters including the reference semimajor axis $a_{\rm ref}$ and the reference eccentricity $e_{\rm ref}$ to characteristic the total angular momentum $G_{\rm tot}$ and the Hamiltonian ${\cal H}$.

Numerical simulations show that (a) the dynamical structures of spin-orbit resonances are distinctly different from the classical structures under the spin-orbit problem for the secondary body and (b) the main structures may significantly change at different reference semimajor axes.

According to pendulum approximations, there are three critical semimajor axes (denoted by $a_{c,1}$ for the synchronous resonance, $a_{c,2}$ for the 2:3 resonance and $a_{c,3}$ for the 2:1 resonance), below and above which the dynamical structures of spin-orbit resonance are totally different. However, it is observed that pendulum approximations of Hamiltonian fail to describe dynamical structures of spin-orbit resonances when the semimajor axis is not far from the critical value.

To address this problem, adiabatic approximation is used to formulate a `general' resonant Hamiltonian for describing spin-orbit coupling. In particular, an adiabatic invariant denoted by $J$ is introduced to describe the resonant dynamics. The conserved quantities including the adiabatic invariant $J$ and the Hamiltonian ${\cal H}$ make the original 2-DOF dynamical model be integrable. As a result, the global structures in phase space can be explored by analysing phase portraits. Results show that there is an excellent agreement between analytical structures arising in phase portraits and numerical structures arising in Poincar\'e sections. By analysing phase portraits, it is possible to produce characteristic curves of resonant centre and saddle points, which can be used to predict the variations of dynamical structures when the reference semimajor axis is changed.

Regarding the spin-orbit coupling problem, \citet{hou2017note} arrived at an intriguing conclusion that the centre of spin-orbit resonance may change when the mass ratio and mutual distance are varied to alternate the sign of the index $S$ ($S=S_1$ for the synchronous resonance, $S=S_2$ for the 3:2 resonance and $S=S_3$ for the 1:2 resonance). In \citet{jafari2023surfing} the author pointed out the change of libration centre is physically impossible through an alternation of the orbital semimajor axis (please see point (4) in their introduction part). Through our analysis, we observe that, for the secondary body in a binary asteroid system, it is indeed physically impossible to change the libration centre through an alternation of semimajor axis. This is because, for the secondary, the sign of $S_{1,2,3}$ remains unchanged at arbitrarily allowed semimajor axes. However, for the primary body, it becomes possible to alternate the resonance centre by changing the semimajor axis when the primary-to-secondary mass ratio is greater than unity.

\backmatter





\bmhead{Acknowledgements}
The author wishes to thank Prof. Xiyun Hou for helpful dicussions on numerical structures arising in Poincar\'e sections, Dr. Jafari-Nadoushan for confirming the expression of mutual gravitational potential for spin-orbit coupling problem and Dr. Shunjing Zhao for his help in deriving high-order spherical harmonic coefficients showin in Appendix. This work is financially supported by the National Natural Science Foundation of China (Nos. 12073011 and 12233003) and the National Key R\&D Program of China (No. 2019YFA0706601).

\section*{Declarations}
The authors declare that they have no Conflict of interest.

\begin{appendices}

\section{High-order dynamical model}
\label{A1}
In the spin-orbit coupling problem, the primary is of ellipsoid. Truncating the mutual potential at the 8th order in $(a_p/r)$, we can write the Hamiltonian governing the evolution of rotation and translation in the following form:
\begin{equation}\label{EqA1}
\begin{aligned}
{\cal H} =& \frac{1}{{2m}}p_r^2 - \frac{{{\cal G}{m_p}{m_s}}}{r} + \left( {\frac{1}{{2m{r^2}}} + \frac{1}{{2{I_3}}}} \right)p_\psi ^2 + \frac{{{G^2_{\rm tot}}}}{{2{I_3}}} + {G_{\rm tot}}\frac{{{p_\psi }}}{{{I_3}}}\\
 &- {\cal G}{m_p}{m_s}\left[ { - \frac{{a_p^2{C_{20}}}}{{2{r^3}}} + \frac{{3a_p^4{C_{40}}}}{{8{r^5}}} - \frac{{5a_p^6{C_{60}}}}{{16{r^7}}} + \frac{{35a_p^8{C_{80}}}}{{128{r^9}}}} \right.\\
 &+ \left( {\frac{{3a_p^2{C_{22}}}}{{{r^3}}} - \frac{{15a_p^4{C_{42}}}}{{2{r^5}}} + \frac{{105a_p^6{C_{62}}}}{{8{r^7}}} - \frac{{315a_p^8{C_{82}}}}{{16{r^9}}}} \right)\cos 2\psi \\
 &+ \left( {\frac{{105a_p^4{C_{44}}}}{{{r^5}}} - \frac{{945a_p^6{C_{64}}}}{{2{r^7}}} + \frac{{10395a_p^8{C_{84}}}}{{8{r^9}}}} \right)\cos 4\psi \\
 &+ \left( {\frac{{10295a_p^6{C_{66}}}}{{{r^7}}} - \frac{{135135a_p^8{C_{86}}}}{{2{r^9}}}} \right)\cos 6\psi \\
&\left. { + \frac{{2027025a_p^8{C_{88}}}}{{{r^9}}}\cos 8\psi } \right]
\end{aligned}
\end{equation}
where the second- and fourth-order spherical harmonic coefficients are given in Sect. \ref{Sect2}. Here we only provide the sixth- and eighth-order coefficients as functions of $C_{20}$ and $C_{22}$ in the following form \citep{balmino1994gravitational}:
\begin{equation}\label{EqA2}
\begin{aligned}
{C_{60}} =& \frac{{125}}{7}\left( {\frac{1}{3}C_{20}^2 + 2C_{22}^2} \right){C_{20}}\\
{C_{62}} =& \frac{{25}}{{21}}\left( {C_{20}^2 + C_{22}^2} \right){C_{22}}\\
{C_{64}} =& \frac{{25}}{{252}}{C_{20}}C_{22}^2,\quad {C_{66}} = \frac{{25}}{{1512}}C_{22}^3
\end{aligned}
\end{equation}
and 
\begin{equation}\label{EqA3}
\begin{aligned}
{C_{80}} &= \frac{{625}}{{11}}\left[ {\frac{1}{3}C_{20}^4 + 2C_{22}^2\left( {2C_{20}^2 + C_{22}^2} \right)} \right]\\
{C_{82}} &= \frac{{625}}{{77}}\left( {\frac{1}{3}C_{20}^2 + C_{22}^2} \right){C_{20}}{C_{22}}\\
{C_{84}} &= \frac{{125}}{{462}}\left( {\frac{1}{2}C_{20}^2 + \frac{1}{3}C_{22}^2} \right)C_{22}^2\\
{C_{86}} &= \frac{{125}}{{16632}}{C_{20}}C_{22}^3,\quad {C_{88}} = \frac{{125}}{{133056}}C_{22}^4
\end{aligned}
\end{equation}
Please refer to Table \ref{TabA1} for the spherical harmonic coefficients for the ellipsoid-shaped primary with semimajor axes $a_p = 1.0$, $b_p=0.95$ and $c_p = 0.85$. 

\begin{table*}
\footnotesize
\centering
\caption{Spherical harmonic coefficients for the ellipsoid primary with semimajor axes $a_p = 1.0$, $b_p=0.95$ and $c_p = 0.85$.}
\begin{tabular*}{\textwidth}{@{\extracolsep{\fill}}lrrrrr@{\extracolsep{\fill}}}
\toprule
{Coefficients}&{$m=0$}&{$m=1$}&{$m=2$}&{$m=3$}&{$m=4$}\\
\midrule
$C_{2,2m}$&$-4.575 \times 10^{-2}$&$4.875 \times 10^{-3}$\\
$C_{4,2m}$&$4.587 \times 10^{-3}$&$-1.593 \times 10^{-4}$&$4.244 \times 10^{-6}$\\
$C_{6,2m}$&$-6.088 \times 10^{-4}$&$1.229 \times 10^{-5}$&$-1.079 \times 10^{-7}$&$1.916 \times 10^{-9}$\\
$C_{8,2m}$&$9.434 \times 10^{-5}$&$-1.306 \times 10^{-6}$&$6.780 \times 10^{-9}$&$-3.984 \times 10^{-11}$&$5.306 \times 10^{-13}$\\
\bottomrule
\end{tabular*}
\label{TabA1}
\end{table*}

\begin{figure*}
\centering
\includegraphics[width=0.49\columnwidth]{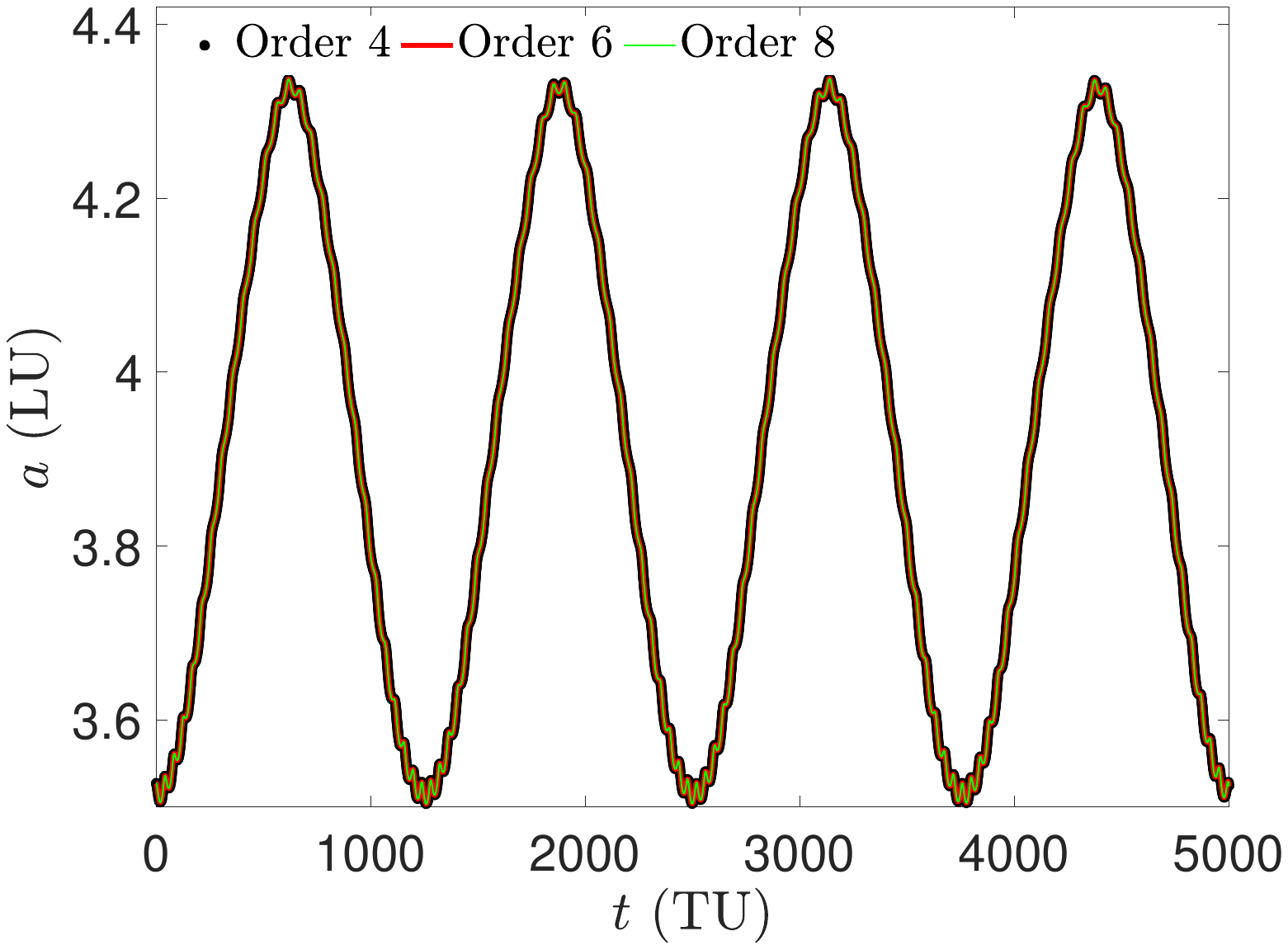}
\includegraphics[width=0.49\columnwidth]{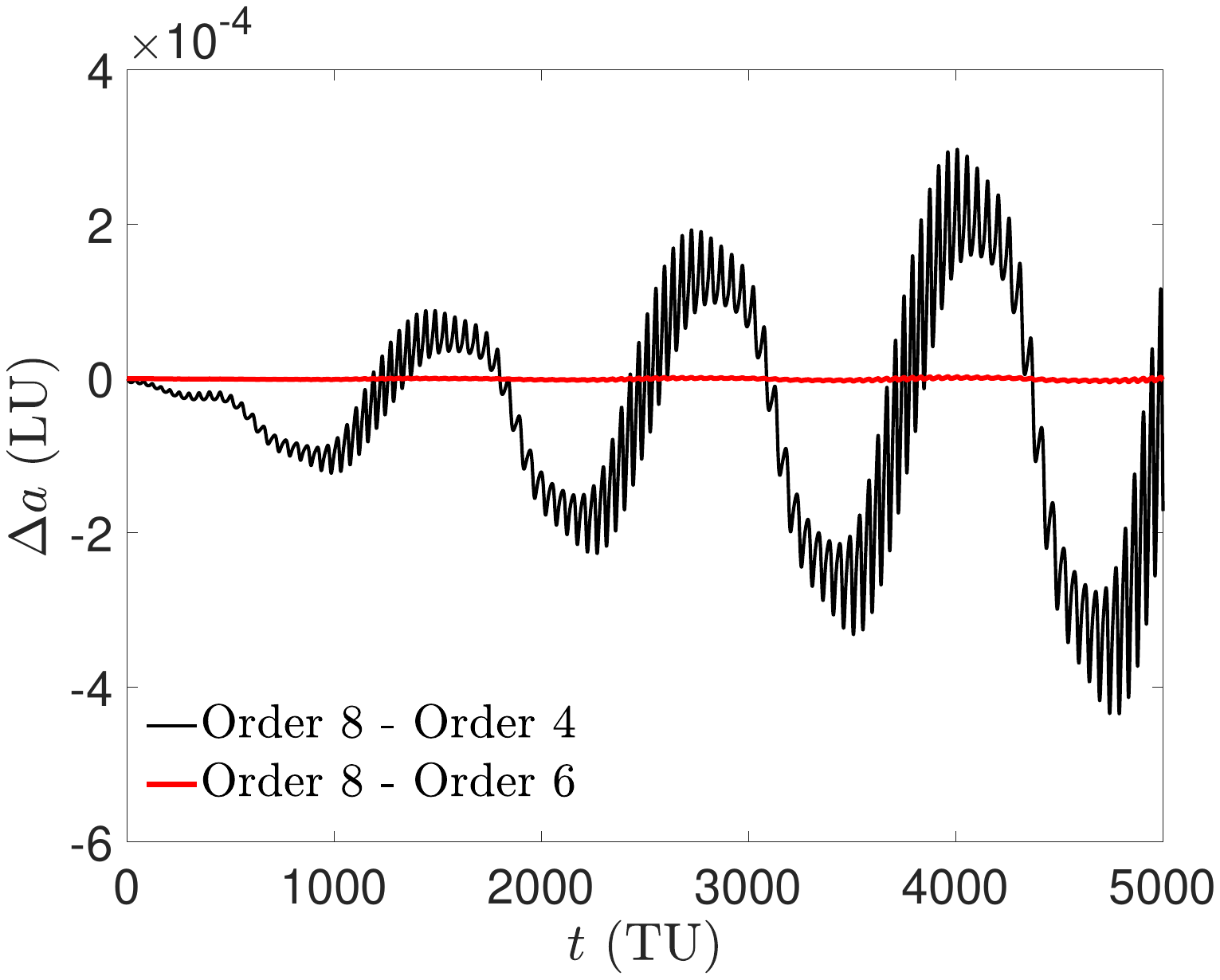}\\
\includegraphics[width=0.49\columnwidth]{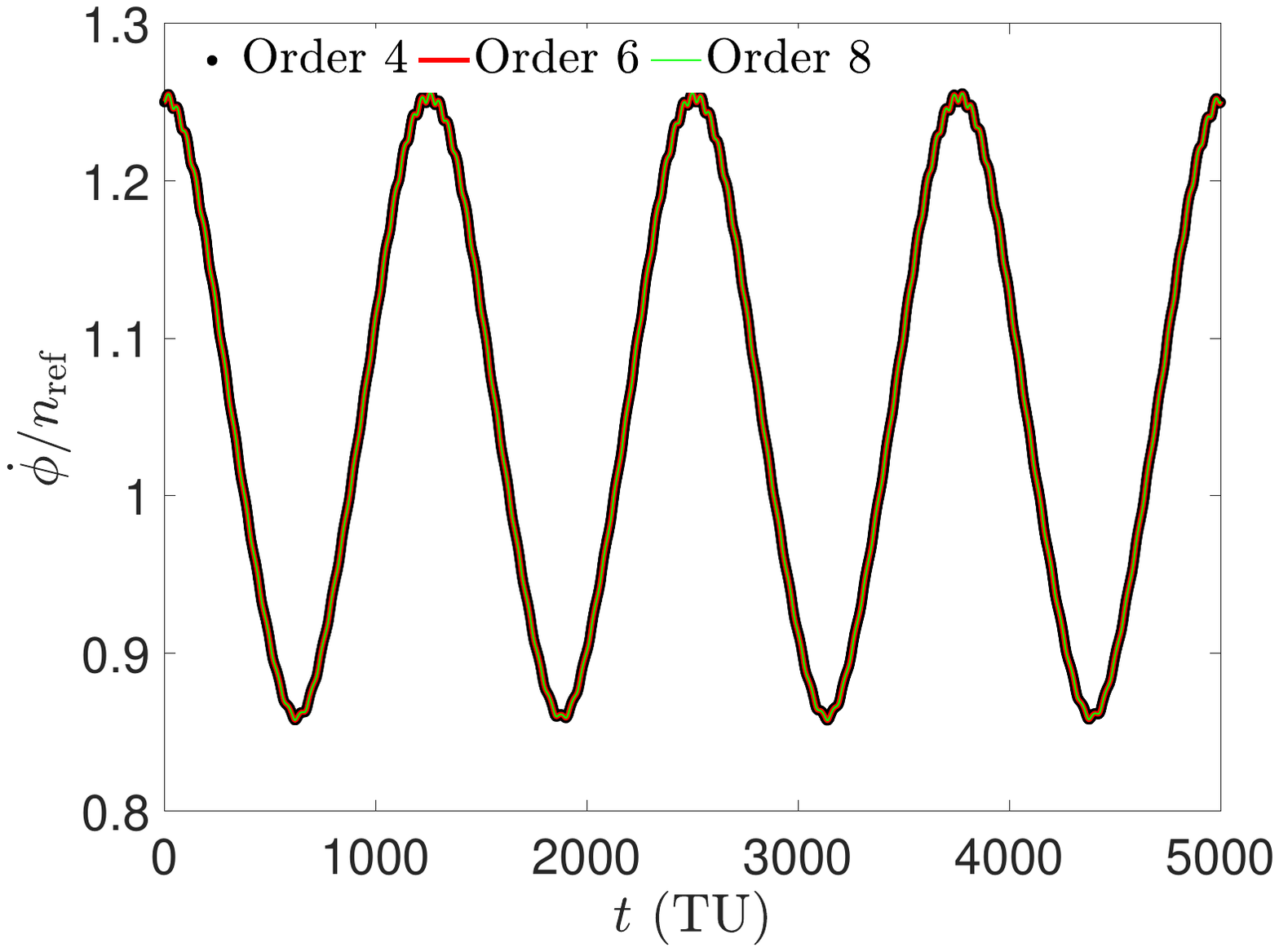}
\includegraphics[width=0.49\columnwidth]{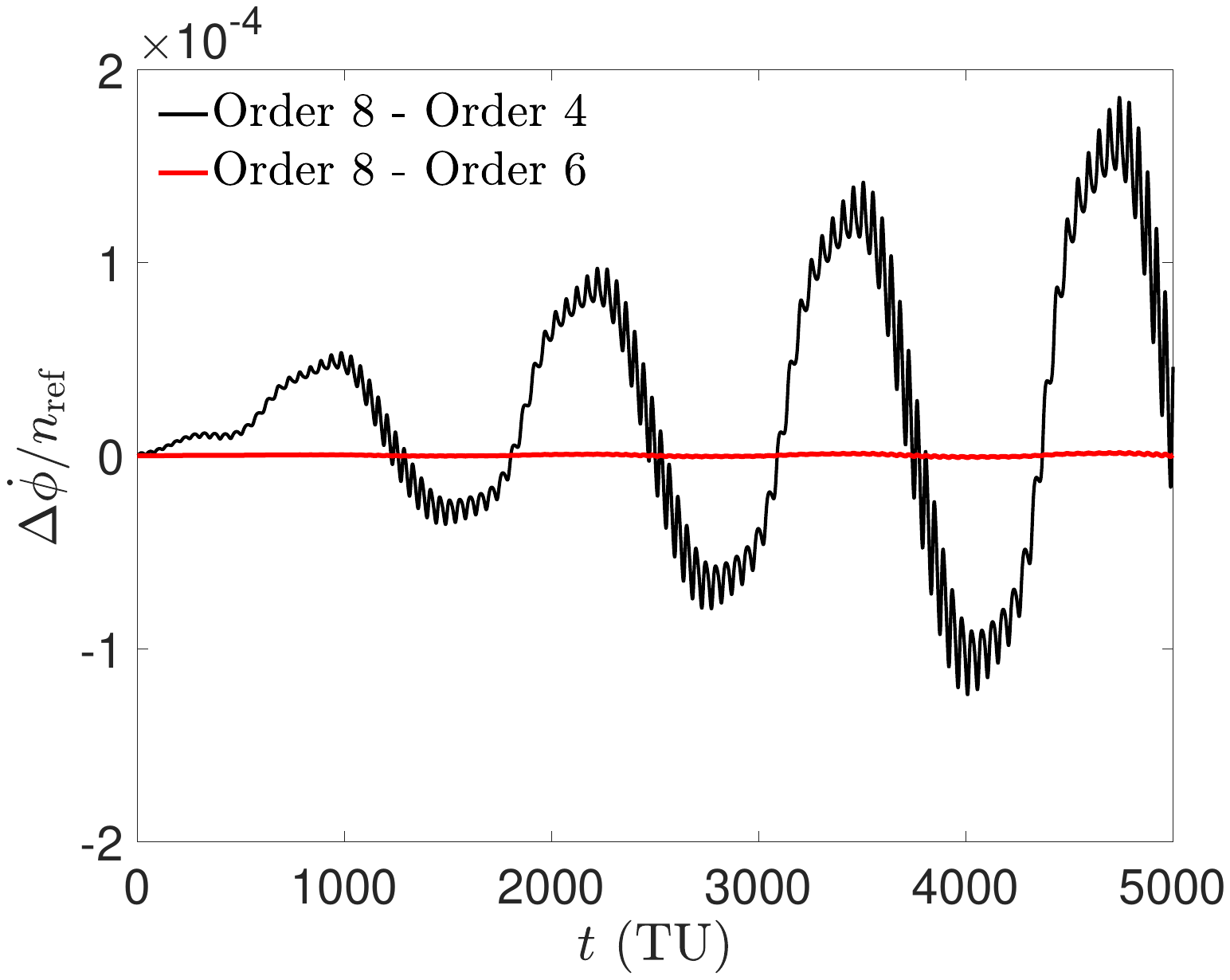}
\caption{Comparison of the numerical trajectories propagated under the Hamiltonian model truncated at different orders in terms of $a_p/r$. The initial condition is the same as the one taken in Fig. \ref{Fig2}.}
\label{FigA1}
\end{figure*}

Replacing the Hamiltonian (\ref{A1}) in Eq. (\ref{Eq2}) leads to the equations of motion for the spin-orbit coupling problem. Then, we integrate the equations of motion with the same initial conditions as the ones adopted by Fig. \ref{Fig2}. Numerical trajectories propagated under the dynamical models truncated at different orders in $a_p/r$ are compared in Fig. \ref{FigA1}. It shows that the deviation between 4th-order and 8th-order models is of $1.0 \times 10^{-4}$, meaning that it is accurate enough to truncate the Hamiltonian at the fourth order in $a_p/r$.

\end{appendices}


\bibliography{mybib}

\begin{thebibliography}{41}
\providecommand{\natexlab}[1]{#1}
\providecommand{\url}[1]{{#1}}
\providecommand{\urlprefix}{URL }
\providecommand{\doi}[1]{\url{https://doi.org/#1}}
\providecommand{\eprint}[2][]{\url{#2}}
 \bibcommenthead

\bibitem[{Balmino(1994)}]{balmino1994gravitational}
Balmino G (1994) Gravitational potential harmonics from the shape of an
  homogeneous body. Celest Mech Dyn Astron 60:331--364.
  \doi{10.1007/BF00691901}

\bibitem[{Celletti(1990{\natexlab{a}})}]{Celletti1990AnalysisII}
Celletti A (1990{\natexlab{a}}) Analysis of resonances in the spin-orbit
  problem in celestial mechanics: Higher order resonances and some numerical
  experiments (part ii). Z Ang Math Phys 41(4):453--479.
  \doi{10.1007/BF00945951}

\bibitem[{Celletti(1990{\natexlab{b}})}]{Celletti1990AnalysisI}
Celletti A (1990{\natexlab{b}}) Analysis of resonances in the spin-orbit
  problem in celestial mechanics: The synchronous resonance (part i). Z Ang
  Math Phys 41(2):174--204. \doi{10.1007/BF00945107}

\bibitem[{Celletti and Chierchia(2000)}]{celletti2000hamiltonian}
Celletti A, Chierchia L (2000) Hamiltonian stability of spin--orbit resonances
  in celestial mechanics. Celest Mech Dyn Astron 76:229--240.
  \doi{10.1023/A:1008341317257}

\bibitem[{Chirikov(1979)}]{chirikov1979universal}
Chirikov BV (1979) A universal instability of many-dimensional oscillator
  systems. Physics reports 52(5):263--379. \doi{10.1016/0370-1573(79)90023-1}

\bibitem[{Correia et~al(2015)Correia, Leleu, Rambaux, and
  Robutel}]{correia2015spin}
Correia AC, Leleu A, Rambaux N, et~al (2015) Spin-orbit coupling and chaotic
  rotation for circumbinary bodies-application to the small satellites of the
  pluto-charon system. A\&A 580:L14. \doi{10.1051/0004-6361/201526800}

\bibitem[{Deprit(1969)}]{1969Canonical}
Deprit A (1969) Canonical transformations depending on a small parameter.
  Celest Mech Dyn Astron 1(1):12--30. \doi{10.1007/BF01230629}

\bibitem[{Flynn and Saha(2005)}]{flynn2005second}
Flynn AE, Saha P (2005) Second-order perturbation theory for spin-orbit
  resonances. Astron J 130(1):295. \doi{10.1086/430410}

\bibitem[{Gkolias et~al(2016)Gkolias, Celletti, Efthymiopoulos, and
  Pucacco}]{gkolias2016theory}
Gkolias I, Celletti A, Efthymiopoulos C, et~al (2016) The theory of secondary
  resonances in the spin--orbit problem. Mon Not R Astron Soc
  459(2):1327--1339. \doi{10.1093/mnras/stw752}

\bibitem[{Gkolias et~al(2019)Gkolias, Efthymiopoulos, Celletti, and
  Pucacco}]{gkolias2019accurate}
Gkolias I, Efthymiopoulos C, Celletti A, et~al (2019) Accurate modelling of the
  low-order secondary resonances in the spin-orbit problem. Commun Nonlinear
  Sci 77:181--202. \doi{10.1016/j.cnsns.2019.04.015}

\bibitem[{Goldreich and Peale(1966)}]{Goldreich1966Spin}
Goldreich P, Peale S (1966) Spin-orbit coupling in the solar system. Astron J
  71(6):425

\bibitem[{Henrard(1990)}]{henrard1990semi}
Henrard J (1990) A semi-numerical perturbation method for separable hamiltonian
  systems. Celest Mech Dyn Astron 49(1):43--67. \doi{10.1007/BF00048581}

\bibitem[{Henrard and Caranicolas(1989)}]{henrard1989motion}
Henrard J, Caranicolas N (1989) Motion near the 3/1 resonance of the planar
  elliptic restricted three body problem. Celest Mech Dyn Astron 47:99--121.
  \doi{doi.org/10.1007/BF00051201}

\bibitem[{Henrard and Lema{\^\i}tre(1987)}]{henrard1987perturbative}
Henrard J, Lema{\^\i}tre A (1987) A perturbative treatment of the 21 jovian
  resonance. Icarus 69(2):266--279. \doi{10.1016/0019-1035(87)90105-9}

\bibitem[{Hori(1966)}]{1966Theory}
Hori GI (1966) Theory of general perturbation with unspecified canonical
  variable. Publ Astron Soc Jpn 18(4):287

\bibitem[{Hou and Xin(2017)}]{hou2017note}
Hou X, Xin X (2017) A note on the spin--orbit, spin--spin, and
  spin--orbit--spin resonances in the binary minor planet system. Astron J
  154(6):257. \doi{10.3847/1538-3881/aa96ab}

\bibitem[{Hughes(1981)}]{hughes1981computation}
Hughes S (1981) The computation of tables of hansen coefficients. Celest Mech
  25:101--107. \doi{10.1007/BF01301812}

\bibitem[{Jafari-Nadoushan(2023)}]{jafari2023surfing}
Jafari-Nadoushan M (2023) Surfing in the phase space of spin--orbit coupling in
  binary asteroid systems. Mon Not R Astron Soc 520(3):3514--3528.
  \doi{10.1093/mnras/stac3624}

\bibitem[{Jafari-Nadoushan and Assadian(2015)}]{jafari2015widespread}
Jafari-Nadoushan M, Assadian N (2015) Widespread chaos in rotation of the
  secondary asteroid in a binary system. Nonlinear Dynam 81:2031--2042.
  \doi{10.1007/s11071-015-2123-0}

\bibitem[{Jafari-Nadoushan and
  Assadian(2016{\natexlab{a}})}]{jafari2016chirikov}
Jafari-Nadoushan M, Assadian N (2016{\natexlab{a}}) Chirikov diffusion in the
  sphere--ellipsoid binary asteroids. Nonlinear Dynam 85:1837--1848.
  \doi{10.1007/s11071-016-2799-9}

\bibitem[{Jafari-Nadoushan and
  Assadian(2016{\natexlab{b}})}]{nadoushan2016geography}
Jafari-Nadoushan M, Assadian N (2016{\natexlab{b}}) Geography of the rotational
  resonances and their stability in the ellipsoidal full two body problem.
  Icarus 265:175--186. \doi{10.1016/j.icarus.2015.10.011}

\bibitem[{Lei(2022)}]{lei2022systematic}
Lei H (2022) A systematic study about orbit flips of test particles caused by
  eccentric von zeipel--lidov--kozai effects. Astron J 163(5):214.
  \doi{10.3847/1538-3881/ac5fa8}

\bibitem[{Lei et~al(2022)Lei, Li, Huang, and Li}]{lei2022zeipel}
Lei H, Li J, Huang X, et~al (2022) The von zeipel--lidov--kozai effect inside
  mean motion resonances with applications to trans-neptunian objects. Astron J
  164(3):74. \doi{10.3847/1538-3881/ac7c6a}

\bibitem[{Lemaitre et~al(2006)Lemaitre, D’Hoedt, and Rambaux}]{lemaitre20063}
Lemaitre A, D’Hoedt S, Rambaux N (2006) The 3: 2 spin-orbit resonant motion
  of mercury. Celest Mech Dyn Astron 95:213--224.
  \doi{10.1007/s10569-006-9032-y}

\bibitem[{Morbidelli(2002)}]{morbidelli2002modern}
Morbidelli A (2002) Modern celestial mechanics: aspects of solar system
  dynamics. Taylor \& Francis, London and New York

\bibitem[{Murray and Dermott(1999)}]{murray1999solar}
Murray CD, Dermott SF (1999) Solar system dynamics. Cambridge university press

\bibitem[{Naidu and Margot(2015)}]{naidu2015near}
Naidu SP, Margot JL (2015) Near-earth asteroid satellite spins under
  spin--orbit coupling. Astron J 149(2):80. \doi{10.1088/0004-6256/149/2/80}

\bibitem[{Peale and Gold(1965)}]{peale1965rotation}
Peale S, Gold T (1965) Rotation of the planet mercury. Nature 206:1240--1241.
  \doi{10.1038/2061240b0}

\bibitem[{Peale(1969)}]{peale1969generalized}
Peale SJ (1969) Generalized cassini's laws. Astron J 74:483

\bibitem[{Peale(1977)}]{peale1977rotation}
Peale SJ (1977) Rotation histories of the natural satellites. In: Planetary
  satellites, p~87

\bibitem[{Pravec et~al(2016)Pravec, Scheirich, Ku{\v{s}}nir{\'a}k, Hornoch,
  Gal{\'a}d, Naidu, Pray, Vil{\'a}gi, Gajdo{\v{s}}, Korno{\v{s}}
  et~al}]{pravec2016binary}
Pravec P, Scheirich P, Ku{\v{s}}nir{\'a}k P, et~al (2016) Binary asteroid
  population. 3. secondary rotations and elongations. Icarus 267:267--295.
  \doi{10.1016/j.icarus.2015.12.019}

\bibitem[{Pravec et~al(2019)Pravec, Fatka, Vokrouhlick{\`y}, Scheirich,
  {\v{D}}urech, Scheeres, Ku{\v{s}}nir{\'a}k, Hornoch, Gal{\'a}d, Pray
  et~al}]{pravec2019asteroid}
Pravec P, Fatka P, Vokrouhlick{\`y} D, et~al (2019) Asteroid pairs: a complex
  picture. Icarus 333:429--463. \doi{10.1016/j.icarus.2019.05.014}

\bibitem[{Saillenfest(2020)}]{saillenfest2020long}
Saillenfest M (2020) Long-term orbital dynamics of trans-neptunian objects.
  Celest Mech Dyn Astron 132:1--45. \doi{10.1007/s10569-020-9954-9}

\bibitem[{Saillenfest et~al(2016)Saillenfest, Fouchard, Tommei, and
  Valsecchi}]{saillenfest2016long}
Saillenfest M, Fouchard M, Tommei G, et~al (2016) Long-term dynamics beyond
  neptune: secular models to study the regular motions. Celest Mech Dyn Astron
  126:369--403. \doi{10.1007/s10569-016-9700-5}

\bibitem[{Scheeres et~al(2006)Scheeres, Fahnestock, Ostro, Margot, Benner,
  Broschart, Bellerose, Giorgini, Nolan, Magri et~al}]{scheeres2006dynamical}
Scheeres DJ, Fahnestock EG, Ostro SJ, et~al (2006) Dynamical configuration of
  binary near-earth asteroid (66391) 1999 kw4. Science 314(5803):1280--1283.
  \doi{10.1126/science.1133599}

\bibitem[{Wang and Hou(2020)}]{wang2020secondary}
Wang H, Hou X (2020) On the secondary’s rotation in a synchronous binary
  asteroid. Mon Not R Astron Soc 493(1):171--183. \doi{10.1093/mnras/staa133}

\bibitem[{Wang et~al(2022)Wang, Xin, Hou, and Feng}]{wang2022stability}
Wang H, Xin X, Hou X, et~al (2022) Stability of the planar synchronous full
  two-body problem—the approach of periodic orbits. Commun Nonlinear Sci
  114:106638. \doi{10.1016/j.cnsns.2022.106638}

\bibitem[{Wisdom(1985)}]{wisdom1985perturbative}
Wisdom J (1985) A perturbative treatment of motion near the 3/1
  commensurability. Icarus 63(2):272--289. \doi{10.1016/0019-1035(85)90011-9}

\bibitem[{Wisdom(2004)}]{wisdom2004spin}
Wisdom J (2004) Spin-orbit secondary resonance dynamics of enceladus. Astron J
  128(1):484. \doi{10.1086/421360}

\bibitem[{Wisdom et~al(1984)Wisdom, Peale, and Mignard}]{wisdom1984chaotic}
Wisdom J, Peale SJ, Mignard F (1984) The chaotic rotation of hyperion. Icarus
  58(2):137--152. \doi{10.1016/0019-1035(84)90032-0}

\bibitem[{Yokoyama(1996)}]{yokoyama1996simple}
Yokoyama T (1996) A simple generalization of wisdom's perturbative method.
  Celest Mech Dyn Astron 64:243--260. \doi{10.1007/BF00728350}

\end{thebibliography}

\end{document}